\documentclass[twocolumn,floats,floatfix,showpacs,prd,superscriptaddress,nofootinbib]{revtex4-2}

\usepackage{graphicx,epsfig}
\usepackage{amssymb,amsmath,amsthm,amsfonts}
\usepackage{bm}
\usepackage{amsmath}
\usepackage{comment}

\usepackage[inline]{enumitem}
\usepackage{tensor}
\usepackage[linktocpage]{hyperref}
\usepackage[caption=false]{subfig}
\usepackage[usenames,dvipsnames]{xcolor}
\usepackage{xspace}
\usepackage{url}
\usepackage[inline]{enumitem}

\usepackage{accents}

\usepackage[normalem]{ulem}

\newcommand{\Ord}[2]{\mathcal O \left(#1^{#2}\right)}

\def\newacronym#1#2#3{\gdef#1{\gdef#1{#2\xspace}#3 (#2)\xspace}}
\newacronym{\amr}{AMR}{adaptive mesh refinement}
\newacronym{\BSSN}{BSSN}{Baumgarte-Shapiro-Shibata-Nakamura}
\newacronym{\bh}{BH}{black hole}
\newacronym{\bbh}{BBH}{binary black hole}
\newacronym{\BL}{BL}{Boyer-Lindquist}
\newacronym{\dCS}{dCS}{dynamical Chern-Simons}
\newacronym{\EFT}{EFT}{effective field theory}
\newacronym{\EHT}{EHT}{Event Horizon Telescope}
\newacronym{\GR}{GR}{General Relativity}
\newacronym{\GW}{GW}{gravitational wave}
\newacronym{\LVK}{LVK}{LIGO-Virgo-KAGRA}
\newacronym{\QBS}{QBS}{quasibound state}
\newacronym{\sGB}{sGB}{scalar Gauss-Bonnet}
\newacronym{\RL}{RL}{refinement level}

\def\Canuda{\textsc{Canuda}\xspace}

\def\CanudadCS{\textsc{Canuda}-d\textsc{CS}\xspace}
\def\ETK{\textsc{Einstein Toolkit}\xspace}
\def\Carpet{\textsc{Carpet}\xspace}
\def\Cactus{\textsc{Cactus}\xspace}
\def\PyCactus{\textsc{PyCactus}\xspace}
\def\PostCactus{\textsc{PostCactus}\xspace}

\def\dif{\textrm{d}}
\def\p{\partial}

\def\rBL{r_{\rm{BL}}}
\def\rBLp{r_{\rm{BL,+}}}
\def\rBLm{r_{\rm{BL,-}}}
\def\rex{r_{\rm{ex}}}

\def\KTheta{K_{\rm \Theta}}
\def\Vth{V({\rm \Theta})}
\def\dVth{V'({\rm \Theta})}
\def\fth{f({\rm \Theta})}
\def\dfth{f'({\rm \Theta})}
\def\ddfth{f''({\rm \Theta})}
\def\Rtf{R^{\rm tf}}
\def\aCS{\alpha_{\rm CS}}
\def\aCSh{\hat{\alpha}}
\def\CS{\,^{\ast}\!RR}
\def\CCS{\mathcal{C}}
\def\dual{\,^{\ast}\!}

\def\Lie{\mathcal{L}}

\def\A{\mathcal{A}}

\def\E{\mathcal{E}}
\def\F{\mathcal{F}}

\def\M{\mathcal{M}}

\def\R{\mathcal{R}}
\def\S{\mathcal{S}}

\def\tgam{\tilde{\gamma}}
\def\tGam{\tilde{\Gamma}}
\def\teps{\tilde{\epsilon}}
\def\tA{\tilde{A}}
\def\tD{\tilde{D}}

\def\tE{\tilde{E}}
\def\tB{\tilde{B}}
\def\tF{\mathcal{\tilde{F}}}

\begin{document}

\title{Black Holes in Massive Dynamical Chern-Simons gravity: scalar hair and quasibound states at decoupling}

\author{Chloe Richards}\email{chloer3@illinois.edu}
\affiliation{Department of Physics and Illinois Center for Advanced Studies of the Universe, University of Illinois at Urbana-Champaign, Urbana, Illinois 61801, USA}

\author{Alexandru Dima}\email{adima@illinois.edu}
\affiliation{Department of Physics and Illinois Center for Advanced Studies of the Universe, University of Illinois at Urbana-Champaign, Urbana, Illinois 61801, USA}

\author{Helvi Witek}\email{hwitek@illinois.edu}
\affiliation{Department of Physics and Illinois Center for Advanced Studies of the Universe, University of Illinois at Urbana-Champaign, Urbana, Illinois 61801, USA}

\begin{abstract}

Black holes have a unique sensitivity to
the presence of ultralight matter fields
or modifications of the underlying theory of gravity.
In the present paper we combine both features by studying an ultralight, dynamical scalar field that is nonminimally coupled to
the gravitational Chern-Simons term.
In particular, we numerically simulate the evolution of such a scalar field around a rotating black hole in the decoupling approximation
and find a new kind of massive scalar hair anchored around the black hole.
In the proximity of the black hole,
the scalar exhibits the typical dipolar structure of hairy solutions in (massless) dynamical Chern-Simons gravity.
At larger distances,  the field transitions to an oscillating scalar cloud that is induced by the mass term.
Finally, we complement the time-domain results with a spectral analysis of the scalar field characteristic frequencies.
\end{abstract}

\maketitle
\tableofcontents

\section{Introduction}\label{sec:intro}

Although possibly being one of the most exotic predictions of \GR, a \bh is an object with a mathematically simple structure.
Famous uniqueness theorems~\cite{Israel:1967wq,Israel:1967za,Carter:1971zc,Wald:1971iw,Robinson:1975bv} state that
a \bh in \GR is completely determined
by the Kerr-Newman metric and
that it is parameterized by only three
numbers:
its mass, angular momentum and electromagnetic charge.
For \bh{s} in \GR,
coupled to ordinary matter,
no other charges (or ``hair'') are allowed~\cite{Bekenstein:1971hc,Bekenstein:1972ky,Pena:1997cy,Sotiriou:2011dz},
and it is hypothesized that astrophysical \bh{s} are uniquely described by the Kerr metric.
We are now in a unique position to test this Kerr hypothesis with astrophysical observations of \bh{s} such as
the shadow of supermassive \bh{s} by the \EHT collaboration~\cite{EventHorizonTelescope:2019dse,
EventHorizonTelescope:2022wkp,EventHorizonTelescope:2022xqj},
stellar-mass \bbh mergers by the \LVK collaboration~\cite{LIGOScientific:2016aoc,LIGOScientific:2021djp,LIGOScientific:2021sio},
or X-ray emission from accretion disks around \bh{s}~\cite{Bambi:2016sac}.
Moreover, \bh{s} provide a new channel to probe for
fundamental fields predicted in beyond Standard Model particle physics or modified theories of gravity,
whose presence would
disprove
the Kerr hypothesis.
That is, we can use astrophysical \bh{s} as cosmic particle detectors
to search for ultralight bosonic fields~\cite{Arvanitaki:2010sy,Kodama:2011zc,Brito:2014wla,Brito:2015oca,Hui:2016ltb,Arvanitaki:2016qwi}
or for scalar fields nonminimally coupled to gravity~\cite{Yunes:2013dva,Berti:2015itd,Sotiriou:2015pka,Barack:2018yly,Sathyaprakash:2019yqt,Barausse:2020rsu,Kalogera:2021bya,Doneva:2022ewd,LISA:2022kgy}.

In this paper we take a step further:
we study an \textit{ultralight} scalar field that is \textit{nonminimally coupled} to gravity.
We investigate its phenomenology
and assess if one can probe for such a field in a ``\bh laboratory.''
Before discussing our setup,
we briefly summarize the phenomenology of the two main ``ingredients,'' namely the scalar's mass term and its nonminimal coupling to gravity.

Let us start with the first: ultralight scalar fields around \bh{s} in \GR{.}
Here, the scalar can form a long-lived \QBS
in the \bh{'s} vicinity,
if the field's Compton wavelength is comparable to the \bh{'s} radius~\cite{Dolan:2007mj}.
These \QBS{s}, or ``scalar clouds,''
can be formed via
the superradiant instability around rotating \bh{s}~\cite{Damour:1976kh,Detweiler:1980uk,Zouros:1979iw,Press:1972zz,Cardoso:2004nk,Dolan:2007mj,Dolan:2012yt,Witek:2012tr,Yoshino:2012kn,Shlapentokh-Rothman:2013ysa,Brito:2015oca,Hui:2022sri,Branco:2023frw},
via accretion~\cite{Barranco:2011eyw,Barranco:2012qs,Okawa:2014nda,Sanchis-Gual:2016jst,Clough:2019jpm,Hui:2019aqm,Bamber:2020bpu,Traykova:2021dua},
or via synchronization of the scalar cloud with the rotating \bh{}
through a process that is qualitatively similar to the tidal lock of the Earth-Moon system~\cite{Cardoso:2011xi,Hod:2013zza,Herdeiro:2014goa,Benone:2014ssa,Herdeiro:2015gia,Chodosh:2015oma}.

Scalar clouds around \bh{s}
can be sufficiently long-lived to
become observable through their interaction with single or binary \bh{s}.
Potentially observable signatures include
\begin{enumerate*}[label={(\roman*)}]
\item a characteristic, near-monochromatic \GW signal~\cite{Okawa:2014nda,Degollado:2014vsa,Arvanitaki:2016qwi,Brito:2017wnc,Brito:2017zvb};
\item gaps in the mass-spin distribution of astrophysical \bh{s}~\cite{Arvanitaki:2010sy,Brito:2014wla,Brito:2015oca,Ficarra:2018rfu},
\item modifications of the \bh shadow~\cite{Cunha:2015yba,Cunha:2019ikd,Creci:2020mfg,Roy:2021uye}
\item modifications to the \GW{s} emitted
during the coalescence of comparable-mass \bbh{s}~\cite{Berti:2019wnn,Zhang:2019eid,Ikeda:2020xvt,Bamber:2022pbs},
their ringdown~\cite{Choudhary:2020pxy}
or by extreme-mass ratio inspirals~\cite{Hannuksela:2018izj,Baumann:2018vus,Baumann:2019ztm,Barsanti:2022vvl,Takahashi:2023flk};
\item effects on the evolution of binary pulsars~\cite{Blas:2016ddr,Blas:2019hxz}.
\end{enumerate*}

Let us now turn to the second
ingredient in our study:
scalar fields nonminimally coupled to gravity.
Such a coupling typically yields \bh{s} that possess scalar hair or that can spontaneously scalarize; see Refs.~\cite{Sotiriou:2015pka,Herdeiro:2015waa,Blazquez-Salcedo:2016yka,Silva:2017uqg,Doneva:2017bvd,Dima:2020yac,HegadeKR:2022xij,Doneva:2022ewd} for reviews.
Well-studied models are
\sGB gravity~\cite{Kanti:1995vq},
\dCS gravity~\cite{Jackiw:2003pm,Alexander:2009tp}
or axi-dilaton gravity~\cite{Kanti:1995cp,Cano:2021rey}.

Here, we focus on
\dCS gravity~\cite{Jackiw:2003pm,Alexander:2009tp}
which extends the Einstein-Hilbert action with a
gravitational
Chern-Simons term
given by the Pontryagin density
coupled to a dynamical pseudoscalar.
It is motivated by theoretical arguments from
particle physics~\cite{Alvarez-Gaume:1983ihn},
string theory~\cite{Green:1984sg,Kanti:1995cp,Polchinski:1998rr},
loop quantum gravity~\cite{Taveras:2008yf,Mercuri:2009vk,Mercuri:2009zi,Mercuri:2009zt} and the \EFT of inflation~\cite{Lue:1998mq,Choi:1999zy,Weinberg:2008hq,Alexander:2004us,Alexander:2004xd}.

In axi-symmetric \bh{} spacetimes, the Pontryagin density
is nontrivial
and thus, it sources the pseudoscalar field
that gives rise to \bh hair~\cite{Grumiller:2007rv,Shiromizu:2013pna,R:2022tqa}.
Such hairy \bh solutions have been obtained
analytically in the slow-rotation and small-coupling limits~\cite{Alexander:2009tp,Konno:2009kg,Yunes:2009hc,Yagi:2012ya,Cardenas-Avendano:2018ocb},
numerically in the small-coupling limit
for arbitrary spins~\cite{Konno:2014qua,Stein:2014xba,McNees:2015srl},
and
in a nonperturbative approach~\cite{Delsate:2018ome}.

In a \bbh coalescence{,} the pseudoscalar \bh hair generates additional scalar radiation that dissipates energy and hence, modifies the binary's dynamics and \GW emission.
The \dCS corrections to the dynamics of coalescing \bbh{s} and
their \GW radiation have been modeled by a combination
of post-Newtonian techniques for the inspiral~\cite{Yagi:2011xp,Loutrel:2018rxs,Loutrel:2018ydv,Loutrel:2022tbk},
numerical relativity simulations
(order-by-order in the \dCS coupling)
for the merger~\cite{Delsate:2014hba,Okounkova:2017yby,Okounkova:2018abo,Okounkova:2018pql,Okounkova:2019dfo,Okounkova:2019zjf},
and \bh perturbation theory for the ringdown~\cite{Yunes:2007ss,Cardoso:2009pk,Molina:2010fb,Kimura:2018nxk,Owen:2021eez,Wagle:2021tam,Li:2022pcy}.

It has been proposed that compact objects and \GW data from \bbh coalescences may constrain the \dCS coupling~\cite{Ali-Haimoud:2011zme,Nakamura:2018yaw,
Yunes:2013dva,Berti:2015itd,Alexander:2017jmt,
Nair:2019iur,Perkins:2021mhb,Alexander:2017jmt,Loutrel:2018rxs,Loutrel:2018ydv,Loutrel:2022tbk,Okounkova:2022grv}.
Indeed, some of the \GW events detected by the \LVK collaboration
have already been used to place observational bounds on the \dCS coupling~\cite{Silva:2022srr,Okounkova:2022grv}.
The most stringent bounds to date~\cite{Silva:2020acr}, however, come from a combination of the GW170817 event observed by the \LVK
collaboration~\cite{LIGOScientific:2017vwq} and NICER data~\cite{Riley:2019yda,Miller:2019cac}.

The \dCS pseudoscalar can acquire a small mass through nonperturbative effects that break the shift-symmetry of the standard \dCS model~\cite{Alexander:inprep2023}.
This mechanism is similar to the one that
yields a mass in nongravitational axion \EFT{s}~\cite{Kallosh:1995hi,Alonso:2017avz} or the string axiverse~\cite{Arvanitaki:2009fg}.
Our previous discussion suggests that
new phenomena arise in \bh spacetimes, and there are first studies in this direction~\cite{Molina:2010fb,Macedo:2018txb,Doneva:2021dcc,Zhang:2021btn,Alexander:2022avt}.
Computations of linear perturbations of the metric and a massive scalar field on a Schwarzschild background
show that \bh{s} in \dCS gravity also support long-lived massive modes~\cite{Molina:2010fb,Macedo:2018txb}.
The spectrum of a massive scalar field
around a slowly rotating \bh in \dCS gravity
was computed in Ref.~\cite{Alexander:2022avt}.

\begin{figure}
    \centering
    \includegraphics[width = \columnwidth]{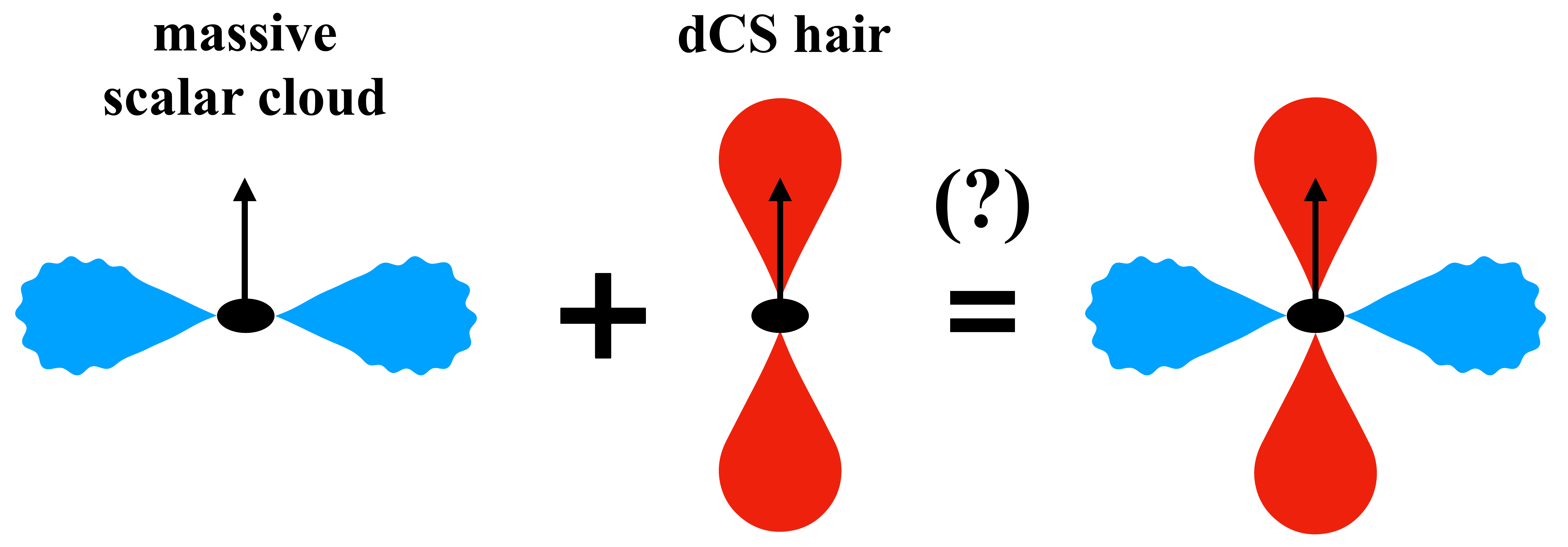}
    \caption{Sketch of a scalar field in massive \dCS gravity around a rotating \bh{.}
    The \bh{'s} axis of rotation is indicated by the arrow.
    The blue clouds depict the \QBS of a massive scalar field that is dominated by an oscillating $l=m=1$ dipole where
    the wavy outline indicates the \QBS{'s} oscillations.
    The red clouds represent the \dCS hair sourced by the Pontragin density and dominated by the $l=1, m=0$ dipole.
    }
    \label{fig:sketch1}
\end{figure}

In this paper, we study the scalar's time evolution
in massive \dCS gravity that combines the two features discussed above.
In particular, one may expect a superposition of the \dCS hair and the massive \QBS
as sketched in Fig.~\ref{fig:sketch1}.
We combine the effect of
(i) a massive scalar cloud
described by an
oscillating \QBS in the equatorial plane
(blue cloud in Fig.~\ref{fig:sketch1}),
with (ii) the dipolar \dCS hair that is sourced by the Pontryagin density and that is aligned with the \bh{'s} spin axis
(red cloud in Fig.~\ref{fig:sketch1}).
We simulate the scalar's growth
in massive \dCS gravity,
identify its final state,
and characterize its spectrum
by conducting a series of numerical relativity simulations for different initial states and mass parameters.

For this study, we have developed the \CanudadCS{} code module~\cite{CanudadCS_repo}
for \Canuda{,} our open-source numerical relativity code for fundamental physics~\cite{Witek:2018dmd,witek_helvi_2023_7791842}.
Specifically, we implemented
the \dCS field equations for
a massive scalar field
in the decoupling approximation,
in which the background is determined by \GR and
the backreaction of the scalar onto the metric is neglected.

In this paper we report a series of results to address three questions:
\begin{enumerate}
    \item How does the nonminimal coupling to curvature affect the
    (equatorial)
    massive scalar cloud?
    \item How does the mass term affect the \dCS hair?
    \item What is the scalar's
    characteristic
    frequency spectrum in massive \dCS gravity?
\end{enumerate}

The bulk of the results are presented in Sec.~\ref{Sec:results}, but let us give you a sneak preview here:
\begin{enumerate}
    \item We find that the evolution of the  massive scalar cloud in the equatorial plane
    (with the leading contribution being the $l=m=1$ multipole;
    see blue cloud in Fig.~\ref{fig:sketch1})
    appears unaffected by the nonminimal coupling to the Pontryagin density far from the \bh, i.e.,
    it is indistinguishable from its \GR counterpart
    (within numerical error).
    See Sec.~\ref{subsec:resultsmassivemode}.
    \item We find that the \dCS hair sourced by the Pontryagin density grows (with the leading contribution being the $l=1,m=0$ dipole; see red cloud in Fig.~\ref{fig:sketch1}),
    and it oscillates
    with a frequency determined by the mass.
    Its amplitude, suppressed by the mass,
    is smaller than in the massless \dCS case.
    See Sec.~\ref{subsec:resultsdCShair}.
    \item We reconstruct the power spectrum of the scalar multipoles by computing the discrete Fourier transform of our data.
    Within the resolution allowed by our simulations and within the decoupling approximation, we observe that the characteristic frequencies of the \QBS{s} are compatible with known results in \GR.
    See Sec.~\ref{subsec:mdCS_spec}.
\end{enumerate}

Our time evolutions monitor the growth of the massive, oscillating \dCS hair and suggest a
new, oscillating steady-state solution as end-state.
A detailed analysis of this \QBS solution and its stability is subject to future work.

The paper is organized as follows:
In Sec.~\ref{Sec:Setup} we present the action and field equations of massive \dCS gravity,
the scalar's time evolution equations and initial conditions, and the background spacetime.
In Sec.~\ref{Sec:NumMeth} we discuss our numerical relativity framework,
introduce our open-source \CanudadCS{} code
~\cite{CanudadCS_repo},
and describe the series of simulations that we conducted.
In Sec.~\ref{Sec:results} we present our results addressing questions $1.$-$3.$ posed above.
We conclude in Sec.~\ref{sec:Conclusion}.
Snapshots of some of our simulations are displayed in App.~\ref{app:snapshots},
and animations of our simulations are available on our Canuda youtube channel~\cite{YoutubeLinkmdCS}.

In this paper, we use natural units $G=c=\hbar=1$ and adopt the mostly-plus signature convention for the metric, $(-,+,+,+)$.

\section{Setting the stage}~\label{Sec:Setup}
\subsection{Action and field equations}\label{ssec:ActionAndEoMs}
We consider \dCS gravity in which a dynamical
pseudoscalar
field $\Theta$
is nonminimally coupled to gravity.
The action is given by~\cite{Jackiw:2003pm,Alexander:2009tp}
\begin{align}
\label{eq:ActiondCS}
S & = \int d^4x \frac{\sqrt{-g}}{16\pi} \left[ R
+\frac{\aCS}{4}f(\Theta)\,^*RR
-\frac{1}{2}(\nabla\Theta)^2
-V(\Theta)
\right]
\,,
\end{align}
where $\aCS$ is a
coupling constant with dimension $[\aCS]=[L]^2$,
$\Vth$ is the field's potential,
and
$f(\Theta)$ is a general function
that couples $\Theta$ to the Pontryagin density
\begin{align}
\label{eq:Pontryagin}
\CS= & \dual R^{abcd}R_{bacd} = -\frac{1}{2}\tensor{\epsilon}{^c^d_e_f}R^{abef}R_{abcd}
\,.
\end{align}
Here, $R_{abcd}$ and $\dual R^{abcd}=\frac{1}{2}\tensor{\epsilon}{^c^d_e_f}R^{abef}$ are
the Riemann and dual Riemann tensors associated to the metric $g_{ab}$.
Note that the Pontryagin density transforms as a pseudoscalar under parity transformations, i.e., it is parity-odd.
Consequently, the field $\Theta$ also transforms as a pseudoscalar field under parity transformations such that the Lagrangian in Eq.~\eqref{eq:ActiondCS}
is parity-even.
That is, the field $\Theta$ is an axionlike particle and in the following we refer to it as ``axion'' or ``\dCS axion.''

We extend the original \dCS action~\cite{Alexander:2009tp} to allow for a general coupling function $\fth$ and a general potential $\Vth$; see Eq.~\eqref{eq:ActiondCS}.
The function $f(\Theta)$ selects different parity-violating theories of gravity.
For example,
nondynamical Chern-Simons theory corresponds to $f=\rm{const}$,
while the traditional \dCS model corresponds to
$f(\Theta)\sim\Theta$ and vanishing potential $V(\Theta)=0$~\cite{Alexander:2009tp}~\footnote{In principle, one could also consider nonlinear coupling functions that might lead to spontaneous scalarization of \bh{s}, similar to the \bh scalarization in scalar Gauss-Bonnet gravity~\cite{Silva:2017uqg,Doneva:2017bvd}.
For instance, Refs.~\cite{Gao:2018acg,Doneva:2021dcc}
study \dCS models with a coupling function respecting a $Z2$ symmetry.
We refer the interested reader to the recent review on scalarization~\cite{Doneva:2022ewd}.}.

In the present work we consider massive \dCS gravity,
determined by
\begin{equation}\label{eq:model}
    f(\Theta) = \Theta
    \,,\quad
    V(\Theta) = \frac{\mu^2}{2}\Theta^2
    \,,
\end{equation}
where $\mu$ is the mass-energy of the scalar field.
We focus on fields that are light enough
to form long-lived \QBS{s} around astrophysical \bh{s}~\cite{Dolan:2007mj,Arvanitaki:2010sy,Brito:2015oca}.
We expect such a scenario when the Compton wavelength of the field is comparable to the \bh{'s} radius.
This corresponds to $2\mu M\sim 1$ in natural units or, equivalently, $\mu \simeq 10^{-10} (M_{\odot}/M)$ eV.

In the following
we make our model selection given by Eq.~\eqref{eq:model} explicit, and refer to App.~\ref{appsec:GeneralDCS} for the general expressions as implemented in our \CanudadCS{} code.
Then, varying the action in Eq.~\eqref{eq:ActiondCS} with respect to the metric and the scalar field, we obtain the field equations
\begin{subequations}
\label{eq:eoms}
\begin{align}
\label{eq:dCSKG}
&\left(\Box - \mu^2\right)\Theta + \frac{\aCS}{4} \CS = 0
\,,\\
\label{eq:dCSTenEoM}
&G_{ab} - \frac{1}{2} T^{\Theta}_{ab} + \aCS ~\CCS_{ab} = 0
\,,
\end{align}
\end{subequations}
where $G_{ab}=R_{ab}-\frac{1}{2}g_{ab}R$ is the Einstein tensor,
and $T^\Theta_{ab}$ indicates the canonical scalar field
stress tensor
\begin{equation}\label{eq:TmnSF}
T^\Theta_{ab}=\nabla_a\Theta\nabla_b\Theta
    -\frac{1}{2}g_{ab} \left( (\nabla\Theta)^2+\mu^2\Theta^2\right)
\,.
\end{equation}
The extension of Einstein's equations
is captured by the C-tensor
\begin{equation}\label{eq:ctensor}
    \CCS^{ab}\equiv \E_c\,\tensor{\epsilon}{^c^d^e^(^a}\nabla_e\tensor{R}{^b^)_d} + \F_{cd}\,^*R^{d(ab)c}
\end{equation}
where the auxiliary tensors $\E_c$ and $\F_{cd}$ are defined as
\begin{align}
\label{eq:CabAuxLinear}
\E_{a} & = \nabla_{a}\Theta
\,,\quad{\textrm{and}}\quad
\F_{ab} = \nabla_{a}\nabla_{b}\Theta
\,.
\end{align}

For completeness, we also list the effective energy-momentum tensor.
Using the convention $G_{ab}= \frac{1}{2} T^{\rm{eff}}_{ab}$ and comparing to Eq.~\eqref{eq:dCSTenEoM},
we find
\begin{align}
\label{eq:Tmneff}
T^{\rm{eff}}_{ab} = T^{\Theta}_{ab} - 2 \aCS \CCS_{ab}
\,,
\end{align}
where
$T^{\Theta}_{ab}$ and $\CCS_{ab}$
are given in Eqs.~\eqref{eq:TmnSF} and~\eqref{eq:ctensor}.

Here, we work in the
decoupling approximation,
i.e., we study the evolution of the massive \dCS field
in a background spacetime determined by Einstein's equations in vacuum.
We apply the decoupling approximation to the field Eqs.~\eqref{eq:eoms},
and rescale $\Theta \rightarrow (\aCS/M^2)\,\Theta$ to obtain~\footnote{Alternatively, one could obtain Eqs.~\eqref{eq:eoms_decoup}
by performing a perturbative expansion
of Eqs.~\eqref{eq:eoms}, power-counting in the coupling constant, and neglecting terms of order $~\Ord{(\aCS/M^2)}{2}$.
However, this linearization can only be applied to models for which $\dfth\neq0$ for all $\Theta$.
Nonlinear coupling functions that can vanish for some value of the scalar, $f'(\Theta_{0}) = 0$, and that may give rise to nonlinear phenomena such as \bh scalarization, are not captured by such a linearization.}
\begin{subequations}
\label{eq:eoms_decoup}
\begin{align}
\label{eq:dCSKG_decoup}
&\Box\Theta - \mu^2\Theta + \frac{\aCSh M^2}{4} \CS = 0
\,,\\
\label{eq:Einstein}
&G_{ab}  = 0
\,.
\end{align}
\end{subequations}
In Eq.~\eqref{eq:dCSKG_decoup} we introduce the dimensionless parameter $\aCSh$ that allows us to switch
the coupling to the Pontryagin density on ($\aCSh=1$) and off ($\aCSh=0$).
This switch parameter enables us to compare against the evolution of a massive scalar field in \GR{.}

\subsection{Spacetime split and background metric}\label{ssec:3p1SplitAndBackground}
To conduct the numerical simulations of the scalar field,
we rewrite Eqs.~\eqref{eq:eoms_decoup} as a time evolution problem.
We obtain the 3+1 formulation by foliating the spacetime $\left(\M,g_{ab}\right)$ into a set of spatial hypersurfaces
$\left(\Sigma_{t},\gamma_{ij}\right)$ with induced metric $\gamma_{ij}$.
Each hypersurface $\Sigma_t$ is labelled by the time parameter $t$ and the $3$-metric is given by $\gamma_{ab}=g_{ab} + n_{a}n_{b}$
where $n^{a}$ is the timelike unit vector normal to the hypersurface,
$\gamma^{a}{}_{b} n^{b} = 0$,
with normalization $n^{a}n_{a} = -1$.
Furthermore, the spatial metric defines a projection operator
\begin{align}
\label{eq:ProjOp}
\gamma^{a}{}_{b} = & \delta^{a}{}_{b} + n^{a} n_{b}
\,.
\end{align}
The line element takes the form
\begin{align}
\label{eq:LineElement3p1}
\dif s^{2} = & g_{ab} \dif x^{a} \dif x^{b}
 \\
        = & - \left(\alpha^{2} - \beta^{k} \beta_{k} \right) \dif t^{2}
            + 2 \gamma_{ij} \beta^{i} \dif t \dif x^{j}
            + \gamma_{ij} \dif x^{i} \dif x^{j}
\,. \nonumber
\end{align}
where $\alpha$ and $\beta^{i}$ are, respectively, the lapse function and shift vector.
We denote the covariant derivative and Ricci tensor associated to the 3-metric $\gamma_{ij}$ as $D_{i}$ and $R_{ij}$.
The extrinsic curvature $K_{ab}$ describes how a hypersurface
is embedded in the spacetime manifold
and is defined as
\begin{align}
\label{eq:DefKij}
K_{ab} = & - \gamma^{c}{}_{a} \gamma^{d}{}_{b} \nabla_{c} n_{d}
        = - \frac{1}{2} \Lie_{n} \gamma_{ab}
\,,
\end{align}
where
$\Lie_{n}$ is the Lie derivative along $n^{a}$.

In this paper, we evolve the scalar field on a fixed, stationary background described by
the Kerr metric in
quasi-isotropic coordinates~\cite{Liu:2009al,Witek:2018dmd}, which we briefly summarize here.
We start from the Kerr metric in
\BL
coordinates $(t,\rBL,\theta,\phi)$,
\begin{align}
\label{eq:KerrBL}
\dif s^2 = & -\left(1 - \frac{2M\rBL}{\Sigma}\right) \dif t^2
    - \frac{4aM\rBL\sin^2\theta}{\Sigma} \dif t \dif\phi
\nonumber\\ &
    + \frac{\Sigma}{\Delta}\dif\rBL^2
    + \Sigma \dif \theta^2
    + \frac{\A}{\Sigma}\sin^2\theta \dif\phi^2
\end{align}
with
\begin{subequations}
\label{eq:KerrMetricFunction}
\begin{align}
\Delta & = (\rBL - \rBLp)(\rBL - \rBLm) \\
\Sigma & = \rBL^2 + a^2 \cos^2\theta \\
\A     & = (\rBL^2+a^2)^2 - \Delta a^2 \sin^2\theta \\
r_{\rm{BL,\pm}} & = M \pm \sqrt{M^2-a^2}
\,,
\end{align}
\end{subequations}
where $\rBL$ is the \BL radial coordinate,
$r_{\rm{BL,\pm}}$ are the inner and outer horizons in BL coordinates,
$M$ is the \bh mass
and $a/M$ its dimensionless spin.
We introduce the quasi-isotropic radial coordinate
$r\in (0,\infty)$.
as~\cite{Liu:2009al,Witek:2018dmd},
\begin{equation}~\label{eq:IsoRad}
    \rBL = r\left(1+\frac{\rBLp}{4r}\right)^2\,.
\end{equation}
The outer horizon in these coordinates sits at $r_{+}=\rBLp/4$,
while the inner horizon and the region inside are not covered.

We apply the coordinate transformation, Eq.~\eqref{eq:IsoRad},
to Eq.~\eqref{eq:KerrBL}.
We write the resulting metric in the 3+1 form of Eq.~\eqref{eq:LineElement3p1},
where the gauge functions and
3-metric are
\begin{subequations}
\label{eq:KQImetric}
\begin{align}
\label{eq:KQI3metric}
\gamma_{ij} = & {\textrm{Diag}}\left[
    \frac{(4 r + \rBLp)^2\Sigma}{16 r^3 (\rBL - \rBLm)},
    \Sigma,
    \frac{\A}{\Sigma}\sin^2\theta\right]
\,,\\
\alpha = & \pm \sqrt{\frac{\Delta\Sigma}{\A}}
\,,\quad
\beta^{i} = \left(0,0,-\frac{2aM\rBL}{\A}\right)
\,.
\end{align}
\end{subequations}
The extrinsic curvature's nonvanishing components are
\begin{align}
K_{r\phi}      = & \alpha\frac{ a M \rBL' \sin^2\theta}{\Delta\Sigma^2}[2\rBL^2(\rBL^2+a^2)+\Sigma(\rBL^2-a^2)]
\,, \nonumber \\
K_{\theta\phi} = & -2\alpha\frac{a^3 M \rBL \sin^3\theta}{\Sigma^2}
\,,
\label{eq:KQIKij}
\end{align}
where $\rBL' = \partial_r \rBL$.
We apply a final coordinate transformation from the quasi-isotropic spherical coordinates
$Y^a=(r,\theta,\phi)$ to
Cartesian coordinates $X^i=(x,y,z)$ via
\begin{equation}
x=r\cos\phi\sin\theta
\,,\quad
y=r\sin\phi\sin\theta
\,,\quad
z=r\cos\theta\,.
\end{equation}
The spatial metric in Cartesian coordinates then reads
\begin{align}~\label{eq:hijIsoCart}
\gamma_{ij}\dif X^i \dif X^j = &
    \psi_0^4 \left[ \eta_{ij} \dif X^i \dif X^j
        + G(x \dif x+y \dif y+z \dif z)^2
\right. \nonumber\\ & \left.
        + a^2 H ( x \dif y - y \dif x)^2 \right]
\,,
\end{align}
where
$\eta_{ij}$ is the flat spatial metric
and we introduced
\begin{align}
\psi_0^4 = \frac{\Sigma}{r^2}
\,,\quad
G=\frac{\rBL}{r^2(\rBL - \rBLm)}
\,,\quad
H=\frac{\Sigma+2M\rBL}{r^2\Sigma^2}
\,.
\end{align}
We obtain the extrinsic curvature and shift vector in Cartesian coordinates
via the coordinate transformation
\begin{align}
K_{ij} = & J_{i}{}^aJ_{j}{}^bK_{ab}
\,,\quad
\beta_i = J_{i}{}^a\beta_a\,.
\end{align}
where $J_{i}{}^a=\partial Y^a/\partial X^i$
is the Jacobian.

\subsection{Time evolution formulation at decoupling}~\label{Sec:EvolutionFormulation}
To derive the axion evolution equations
(as a set of first order in time differential equations)
we introduce the momentum of the scalar field as
\begin{align}
\label{eq:DefKTheta}
\KTheta = & - \Lie_{n} \Theta
\,.
\end{align}
Inserting
Eq.~\eqref{eq:DefKTheta} into the field equation,  Eq.~\eqref{eq:dCSKG_decoup},
and rewriting it in terms of the
3+1 variables, $(\gamma_{ij},K_{ij},\alpha,\beta^{i})$,
we obtain the evolution equations
\begin{subequations}
\label{eq:dtdCSKG_decoup}
\begin{align}
\label{eq:dtThetaADM}
\dif_{\rm t} \Theta   = & -   \alpha \KTheta
\,,\\
\label{eq:dtKThetaADM}
\dif_{\rm t} \KTheta  = & - \alpha  D^{i} D_{i}\Theta - D^{i}\alpha D_{i} \Theta
        \nonumber\\
        & + \alpha\left( K \KTheta + \mu^2\Theta - \frac{\aCSh M^2}{4}\CS \right)
\,.
\end{align}
\end{subequations}
where $\dif_{\rm t} = \left(\p_{t} - \Lie_{\beta}\right)$,
and $\Lie_{\beta}$ is the Lie derivative along the shift vector.
In the decoupling limit, the scalar is sourced by the Pontryagin density in Eq.~\eqref{eq:Pontryagin}, evaluated on the \GR background
\begin{align}
\label{eq:PontinEB}
    \CS=\,^*W_{abcd}W^{bacd} = -16E^{ij}B_{ij}\,,
\end{align}
where
$^{\ast}W_{abcd}=\frac{1}{2}\epsilon_{cd}{}^{ef} W_{abef}$ is the dual Weyl tensor
and we introduce
its gravito-electric,
$E_{ij}=\gamma^{a}{}_{i} \gamma^{b}{}_{j} n^c n^d W_{acbd}$,
and gravito-magnetic,
$B_{ij}=\gamma^{a}{}_{i} \gamma^{b}{}_{j} n^c n^d \,^{\ast}W_{acbd}$,
components.
We compute the gravito-electric and gravito-magnetic components of the Weyl tensor in terms of the 3+1 variables, given by
\begin{subequations}
\label{appeq:EijBijInGR}
\begin{align}
\label{eq:EijInGR}
E_{ij} = & R^{\rm tf}_{ij} + \frac{1}{3} A_{ij} K - A_{i}{}^{k}A_{jk} + \frac{1}{3}\gamma_{ij} A_{kl}A^{kl}
\,,\\
\label{eq:BijInGR}
B_{ij} = & -\epsilon_{(i|}{}^{kl}D_{l}A_{|j)k}
\,,
\end{align}
\end{subequations}
where
$\Rtf_{ij}$ is the trace-free part of the spatial Ricci tensor associated to $\gamma_{ij}$,
$K=\gamma^{ij}K_{ij}$ is the trace of the extrinsic curvature,
and $A_{ij}=K_{ij}-\frac{1}{3}\gamma_{ij}K$
is its trace-free part.
The explicit 3+1 form of the scalar's effective energy-momentum tensor in Eq.~\eqref{eq:Tmneff} is given in App.~\ref{appsec:GeneralDCS}.

We re-write the evolution equations, Eqs.~\eqref{eq:dtdCSKG_decoup},
in terms of the
\BSSN variables for the metric
given by
\begin{subequations}
\label{eq:BSSNvars}
\begin{align}
W = & \gamma^{-\frac{1}{6}}
\,,\quad
\tgam_{ij} = W^2 \gamma_{ij}
\,,\\
K = & \gamma^{ij} K_{ij}
\,,\quad
\tA_{ij} = W^2\left(K_{ij}-\frac{1}{3}\gamma_{ij} K\right)
\,,\\
\tGam^{i} = & \tgam^{kl} \tGam^{i}{}_{kl}
\,.
\end{align}
\end{subequations}
The resulting evolution equations are given by
\begin{subequations}
\label{eq:dtdCSKG_decoupBSSN}
\begin{align}
\label{eq:dtThetaADM}
\dif_{\rm t} \Theta   = & -\alpha \KTheta
\,,\\
\label{eq:dtKThetaADM}
\dif_{\rm t} \KTheta  = & - W^2 \tD^{i}\alpha \tD_{i} \Theta - \alpha  \left(W^2 \tD^{i} \tD_{i}\Theta - W \tD^i \Theta \tD_i W \right. \nonumber \\
& \left. - K \KTheta - \mu^2 \Theta + \frac{\aCSh M^2}{4}\CS\right)
\,
\end{align}
\end{subequations}
where
we raise indices with the conformal metric $\tgam^{ij}$.
The Pontryagin density
becomes
\begin{align}
\label{eq:PontInEBBSSN}
\CS = & - 16 \tgam^{ia}\tgam^{jb}\tE_{ab} \tB_{ij}
\end{align}
where the gravito-electric and gravito-magnetic components of the Weyl tensor in BSSN variables  $\tE_{ij}$ and $\tB_{ij}$ are given by Eqs.~\eqref{eq:tEtBinBSSN}
in App.~\ref{appsec:GeneralDCS}.

\subsection{Scalar field initial data}\label{subsec:IDSF}
In our numerical simulations we perform a series of runs with different types of initial data for the scalar field. This includes
a trivial field,
a Gaussian type perturbation,
the solution for the massless \dCS axion,
and the \QBS solution for massive fields in \GR{.}

\noindent{\bf{Initial data 1 (ID1): Zero scalar field.}}
The simplest initial data that we consider is the trivial one, in which we set both the scalar field and its momentum to zero,
\begin{equation}\label{eq:ID1}
    \Theta(t=0) = 0 = K_\Theta(t=0)\,.
\end{equation}
Although this data is not a solution of the field equation, Eq.~\eqref{eq:dCSKG_decoup},
(except for a Schwarzschild \bh{)},
it allows us to follow the formation of the massive \dCS hair around a Kerr \bh
until it forms a steady-state configuration.

\noindent{\bf{Initial data 2 (ID2): Gaussian.}}
A second choice of initial data for the scalar field $\Theta$ is
that of a Gaussian shell centered around $r_0$ given by
\begin{eqnarray}\label{eq:ID2}
\Theta(t=0) = A\,\exp\left\{-\frac{(r-r_0)^2}{\sigma^2}\right\} \Sigma_{lm}(\theta,\phi),
\end{eqnarray}
where $A$ is the amplitude, $\sigma$ is the width, and $\Sigma_{lm}$ is a superposition of spherical harmonics.
In the present paper, we typically choose $\Sigma_{11} = Y_{1-1}- Y_{11}$.
We also set a vanishing scalar momentum,
$\KTheta(t=0)=0$.
For our choice of angular profile, this solution is exact
\begin{align}
\KTheta(t=0) = & - \frac{1}{\alpha}\left(\p_{t} - \beta^{a}\p_{a}\right)\Theta|_{t=0}
    = \frac{\beta^{\phi}}{\alpha}\p_{\phi}\Theta|_{t=0}
    = 0
\,,
\end{align}
as follows from the metric in Eq.~\eqref{eq:KQImetric},
the definition from Eq.~\eqref{eq:DefKTheta}, and
the profile from Eq.~\eqref{eq:ID2} with $\Sigma_{lm}=\Sigma_{11}$.

\noindent{\bf{Initial data 3 (ID3): \dCS hair.}}
As a third choice of initial data, we implement the small-spin, small-coupling perturbative solution for the \bh hair in (massless) \dCS gravity~\cite{Yagi:2012ya,Cano:2019ore,Cano:2021rey}
\begin{align}\label{eq:ID3}
\Theta = & \aCSh \frac{a}{M}
    \left( \frac{5M^2}{8\rBL^2} + \frac{5M^3}{4\rBL^3} +  \frac{9M^4}{4\rBL^4}\right)\cos\theta
\\ &
    - \aCSh \frac{a^3}{M^3} \left[
        \left(\frac{M^2}{16\rBL^2} + \frac{M^3}{8\rBL^3} + \frac{3M^4}{20\rBL^4} + \frac{M^5}{10\rBL^5}\right)\cos\theta
    \right. \nonumber\\ &\left.
        + \left( \frac{3M^4}{4\rBL^4} + \frac{3M^5}{\rBL^5} + \frac{25M^6}{3\rBL^6} \right)\cos^3\theta
    \right]
    + \Ord{\left(\frac{a}{M}\right)}{5}
\,,\nonumber
\end{align}
where $a/M$ is the dimensionless \bh spin
and $M$ is the \bh mass.
We note that this solution is accurate to first-order in the \dCS coupling and accurate to third-order in the spin.
Similar to the construction of ID2, we can set
\begin{align}
\KTheta(t=0) = \frac{\beta^{\phi}}{\alpha} \p_{\phi}\Theta|_{t=0}
    = 0
\,.
\end{align}

\noindent{\bf{Initial data 4 (ID4): Quasibound state.}}
Finally, we implement initial data given by a monochromatic \QBS for massive scalar fields in \GR{,} following Ref.~\cite{Dolan:2007mj}.
We take a mode ansatz for the scalar field,
\begin{align}
\label{eq:QBSansatz}
\Theta_{lm} = \exp(-i \omega t) \exp(i m \phi) \S_{lm}(\theta) \R_{lm}(\rBL)
\,,
\end{align}
where
$\omega=\bar{\omega}+\imath \bar{\nu}$ is the complex frequency,
$\S_{lm}(\theta)$ are the $s=0$ spheroidal harmonics~\cite{Berti:2005gp},
$\R_{lm}(\rBL)$ is the radial profile,
and $\rBL$ is the \BL radial coordinate that is related to the numerical radial coordinate via Eq.~\eqref{eq:IsoRad}.
We compute the spheroidal harmonics using the continued fraction method with a three-term recurrence relation
and implement
Eqs.~(2.2)~--~(2.9) of Ref.~\cite{Berti:2005gp}.

To compute the radial function $\R_{lm}$ we implement the scheme laid out in Ref.~\cite{Dolan:2007mj}.
Because we are interested in \QBS solutions,
we set boundary conditions that are ingoing at the horizon and
vanish
at infinity.
This behaviour is incorporated in the ansatz for the radial function
\begin{align}
\label{eq:QBSRadansatz}
\R_{lm}(\rBL) & =
    \Delta_{+}^{-i\eta}
    \Delta_{-}^{i\eta + \chi - 1}
    e^{q\,\rBL}
    \sum_{n=0}^{\infty} a_{n} \left(\frac{\Delta_{+}}{\Delta_{-}}\right)^{n}
\,,
\end{align}
where
\begin{subequations}
\label{eq:QBSAuxFuncts}
\begin{align}
\Delta_+ & = \rBL - \rBLp
\,,\quad
\Delta_{-} = \rBL - \rBLm
\,,\\
\eta & = \frac{2\rBLp(\omega-\omega_{\rm{c}})}{\rBLp - \rBLm}
\,,\quad
\chi    = \frac{\mu^2 - 2 \omega^2}{q}
\,,\\
q & = \pm \sqrt{\mu^2-\omega^2}
\,,
\end{align}
\end{subequations}
and $\omega_{\rm{c}}=m\Omega_{\rm{H}} = m \frac{a}{2M\rBLp}$ is the critical frequency for superradiance.
The coefficients $a_{n}$ in Eq.~\eqref{eq:QBSRadansatz} are solved for numerically
by adopting Leaver's continued fraction method~\cite{Leaver:1986gd} and using a three-term recurrence relation;
see Eqs.~(35)~--~(45) of Ref.~\cite{Dolan:2007mj}.

Finally, the initial scalar field is
$\Theta_{lm}(t=0)= \exp(i m \phi) \S_{lm}(\theta) \R_{lm}(\rBL)$
with the spheroidal harmonics and radial function solved for numerically as described above.
As a simplifying assumption to compute the scalar's initial momentum, we set the initial gauge functions to $\alpha=1$, $\beta^{i}=0$, and obtain
\begin{align}
\label{eq:QBSKTheta}
\KTheta{}_{,lm}(t=0) & = i \omega \Theta_{lm}(t=0)
\,.
\end{align}

\section{Numerical relativity framework}~\label{Sec:NumMeth}
In this section, we describe the numerical relativity framework and the simulations that we have performed.

\subsection{Code description}

To perform our numerical experiments, we
employ
the \ETK~\cite{Loffler:2011ay,Zilhao:2013hia,EinsteinToolkit:2022_11},
an open-source numerical relativity code for computational astrophysics,
with our open-source \Canuda  code~\cite{witek_helvi_2023_7791842,Okawa:2014nda,Witek:2018dmd,Silva:2020acr}
for fundamental physics.
The \ETK is based on the \Cactus computational toolkit~\cite{Cactuscode:web,Goodale:2002a},
and the \Carpet driver~\cite{Schnetter:2003rb,CarpetCode:web} to provide boxes-in-boxes
\amr and MPI parallelization.

Here, we focus on a stationary \bh background given by the Kerr metric in quasi-isotropic coordinates
as described in Sec.~\ref{ssec:3p1SplitAndBackground}.
The metric is implemented in the {\textsc{KerrQuasiIsotropic}} thorn (as code modules in the \ETK are called) of \Canuda{.}

To simulate the massive \dCS field,
we implement
the new arrangement \CanudadCS{}~\cite{CanudadCS_repo}.
The latter is a collection of thorns that provide the scalar field initial data and the implementation of the scalar's time evolution equations.
Since the scalar field initial data types ID1~--~ID3 in Sec.~\ref{subsec:IDSF}
are analytical, we simply assign the indicated function values in quasi-isotropic coordinates; see Eq.~\eqref{eq:IsoRad}.
For the \QBS initial data (ID4 in Sec.~\ref{subsec:IDSF}0,
we implement Leaver's continued fraction method to solve for the spheroidal harmonics and the radial function.

We implement the scalar field evolution equations
in terms of the \BSSN metric variables, Eqs.~\eqref{eq:BSSNvars},
and solve them using the methods of lines.
Spatial derivatives are realized by fourth-order finite differences, and we use a fourth-order Runge-Kutta scheme for the time integration.
Fifth order Kreiss-Oliger dissipation is employed as provided by the \ETK to reduce the high-frequency, numerical noise produced by the refinement boundaries.
At the outer boundary we apply radiative boundary conditions via the \textsc{NewRad} thorn~\cite{Alcubierre:2002kk}.

To analyze our data, we compute the $(l,m)$ multipoles of $\Theta$ using the {\textsc{Multipole}} thorn.
In particular, the field $\Theta$
is interpolated onto spheres of fixed extraction radii $\rex$ and projected onto $s=0$ spherical harmonics $Y_{lm}(\theta,\phi)$,
\begin{align}
\label{eq:ThetaYlmProjection}
\Theta_{lm}(t,\rex) = & \int\dif\Omega\, \Theta(t,\rex,\theta,\phi) Y^{\ast}_{lm} (\theta,\phi)
\,.
\end{align}
We visualize our numerical data with \PyCactus or \PostCactus~\cite{2021ascl.soft07017K}, a set of tools for post-processing data generated with the \ETK{.}

\subsection{Simulations}\label{Simulations}
We perform a series of simulations to study the evolution of a massive scalar field, nonminimally coupled to gravity through the Pontryagin density
to address the questions posed in Sec.~\ref{sec:intro}.
We list our simulations in Table~\ref{tab:simulations},
where we indicate the name of the runs,
the axion's dimensionless mass parameter $\mu M$,
the dimensionless \bh spin $a/M$,
and the type of initial data ``ID.''
We also indicate the switch parameter $\aCSh$, where $\aCSh=0$ and $\aCSh=1$ refer to evolutions in \GR and \dCS gravity, respectively.

\begin{table}[h!]
    \centering
    \begin{tabular}{|c|c|c|c|c|}
    \hline
        Run       &
        $\mu M$   &
        $a/M$     &
        $\aCSh$   &
        ID
        \\
    \hline
    \hline
    a01\_SF11\_mu$00$ & $0.0$ & $0.1$ & $1$ & ID2\\
    a01\_SF11\_mu$01$ & $0.1$ & $0.1$ & $1$ & ID2\\
    a01\_SF11\_mu$03$ & $0.3$ & $0.1$ & $1$ & ID2\\
    a01\_SF11\_mu$10$ & $1.0$ & $0.1$ & $1$ & ID2\\
    \hline
    a07\_SF11\_mu$00$ & $0.0$ & $0.7$ & $1$ & ID2\\
    a07\_SF11\_mu$01$ & $0.1$ & $0.7$ & $1$ & ID2\\
    a07\_SF11\_mu$0187$ & $0.187$ & $0.7$ & $1$ & ID2\\
    a07\_SF11\_mu$10$ & $1.0$ & $0.7$ & $1$ & ID2\\
    \hline
    a099\_SF11\_mu$00$ & $0.0$ & $0.99$ & $\lbrace 0, 1\rbrace$ & ID2\\
    a099\_SF11\_mu$01$ & $0.1$ & $0.99$ & $1$ & ID2\\
    a099\_SF0\_mu$042$ & $0.42$ & $0.99$ & $1$ & ID1 \\
    a099\_SF11\_mu$042$ & $0.42$ & $0.99$ & $\lbrace 0, 1\rbrace$ & ID2\\
    a099\_dCSHair\_SF11\_mu$042$ & $0.42$ & $0.99$ & $1$ & ID2 + ID3 \\
    a099\_dCSHair\_mu$042$ & $0.42$ & $0.99$ & $1$ & ID3 \\
    a099\_SF\_QBS\_mu$042$ & $0.42$ & $0.99$ & $\lbrace 0, 1\rbrace$ & ID4 \\
    a099\_SF11\_mu$10$ & $1.0$ & $0.99$ & $1$ & ID2\\
    \hline
    \end{tabular}
    \caption{Summary of simulations of a scalar field with dimensionless mass parameter $\mu M$ around a Kerr \bh with dimensionless spin $a/M$.
    We set $\aCSh=1$ ($\aCSh=0$) for simulations in \dCS gravity (\GR{}).
    We indicate the choice of initial data, ID, as described in Sec.~\ref{subsec:IDSF}.
    }
    \label{tab:simulations}
\end{table}

In the majority of our simulations, we set $\aCSh=1$ and initialize the massive scalar field as a Gaussian,
ID2 in Sec.~\ref{subsec:IDSF},
with angular distribution $\Sigma_{lm}=\Sigma_{11}$,
width $\sigma = 1M$, amplitude $A = 1$
and centered around $r_{0}=6M$;
see Eq.~\eqref{eq:ID2}.
Note that in the absence of the coupling to the Pontryagin density, the $l=m=1$ scalar multipole is the fastest growing superradiant mode.
In the majority of our simulations we set $\Theta_{10}(t=0)=0$ to follow the growth of this mode that,
in the absence of the mass term,
would be the leading contribution to the \dCS hair.

To check the robustness of our results, we perform a set of complementary simulations with different scalar field initial data
in the background of a \bh with spin $a/M=0.99$.
In particular, we compare our results against a scalar field initialized as a \QBS in the $l=m=1$ multipole with $\mu M=0.42$, ID4 in Sec.~\ref{subsec:IDSF}, both in \GR and in \dCS gravity.
We also initialize the $l=1,m=0$ scalar multipole as the massless \dCS hair
given in Eq.~\eqref{eq:ID3}.
This choice of initial data allows us to explore the evolution of the
\dCS hair in the presence of a mass term.

Our simulations are performed on three-dimensional Cartesian grids with the outer boundary located at $256M$.
We use the boxes-in-boxes \amr grid structure provided by \Carpet~\cite{Schnetter:2003rb,CarpetCode:web}
with seven refinement levels.
The level's ``radii'' (half of the boxes' lengths) are given by
$\{256, 128, 32, 6, 3, 1.5, 0.75\}$.
On the outermost domain we set the grid spacing to $\dif x=1M$, which results in a spacing of
$\dif x / 2^{\rm{RL}-1} = \dif x / 2^6 = 1/64M$
on the innermost level that contains the \bh{.}
We set the time step to $\dif t = 0.25 \dif x$ on each refinement level such that the Courant-Friedrichs-Lewy condition is satisfied.

To assess the numerical error of our simulations,
we perform a convergence analysis of
the scalar's $l=1,m=0$ and $l=m=1$ multipoles for
Run
{\text{a099\_SF11\_mu042}} in Table~\ref{tab:simulations}.
For this purpose, we simulate
{\text{a099\_SF11\_mu042}}
with the additional coarse resolution
$\dif x_{c} = 1.1 M$
and the high resolution
$\dif x_{h} = 0.9 M$.
The default choice for our simulations,
$\dif x = 1.0M$, corresponds to the medium resolution run.
At intermediate times of $t \sim 500M$, we find a relative error of
$\Delta\Theta_{10}/\Theta_{10,h}\lesssim 10\%$ for the $l=1,m=0$ mode
and
$\Delta\Theta_{11}/\Theta_{11,h}\lesssim 0.2\%$ for the $l=m=1$ mode.
At late times of $t \sim 1000M$, we find a relative error of
$\Delta\Theta_{10}/\Theta_{10,h}\lesssim 12\%$ for the $l=1,m=0$ mode
and
$\Delta\Theta_{11}/\Theta_{11,h}\lesssim 0.1\%$ for the $l=m=1$ mode.
The difference of the relative numerical error for the different multipoles
can be understood by the absolute value of $\Theta_{10}$ being about one order of magnitude smaller than that of $\Theta_{11}$, while
$\Delta\Theta$ between the coarse and high resolutions are comparable.
Details of the convergence study are presented in App.~\ref{App:ConvTest} and illustrated in Fig.~\ref{Fig:ConvTest}.

\section{Results}~\label{Sec:results}
In this section, we present the results
of our simulations of the massive \dCS field, $\Theta$,
that we evolve in the background of a Kerr \bh{.}
They aid us in addressing the questions posed in the introduction, namely
\begin{enumerate*}[label={(\roman*)}]
    \item How does the nonminimal coupling to curvature affect the
    (equatorial)
    massive scalar cloud?
    \item How does the mass term affect the \dCS hair?
    \item What is the scalar's
    characteristic
    frequency spectrum in massive \dCS gravity?
\end{enumerate*}

We illustrate our setup in the
sketch
shown in Fig. \ref{fig:sketch1}.
Here, the blue cloud represents the oscillatory, massive scalar cloud
dominated by a $l=m=1$ mode,
and the red cloud represents the
static \dCS hair dominated by a $l=1,m=0$ mode.
For small spins
these modes essentially decouple,
as we demonstrate in App.~\ref{app:SphHamDecPont}.
Therefore, we first study the impact of the coupling to the Pontryagin density on the
$l=m=1$ mode in Sec.~\ref{subsec:resultsmassivemode}, then we study the impact of the mass term on the $l=1,m=0$ mode in Sec.~\ref{subsec:resultsdCShair},
and finally we present a spectral analysis of the massive \dCS field in Sec.~\ref{subsec:mdCS_spec}.

\begin{figure*}[h!t]
    \centering
    \includegraphics[width = \textwidth]{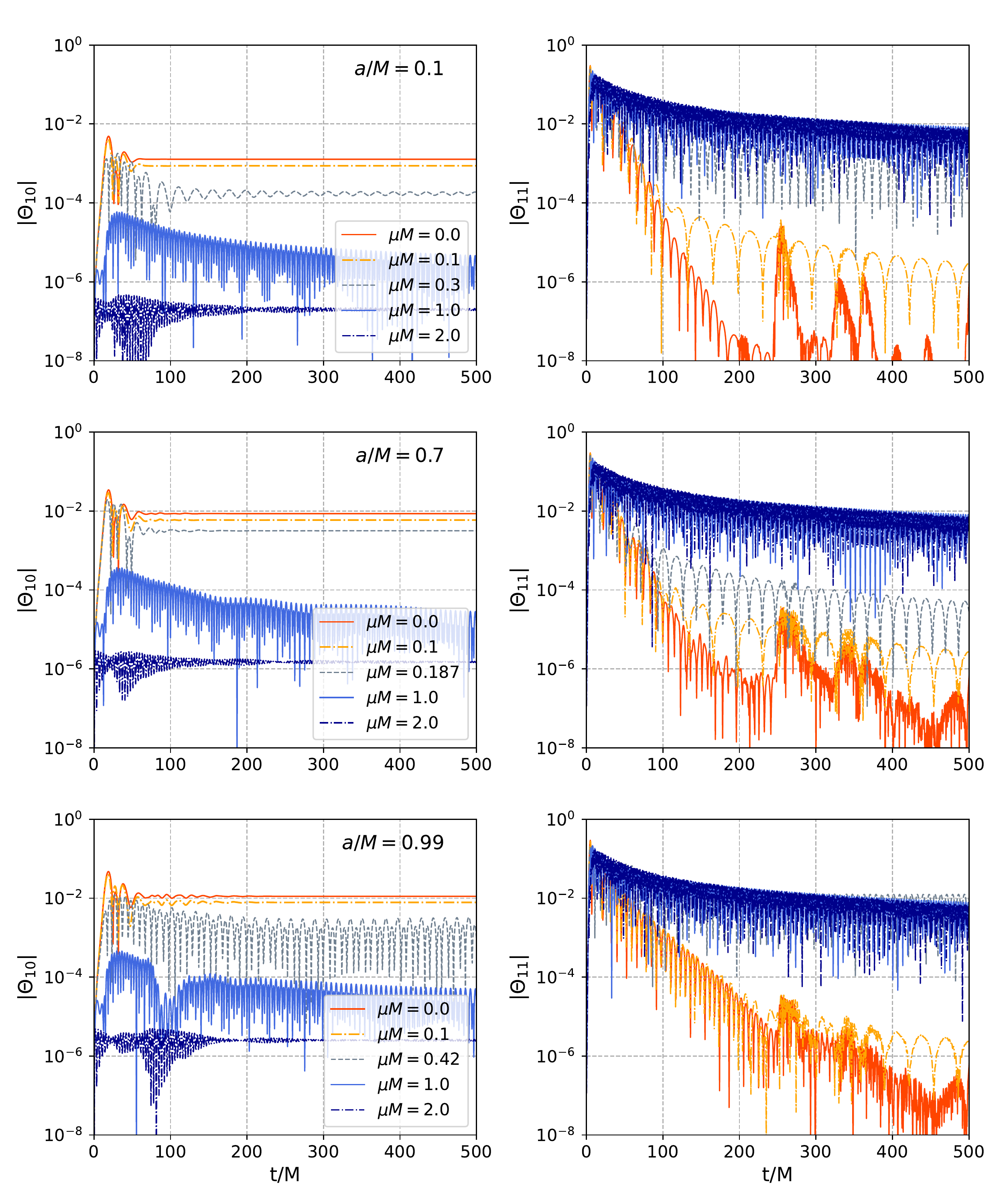}
    \caption{Evolution of the $l = 1, m = 0$ (left) and $l = m = 1$ (right) multipoles of the massive \dCS field $\Theta$,
    measured at $\rex=10$M,
    in the background of a rotating \bh
    with dimensionless spin $a/M=0.1$ (top), $a/M=0.7$ (middle) and $a/M=0.99$ (bottom).
    The field is initialized as a Gaussian with $\Sigma_{lm}=\Sigma_{11}$ in Eq.~\eqref{eq:ID2}.
    We vary the field's mass parameter $\mu M$ as listed in the legends.
    The $l=1,m=0$ multipole (left panel)
    approaches the \dCS solution for $\mu M=0.0$.
    For $\mu M \neq 0$, the \dCS hair is suppressed, and for a range of small mass parameters, the \dCS hair exhibits an oscillatory behaviour.
    The $l=m=1$ mode (right panel)
    exhibits the quasinormal ringdown for $\mu M=0$, intermediate or late-time massive power-law tails for small or large $\mu M$, or a \QBS for intermediate $\mu M$.
}
    \label{fig:ID2_all_spin_all_mass}
\end{figure*}

\begin{figure*}[h!t]
    \centering\includegraphics[width = \textwidth]{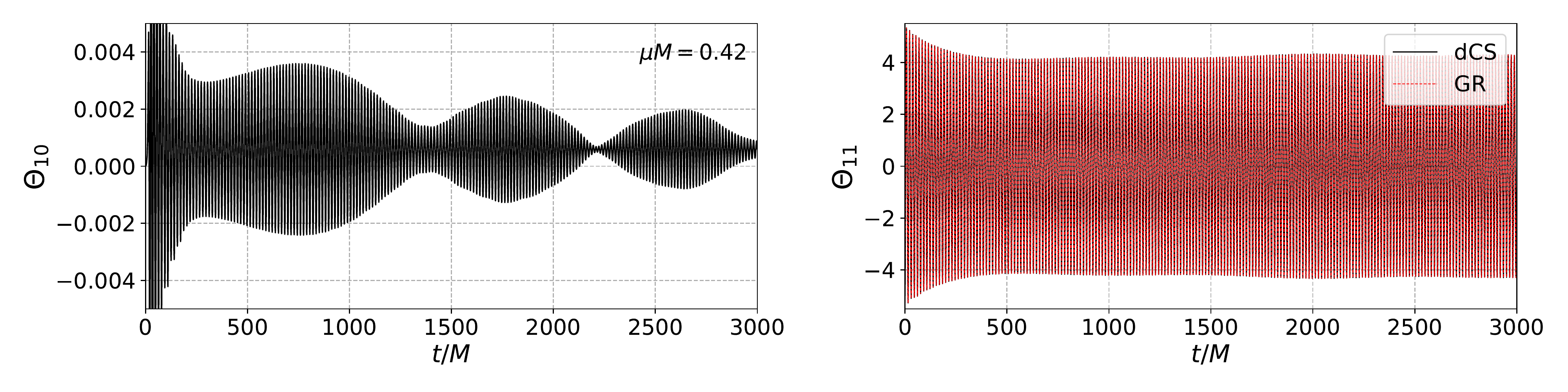}
    \caption{Same as Fig.~\ref{fig:ID2_all_spin_all_mass} but for the field $\Theta$ initialized as the massive,
    $l=m=1$
    \QBS, ID4 in Sec.~\ref{subsec:IDSF}
    with $\mu M=0.42$ around a \bh with $a/M=0.99$.
    We show the $l=1,m=0$ (left) and the $l=m=1$ (right) multipole,
    extracted at $\rex = 10M$.
    We compare the \dCS field (black solid lines) to the evolution in \GR (red dashed line).
    The $l=m=1$ mode (right panel) remains a nearly constant-in-time \QBS{,}
    and the \dCS and \GR case appear indistinguishable.
    Instead, the $l=1,m=0$ \dCS hair (left panel) oscillates with a frequency induced by the massive $\Theta_{11}$ mode.
    }
    \label{fig:ID_SF_BS}
\end{figure*}

Figs.~\ref{fig:ID2_all_spin_all_mass}
and~\ref{fig:ID_SF_BS}
give an overview of our results,
where we show the time evolution of the
$\Theta_{10}$ (left panels)
and $\Theta_{11}$ (right panels) multipoles in
massive \dCS gravity.
In Fig.~\ref{fig:ID2_all_spin_all_mass}
we initialize the field as a $l=m=1$ Gaussian, see Eq.~\eqref{eq:ID2}.
We consider a series of mass parameters $\mu M$ for the scalar field and three different spins $a/M=0.1$ (top), $a/M=0.7$ (middle), and $a/M=0.99$ (bottom) for the \bh in the background.
In Fig.~\ref{fig:ID_SF_BS} we initialize the scalar field as a \QBS with $\mu M=0.42$, and
we concentrate on a \bh with dimensionless spin $a/M=0.99$.
Independently of the initial data type,
\bh spin and mass parameter,
the time evolution of the massive mode, $\Theta_{11}$,
appears largely unaffected by the coupling to the Pontryagin density as is discussed in Sec.~\ref{subsec:resultsmassivemode}.
In Sec.~\ref{subsec:resultsdCShair},
we demonstrate that the mass term modifies the evolution of the \dCS hair, $\Theta_{10}$.
In Figs.~\ref{fig:xyxz_t250_GRvsmdCS_out} and~\ref{fig:xyxz_t250_dCSvsmdCS_out} we present snapshots
of the massive \dCS scalar with $\mu M=0.42$ and $a/M = 0.99$
in the $xy$ and the $xz$ planes,
taken at $t=250M$.
These are compared, respectively, against \GR ($\hat{\alpha}=0$) and massless \dCS gravity ($\hat{\alpha}=1$ and $\mu M = 0$).
In Figs.~\ref{fig:xyxz_t250_GRvsmdCS_in} and  ~\ref{fig:xyxz_t250_dCSvsmdCS_in}
we provide a close-up view on the scalar's structure near the \bh{.}
Full 2D animations of the scalar field evolution are available
on the \Canuda{} code's youtube channel
~\cite{YoutubeLinkmdCS}.


\subsection{Effect of the Pontryagin density on the evolution of a massive quasibound state }\label{subsec:resultsmassivemode}
In this section we focus on the
$\Theta_{11}$ multipole and investigate how its evolution is affected by the coupling to the Pontryagin density.
To explore this effect, we initialize the $\Theta_{11}$ mode either as a Gaussian with $\Sigma_{11}$ in Eq.~\eqref{eq:ID2}
or as a \QBS{,} see Eq.~\eqref{eq:QBSansatz},
while we set $\Theta_{10}(t=0)=0$.
The main results are shown in the right panels of Fig.~\ref{fig:ID2_all_spin_all_mass},
where we present the time evolution of $\Theta_{11}$ for a variety of mass parameters $\mu M$ and for \bh spins $a/M=0.1$ (top), $a/M=0.7$ (middle), and $a/M=0.99$ (bottom).
We see that the massless field (solid red lines) decays exponentially in time
consistent with a quasinormal ringdown.
Instead, the field with a small mass $\mu M=0.1$ (dashed-dot orange lines) exhibits a short ringdown phase followed by an oscillatory power-law tail
consistent with an intermediate time massive tail~\cite{Hod:1998ra}.
For intermediate masses,
$\mu M=0.187,0.3,0.42$ (dashed gray lines),
the field appears to transition to a long-lived \QBS{,} consistent with the solutions of Ref.~\cite{Dolan:2007mj} and the simulations in Ref.~\cite{Witek:2012tr}.
Finally, fields with a larger masses,
$\mu M=1.0$ (solid blue) and $\mu M=2.0$ (dash-dotted dark blue),
oscillate with a frequency $\bar{\omega}\sim\mu$ and decay with a universal power-law tail
that is consistent with the very late-time tail of massive fields~\cite{Koyama:2001qw,Koyama:2001ee}.

\begin{figure*}[htpb!]
    \centering
    \includegraphics[width = 0.329\textwidth]{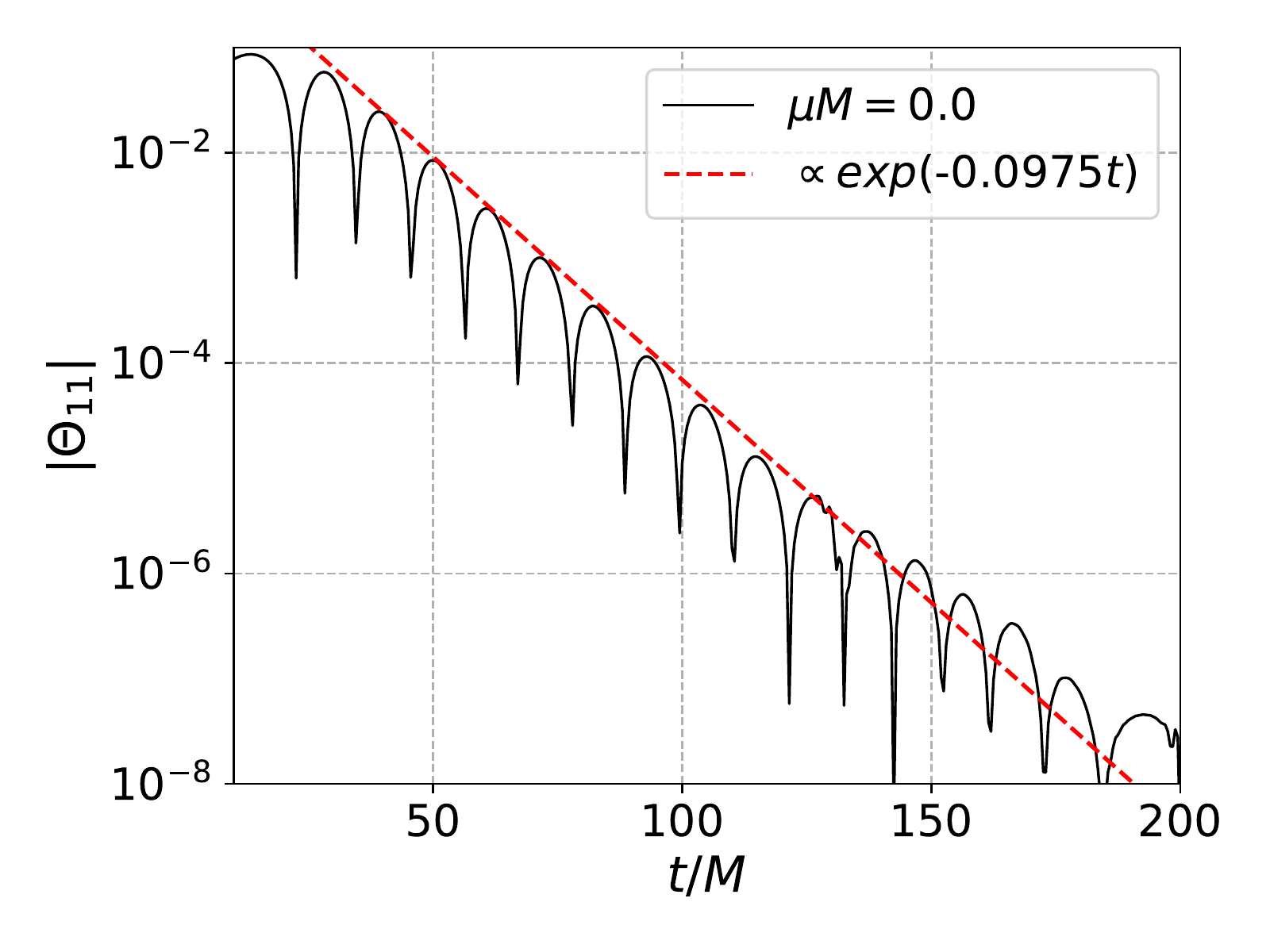}
    \includegraphics[width = 0.329\textwidth]{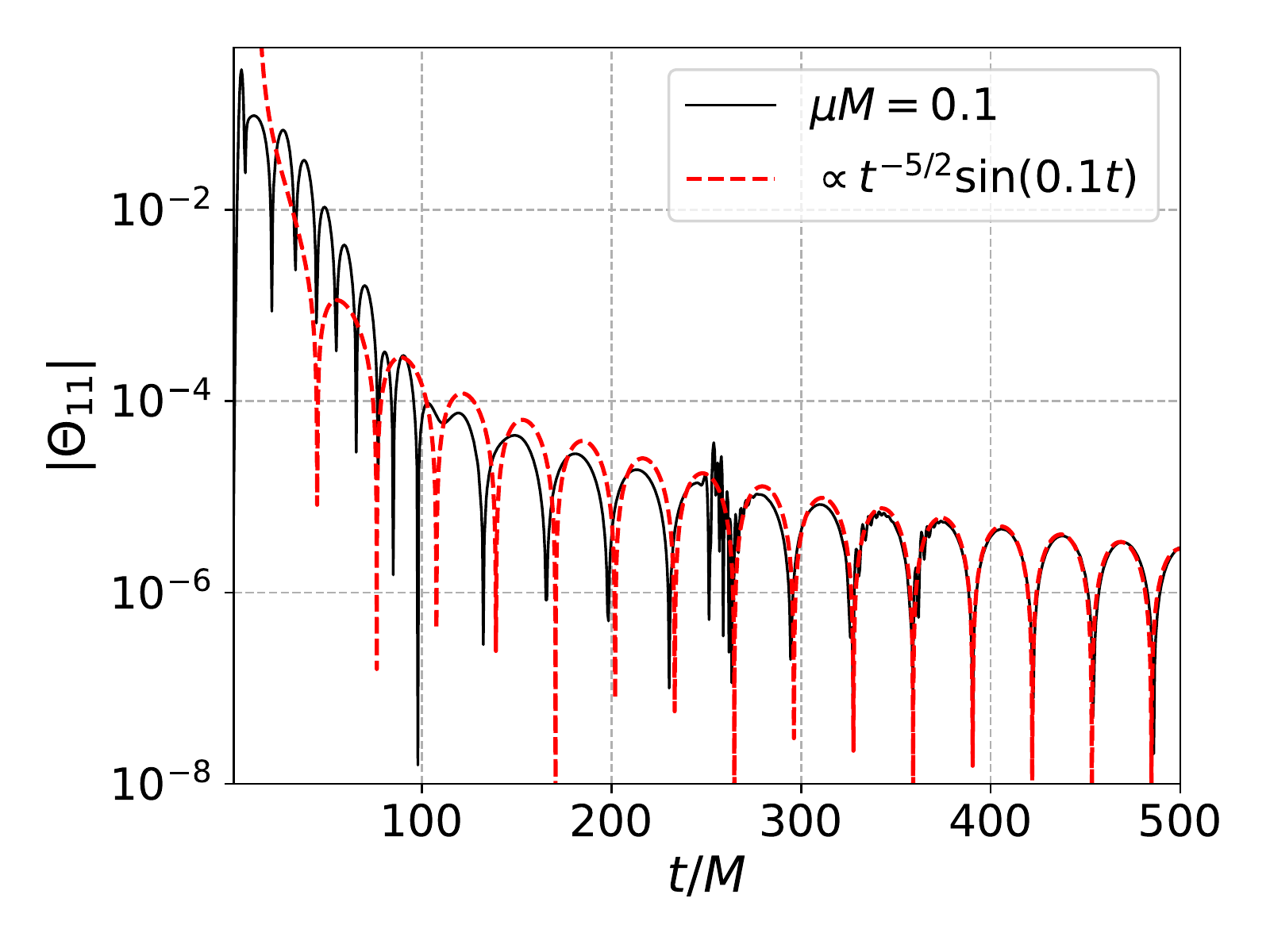}
    \includegraphics[width = 0.329\textwidth]{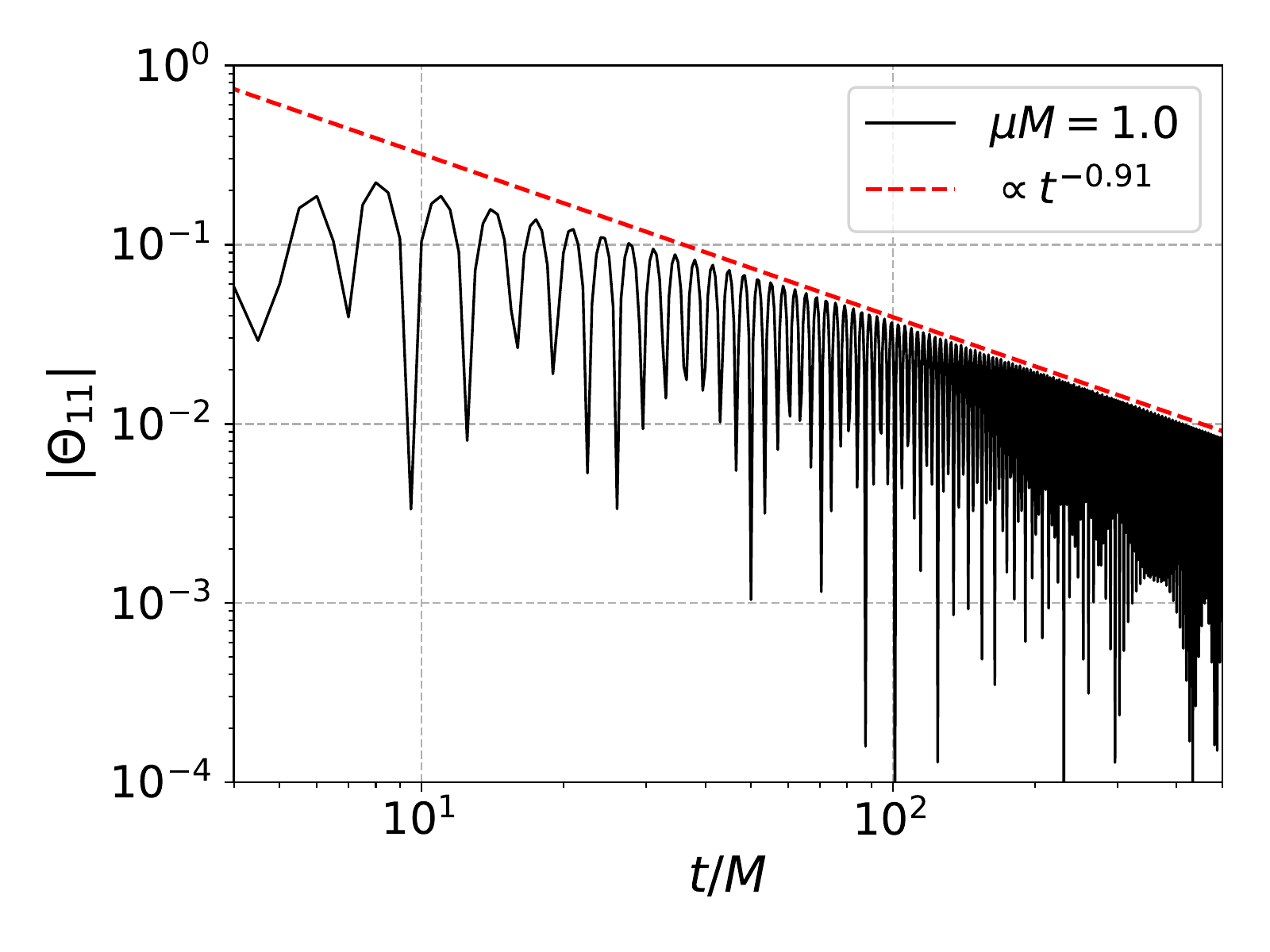}
    \caption{Evolution of the $\Theta_{11}$ mode,
    measured at $\rex=10$M, with
    $\mu M=0.0$ (left),
    $\mu M=0.1$ (middle),
    $\mu M=1.0$ (right)
    around a \bh with
    spin $a/M=0.1$.
    We show our numerical data (solid black lines) and compare it to the analytic solutions computed in \GR (dashed red lines).
    Left: The massless field follows a quasinormal ringdown with decay rate $M\bar{\nu} = -9.75\times 10^{-2}$.
    Middle: The field with $\mu M=0.1$ exhibits a short ringdown followed by an oscillating, intermediate-time power-law tail with frequency $\bar{\omega}\sim\mu$.
    Right: The field decays at a rate consistent with the very-late time power-law tail $\sim t^{-5/6}$.
}
    \label{fig:a01_fits}
\end{figure*}

\begin{figure*}[!htp]
    \centering
    \includegraphics[width = 0.329\textwidth]{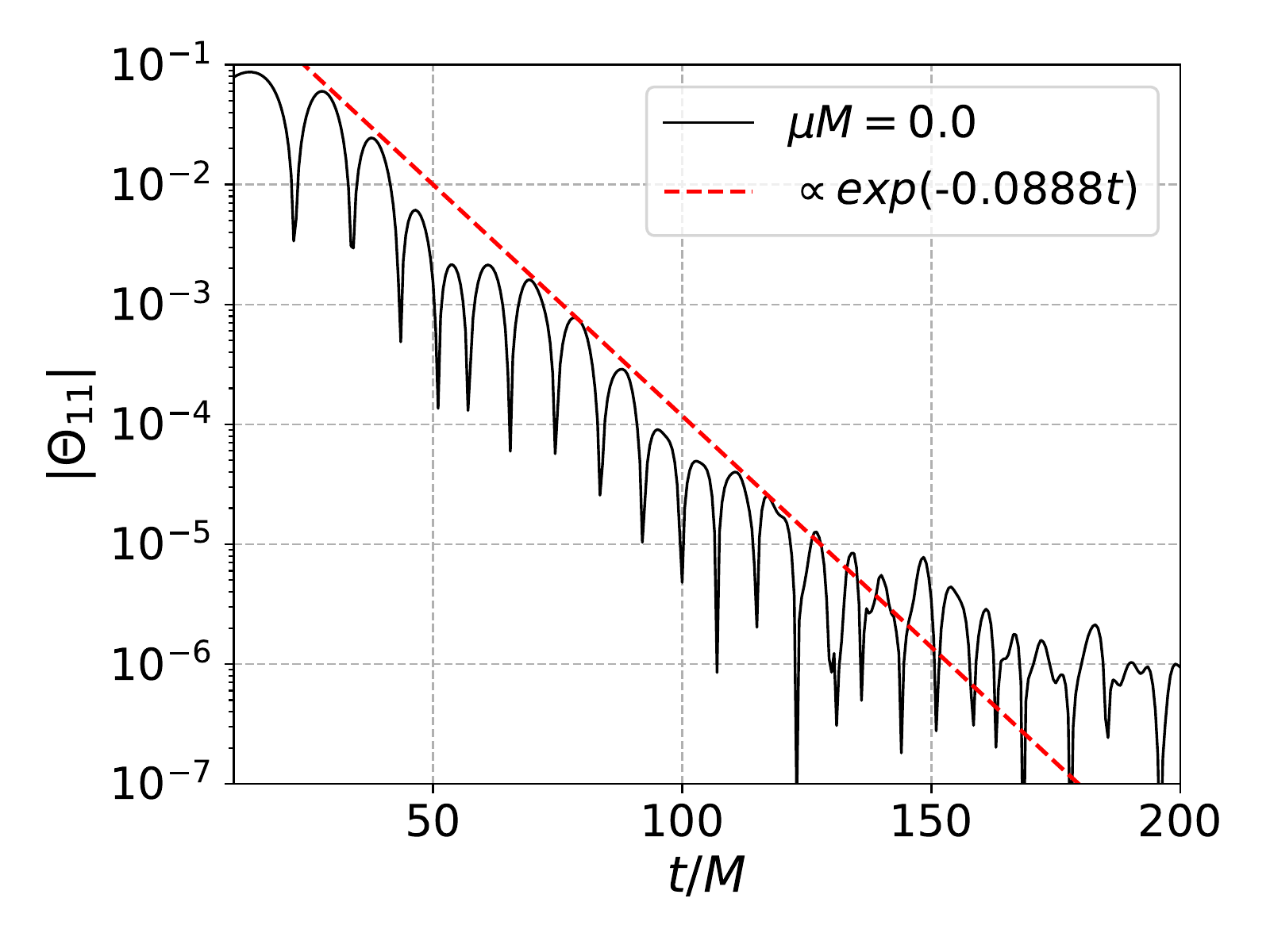}
    \includegraphics[width = 0.329\textwidth]{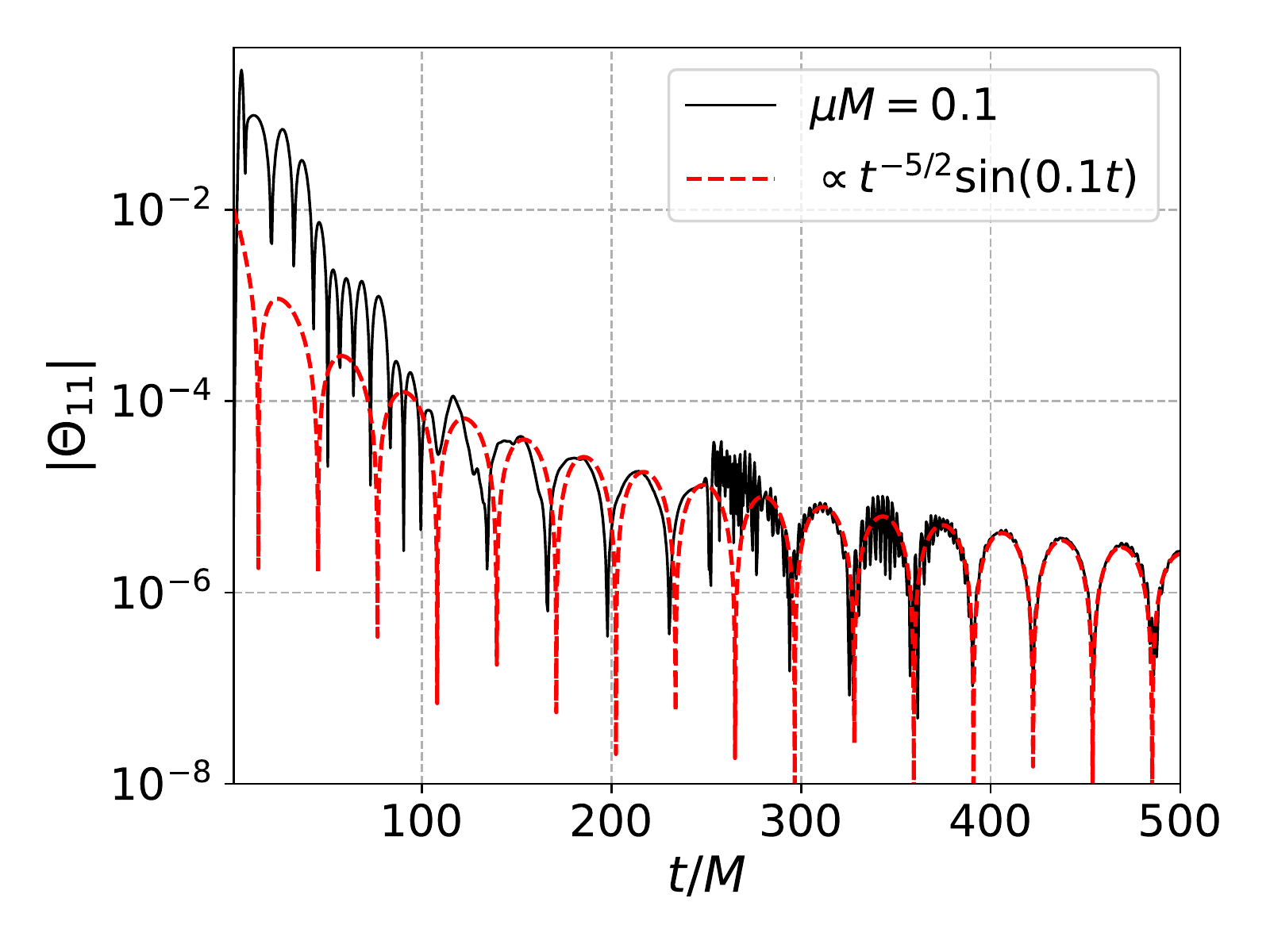}
    \includegraphics[width = 0.329\textwidth]{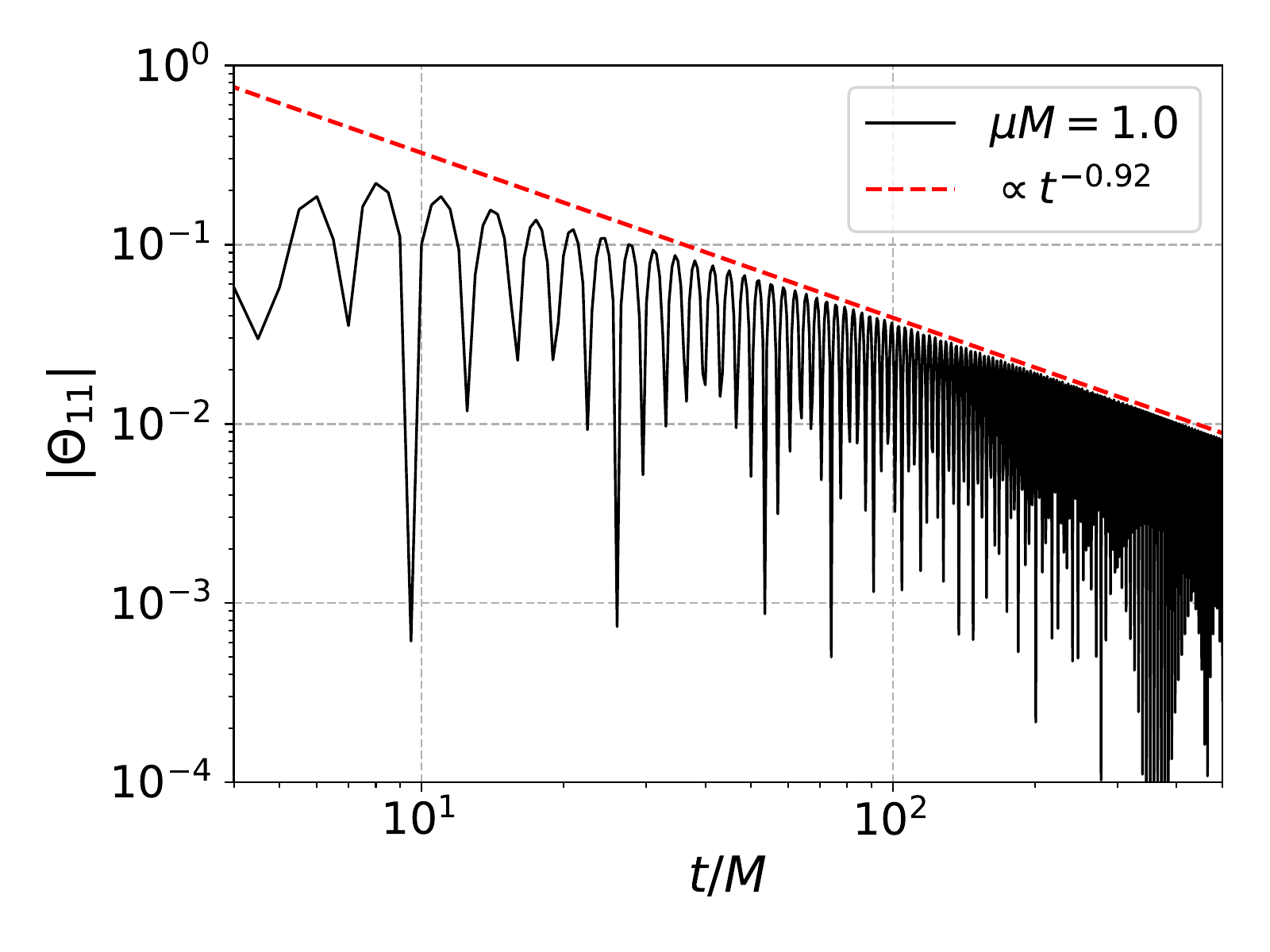}
    \caption{Same as Fig. \ref{fig:a01_fits} but for a BH with dimensionless spin $a/M = 0.7$.}
    \label{fig:a07_fits}
\end{figure*}

\begin{figure*}[!htp]
    \centering
    \includegraphics[width = 0.329\textwidth]{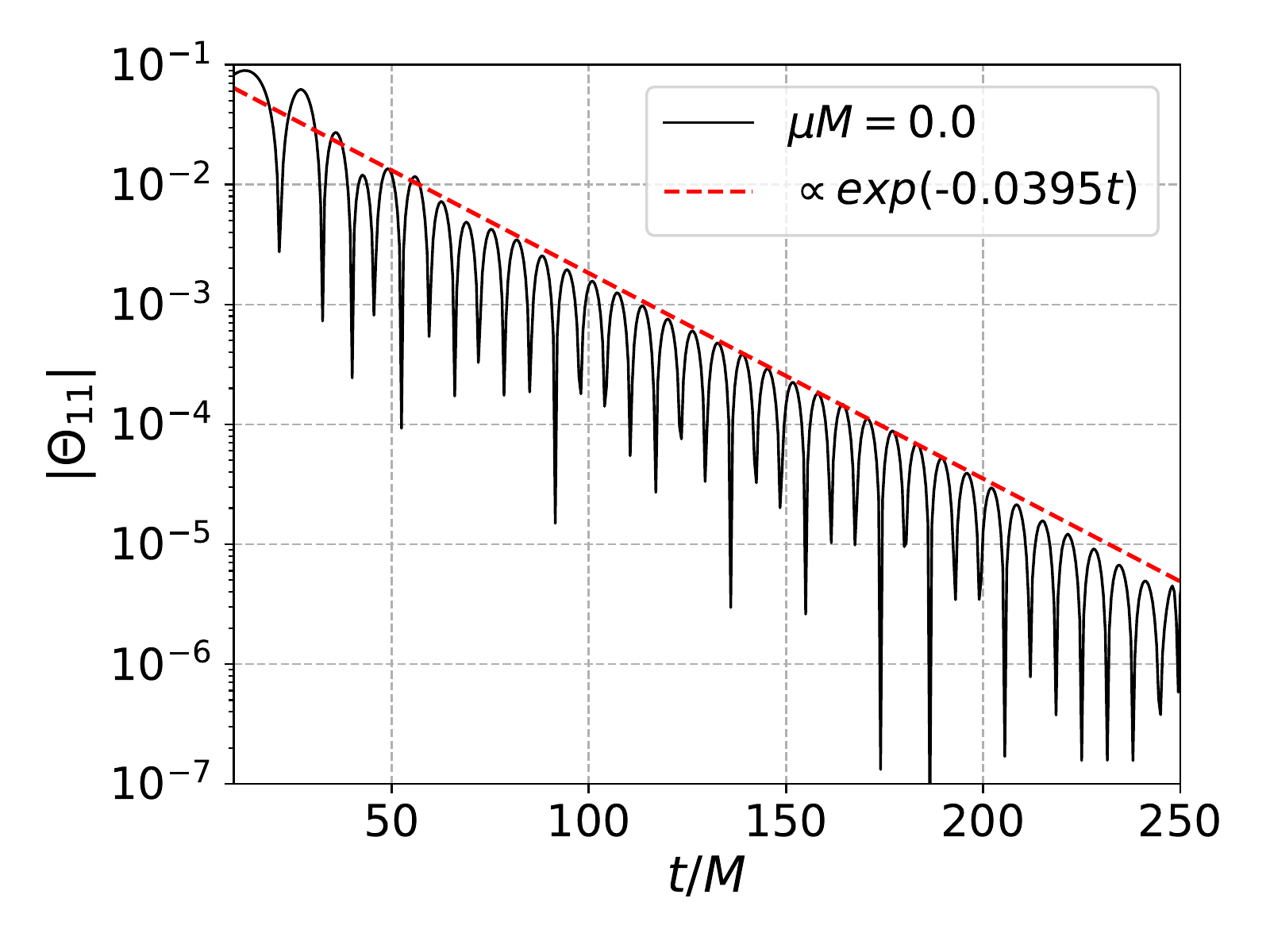}
    \includegraphics[width = 0.329\textwidth]{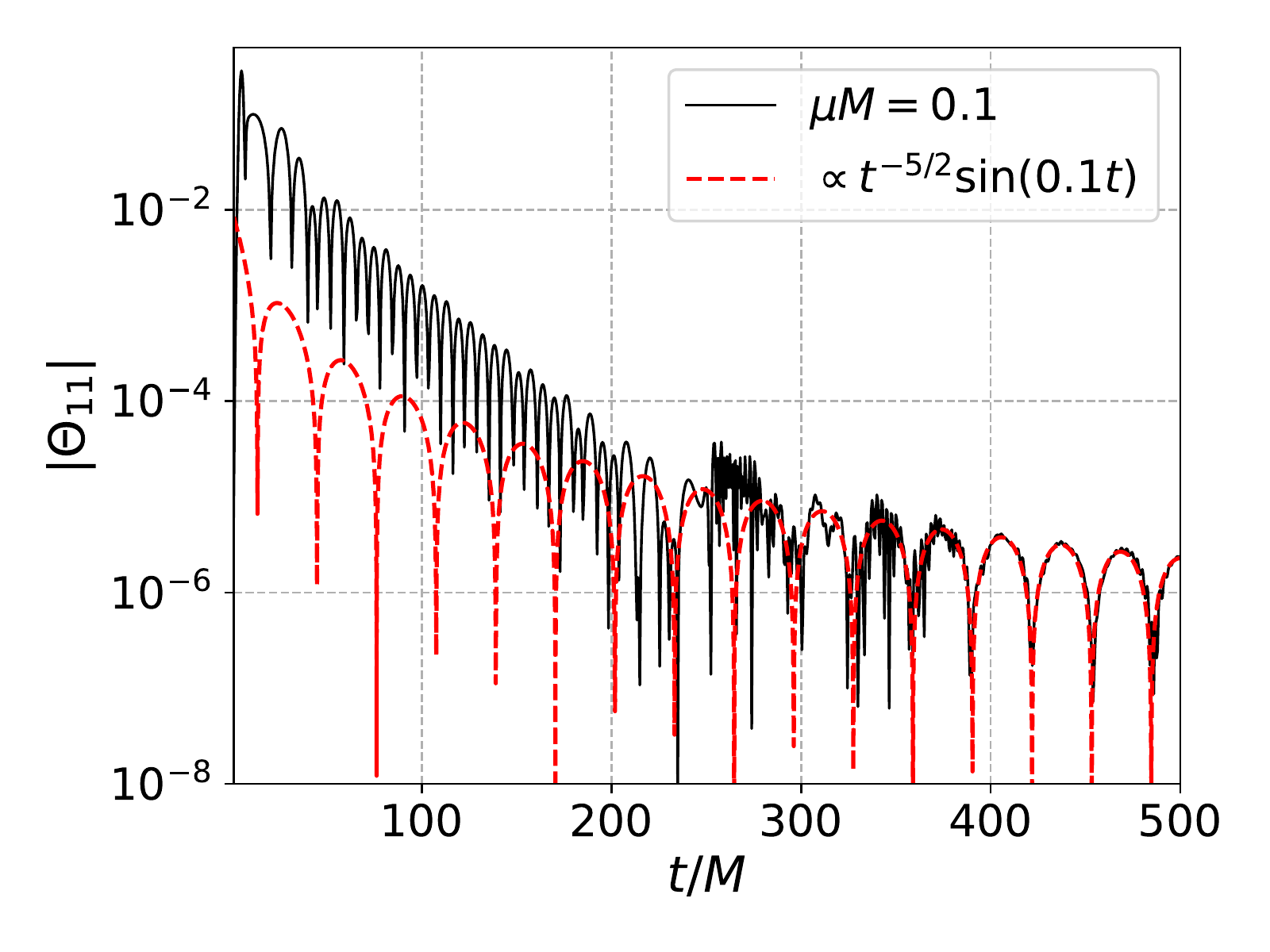}
    \includegraphics[width = 0.329\textwidth]{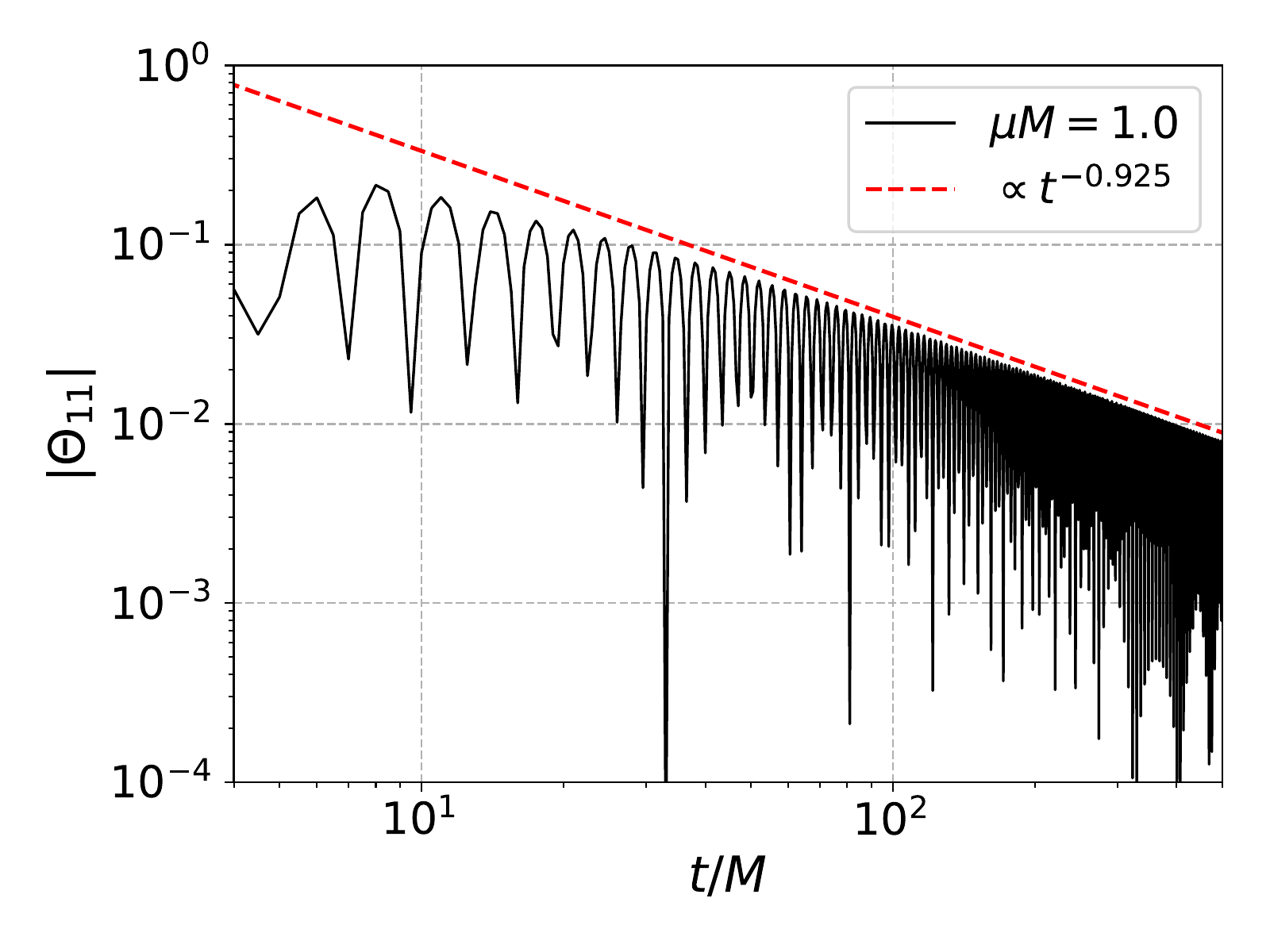}
    \caption{Same as Fig. \ref{fig:a01_fits} but for a \bh with dimensionless spin $a/M = 0.99$.}
    \label{fig:a99_fits}
\end{figure*}

This qualitative behaviour is
compatible
with that of massive scalar fields evolving around Kerr \bh{s} in \GR{.}
This begs the questions:
Are the decay rates for the ringdown, the \QBS, and the late time tails modified by the coupling to the Pontryagin density?
Does the Pontryagin density modify the scalar's profile near the \bh where the curvature is largest?
To answer the first question we compare the evolution of the $l=m=1$ \QBS in \GR and in \dCS gravity in Fig.~\ref{fig:ID_SF_BS}
and present fits of the ringdown and tails, respectively, in Figs.~\ref{fig:a01_fits}
-~\ref{fig:a99_fits}.
To address the second question, we inspect
the two-dimensional snapshots in
Fig.~\ref{fig:xyxz_t250_GRvsmdCS_in}
showing the scalar field profile near the \bh{.}

Let us start by analyzing the \QBS{.}
For this purpose, we
initialize the field $\Theta$ as a \QBS with $l=m=1$, mass parameter $\mu M=0.42$,
and oscillation frequency $M\bar{\omega}=0.409$
in the background of a Kerr \bh of spin $a/M=0.99$ using the solution of Ref.~\cite{Dolan:2007mj};
see ID4 in Sec.~\ref{subsec:IDSF}.
The $\Theta_{10}$ multipole is initialized as zero, but it assumes nonzero values during the evolution because it is sourced by the Pontryagin density;
see left panel of Fig.~\ref{fig:ID_SF_BS}.

We evolve two cases for this \QBS
and show its $l=m=1$ multipole measured at $\rex=10$M in the right panel of Fig.~\ref{fig:ID_SF_BS}.
In the first case, we set $\aCSh=0$, such that the field's equation of motion~\eqref{eq:dCSKG_decoup}
reduces to the Klein-Gordon equation in \GR (red dashed line in Fig.~\ref{fig:ID_SF_BS}).
In the second case, we set $\aCSh=1$ in Eq.~\eqref{eq:dCSKG_decoup}, such that the massive field is also sourced by the Pontryagin density
(black solid line in Fig.~\ref{fig:ID_SF_BS}).
Note, that we are not attempting to follow the evolution of the  superradiant instability as the timescales for massive scalars are prohibitively long for $3+1$ simulations
(with the shortest e-folding timescale being $\tau\sim10^7$M~\cite{Dolan:2007mj}).
Instead, our goal is to compare the evolution in the presence and absence of the Pontryagin term.
Both curves
exhibit an oscillatory signal which, after a short transition period, remains almost constant in time.
More importantly, both curves appear
indistinguishable.
Indeed, a quantitative analysis shows that
the curves
differ by
$|\Theta_{11,{\rm{GR}}}-\Theta_{11,{\rm{dCS}}}|/\Theta_{11,{\rm{GR}}}\sim 4\times 10^{-15}$, consistent with round off error.
In summary,
in the wave-zone,
the \QBS of a massive \dCS field around a highly spinning Kerr \bh is the same as that of a massive scalar in \GR{,} within numerical error.

We next analyze the quasinormal ringdown or massive power-law tail of
massive \dCS fields around a \bh with dimensionless spin
$a/M=0.1$ in Fig.~\ref{fig:a01_fits},
$a/M=0.7$ in Fig.~\ref{fig:a07_fits},
and
$a/M=0.99$ in Fig.~\ref{fig:a99_fits}.
We display the numerical data of $\Theta_{11}$ (black solid lines)
and a fit to its functional behaviour (red dashed lines).
In these figures, we present results for fields with
$\mu M=0.0$ (left panels),
$\mu M=0.1$ (middle panels)
and
$\mu M=1.0$ (right panels).

The massless field
(left panels of Figs.~\ref{fig:a01_fits}--~\ref{fig:a99_fits})
exhibits
a quasinormal ringdown pattern
$\propto \exp(-\imath\, \omega t)=\exp(-\imath\bar{\omega} t)\exp(\bar{\nu}t)$
where
$\bar{\omega}$ is the mode's oscillation frequency
and $\bar{\nu}$ indicates its decay (or growth) rate.
For the \bh with spin $a/M=0.1$
our numerical data gives
$M\omega= 0.29 -\imath\, 9.75\times 10^{-2}$,
and for a
\bh with $a/M=0.7$ we find
$M\omega = 0.395 - \imath\,8.88\times 10^{-2}$.
Both results are in good agreement,
within $4.2 \%$,
with frequency-domain computations in \GR{}~\cite{Berti:2009kk,Berti_ringdown,Berti:2004md}.
We note that the slight modulation of $\Theta_{11}$ around the
\bh with $a/M=0.7$,
seen in the left panel of Fig.~\ref{fig:a07_fits},
is likely due the superposition with an overtone close to the fundamental mode.
For the highly spinning \bh with $a/M = 0.99$, we find
$M\omega = 0.49 - \imath\,3.95\times 10^{-2}$,
which agrees with the frequency-domain result in \GR within $7.6\%$~\cite{Berti_ringdown}.
This is compatible with the numerical error
around highly spinning \bh{s}; see, e.g., Ref.~\cite{Witek:2012tr} and App.~\ref{App:ConvTest}.

The scalar with $\mu M = 0.1$, shown in the middle panels of Figs.~\ref{fig:a01_fits}--~\ref{fig:a99_fits},
exhibits a quasinormal ringdown followed by an oscillatory power-law tail.
We fit the latter with the functional behaviour
$\Theta_{11}\propto t^{-(l+3/2)}\sin(\mu t) = t^{-5/2} \sin(0.1 t)$ which is consistent with the intermediate time massive tails in \GR~\cite{Hod:1998ra}.
Here, ``Intermediate time'' refers to the time interval $1 < t/M < 1/(\mu M)^3 = 1000$.
Again, we find no measurable deviation from the \GR value
even around the highly rotating \bh{.}

For a larger mass, $\mu M=1.0$,
shown in the right panels of Figs.~\ref{fig:a01_fits}--~\ref{fig:a99_fits}, we find an oscillating power-law decay $\propto t^{-p}\sin(\mu t)$ with
$p= 0.910$ for $a/M =0.1$,
$p= 0.920$ for $a/M =0.7$, and
$p= 0.925$ for $a/M =0.99$.
This is within $\sim10\%$
of the universal, very late-time power-law tail
of $p=5/6$ known for massive fields in \GR~\cite{Koyama:2001ee,Koyama:2001qw,Burko:2004jn,Konoplya:2006gq,Hod:2013gmq}.
We note that we find comparable results for a mass parameter $\mu M=2.0$ (not displayed here).

\begin{figure*}
    \centering
    \includegraphics[width = 0.93\textwidth]{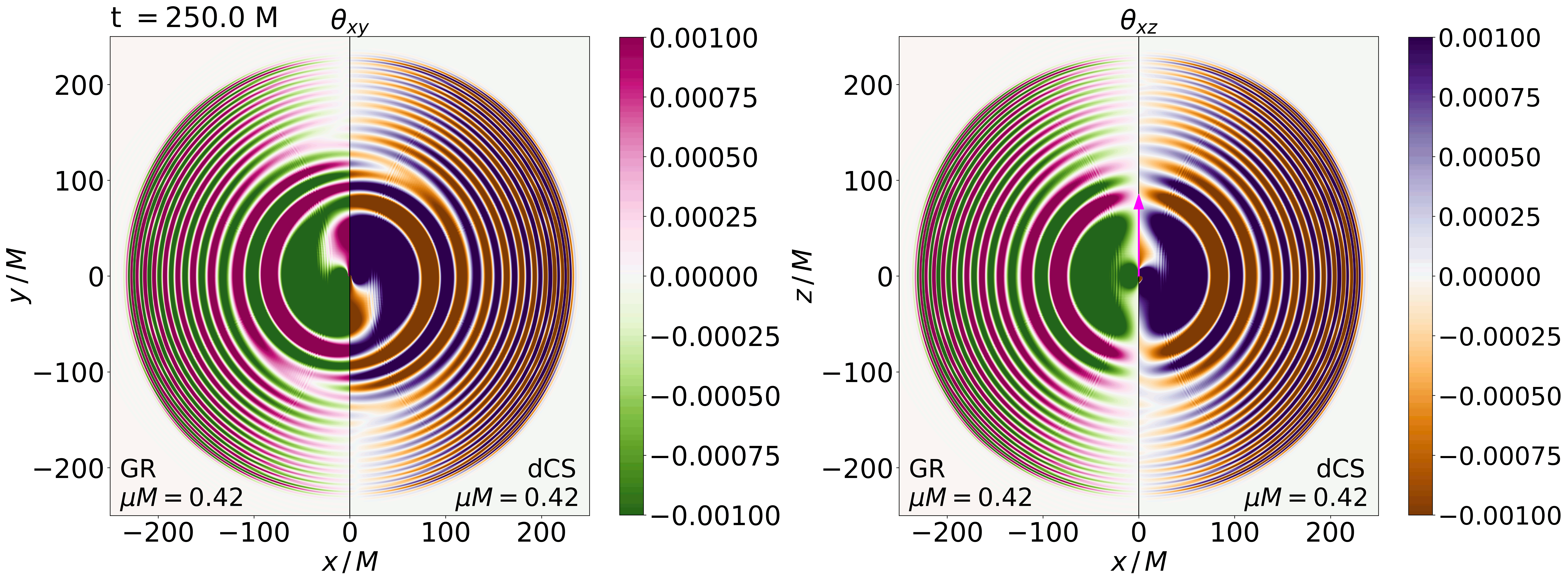}
    \caption{2D snapshots of the massive scalar field $\Theta$ at $t=250M$.
    The field with $\mu M=0.42$ is evolved around a \bh with $a/M=0.99$.
    We show the field in the equatorial plane (left panel) and in the $xz$ plane (right panel).
    The latter also indicates the \bh{'s} rotation axis with a magenta arrow.
    We compare the evolution of a massive scalar field in \GR ($\aCSh=0$, left halves) and in \dCS gravity ($\aCSh=1$, right halves).
    At large distances, the scalars' profiles appear almost identical.}
    \label{fig:xyxz_t250_GRvsmdCS_out}
\end{figure*}

\begin{figure*}
    \centering
    \includegraphics[width = 0.9\textwidth]{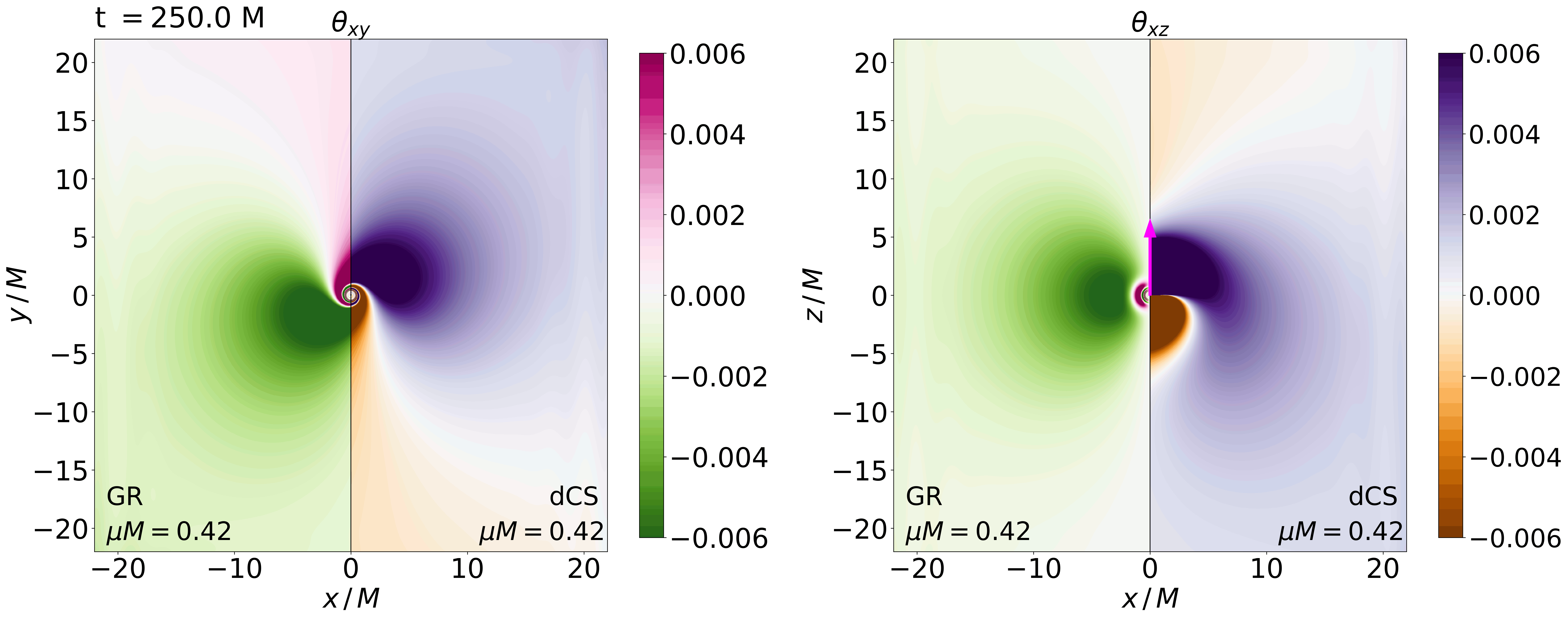}
    \caption{Same as Fig.~\ref{fig:xyxz_t250_GRvsmdCS_out}, but zoomed in near the \bh{.}
    There is no visible evidence of the Pontryagin density affecting the scalar cloud in the $xy$ plane (left panel).
    In the $xz$ plane (right panel), however, we see that the Pontryagin density sources the $l=1,m=0$ scalar multipole,
    absent in \GR{,}
    that exhibits the same features as the massless \dCS hair near the \bh and then transitions to an oscillating \QBS configuration at large distances.
    }
    \label{fig:xyxz_t250_GRvsmdCS_in}
\end{figure*}

To aid our analysis,
we present 2D snapshots of the scalar field with $\mu M=0.42$ evolved around a \bh with spin $a/M=0.99$ at time
$t=250M$.
Fig.~\ref{fig:xyxz_t250_GRvsmdCS_out} shows a 
color map of the scalar's amplitude in the equatorial plane (left panel) and in the  $xz$ plane (right panel) on the entire numerical domain.
Fig.~\ref{fig:xyxz_t250_GRvsmdCS_in} zooms in close to the \bh{.}
The left half of each panel shows the massive scalar field in \GR (i.e., with $\aCSh=0$),
while the right half shows the massive field in \dCS gravity (i.e., with $\aCSh=1$).

Comparing the two cases, we see that the scalar cloud's profile in the equatorial plane is essentially unaffected by the nonminimal coupling to gravity
via the Pontryagin density
both in the far region (see Fig.~\ref{fig:xyxz_t250_GRvsmdCS_out}) and near the \bh (see Fig.~\ref{fig:xyxz_t250_GRvsmdCS_in}).
One is brought to similar conclusions about the far-region when considering the snapshots in the $xz$ plane:
the oscillatory pattern of the nonminimally coupled massive scalar field matches that of the minimally coupled one {\textit{at large distances}}.
This pattern changes close to the \bh as can be seen in the right panel of Fig.~\ref{fig:xyxz_t250_GRvsmdCS_in}.
Here, the spacetime curvature and, hence, the Pontryagin density is largest and it sources an axi-symmetric, dipolar scalar structure along the $z$-axis.
This dipole connects smoothly to the oscillation pattern at distances larger than $\sim 1/(\mu M)^2$.
Note, that in the absence of the mass-term this dipole would correspond to the \dCS hair given in Eq.~\eqref{eq:ID3}.
We also note, that such a $l=1,m=0$ structure is absent if there is no coupling to the Pontryagin density; see left half of the right panel in Fig.~\ref{fig:xyxz_t250_GRvsmdCS_in}.


In summary, the evolution of the massive $l=m=1$ dipole
around rotating \bh{s} in \dCS gravity appears consistent with that of massive fields in \GR{} at sufficiently large distances.
It only differs near the \bh along the axis of rotation,
where $m=0$ scalar multipoles are sourced by the
Pontryagin density.
We have demonstrated that this result is robust
against different choices of initial data for $\Theta_{10}$ and $\Theta_{11}$,
and that it holds also for highly spinning \bh{s}.
The simulations indicate that, at late times, the scalar field approaches an oscillating,
quasistationary dipole configuration.

\subsection{Effect of the mass term on the evolution of the dynamical Chern-Simons hair}\label{subsec:resultsdCShair}
In this section we analyze how
the growth and evolution of
the \dCS hair is affected by the
mass term.
Therefore, we evolve the massive \dCS field
around a single rotating \bh
according to Eqs.~\eqref{eq:dtdCSKG_decoupBSSN}.
We typically initialize $\Theta_{10}=0$ and
$\Theta_{11}$ as a Gaussian or as a \QBS{.}

In the left panels of Fig.~\ref{fig:ID2_all_spin_all_mass}
we present the time evolution of $\Theta_{10}$,
measured at $\rex=10M$,
for a variety of the axion's mass parameter $\mu M$ and for \bh spins $a/M=0.1$ (top), $a/M=0.7$ (middle), and $a/M=0.99$ (bottom).
The massless field (solid red lines) grows over
$t\sim100$M until it reaches a constant value
that corresponds to the \dCS hair at this extraction radius; see Eq.~\eqref{eq:ID3}.
The hair's final magnitude $|\Theta_{10}|$ increases with increasing \bh spin, in agreement with the solutions of Ref.~\cite{Stein:2014xba}.\\

As the mass parameter $\mu M$ of the scalar field increases, the magnitude of the \dCS hair, $\Theta_{10}$, decreases.
Specifically, for small masses $\mu M=0.1$ (dashed-dotted orange lines), the \dCS hair
approaches a constant magnitude that is smaller than in the massless case.
For intermediate masses, $\mu M\sim0.3\ldots0.4$ (dashed gray lines),
the \dCS hair reaches a smaller magnitude and, more importantly, it oscillates.
The oscillation frequencies of the
$\Theta_{10}$ and $\Theta_{11}$ modes are determined by the mass parameter.
They are consistent with the \QBS frequencies in \GR~\cite{Detweiler:1980uk,Dolan:2007mj,Baumann:2019eav},
that are given by
\begin{equation}
\omega_{lmn}\simeq\mu\left(1-\frac{(\mu M)^2}{2(l+n+1)^2}\right)
\,,
\end{equation}
in the limit $\mu M\ll1$.
We refer to
Sec.~\ref{subsec:mdCS_spec}
for the detailed spectral analysis.
This response of the $\Theta_{10}$ mode is present for both Gaussian and \QBS initial data of $\Theta_{11}$; compare the left panels of Figs.~\ref{fig:ID2_all_spin_all_mass} and~\ref{fig:ID_SF_BS}.
As the field's mass parameter exceeds $\mu M\gtrsim1.0$,
the magnitude of $\Theta_{10}$ is reduced
or entirely suppressed,
as illustrated in the left panel of Fig.~\ref{fig:ID2_all_spin_all_mass}
for $\mu M=1.0, 2.0$.

\begin{figure}[t]
    \centering
    \includegraphics[width = \columnwidth]{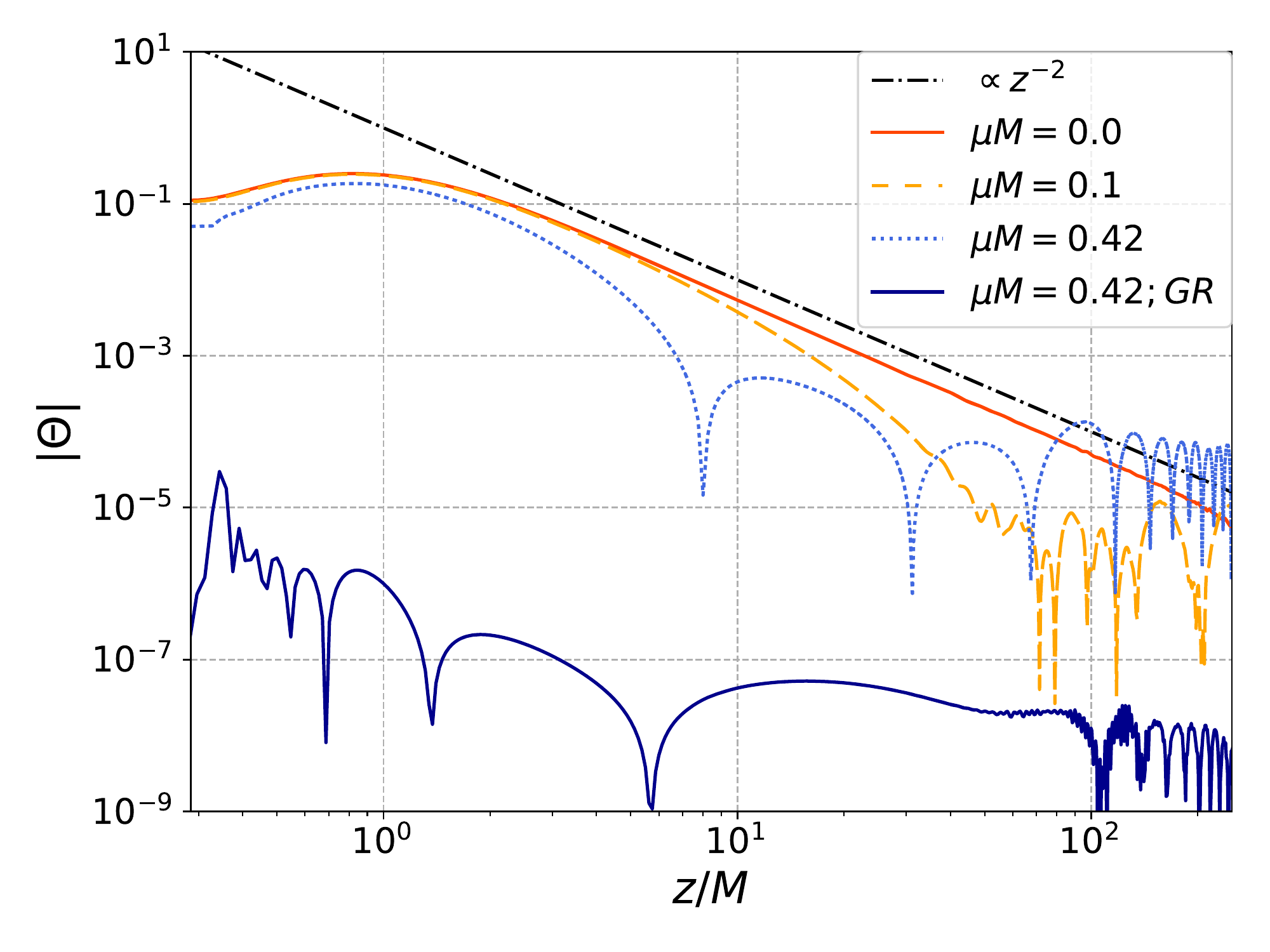}
    \caption{Profile of the massive \dCS field $\Theta$ with mass parameters
    $\mu M=0.0$ (solid red line),
    $\mu M=0.1$ (dashed yellow line),
    $\mu M=0.42$ (dotted blue line)
    along the rotation axis,
    evolved around a \bh with $a/M=0.99$ at $t=500$M.
    For comparison, we display the expected fall-off of the \dCS hair with vanishing mass (dash-dotted black line)
    and the profile of a scalar field with $\mu M=0.42$ in \GR  (solid dark blue line).
    }
    \label{fig:Th_vs_z}
\end{figure}

To understand this behaviour, we plot the late-time profile of the field $\Theta$,
evolved around a \bh with spin $a/M=0.99$
for different $\mu M$,
along the \bh{'s} rotation axis in Fig.~\ref{fig:Th_vs_z}.
For a massless field (solid red line),
this late-time solution is the \dCS hair sourced by the Pontryagin density.
For comparison we also display the $\sim 1/z^2$ fall-off
(dash-dotted black line)
found for analytical solutions in the small-spin approximation~\cite{Yunes:2009hc},
and we find excellent agreement.
For a massive scalar field with a small mass $\mu M=0.1$ (dashed orange line), we find that it behaves like the massless case until about $z\sim 100M$ from the \bh{.}
At larger distances from the \bh{}
the field oscillates.
Both the location and the frequency of this oscillation is determined by the mass parameter.
The position of this transition appears consistent with the peak of the \QBS found in \GR
for small mass parameters
given by~\cite{Arvanitaki:2010sy,Yoshino:2013ofa,Brito:2015oca}
\begin{equation}
\label{eq:SRCloudRmax}
r_{\rm{\QBS, max}} \sim \frac{l(l+1)}{(\mu M)^2} M
\,.
\end{equation}
For a field with $l=1$ and
$\mu M = 0.1, 0.42, 1.0$
this corresponds to
$r_{\rm{\QBS, max}} \sim 200M, 11.3 M, 2 M$.
We find consistent results for the \dCS field with mass $\mu M = 0.42$ (dotted light blue line in Fig.~\ref{fig:Th_vs_z}):
it follows a $\sim1/z^2$ fall-off close to the \bh and begins to oscillate around $z\sim 10M$
where its frequency determined by $\mu M$
at larger distances.
For comparison  we also show the late-time profile of a field with $\mu M =0.42$ evolved in \GR (solid blue line),
and demonstrate that, along the z-axis, it remains several orders of magnitude smaller than its counterpart in \dCS gravity.
We interpret this as a further indication that the oscillation of the \dCS hair is indeed induced by the mass term.

\begin{figure*}[]
    \centering
   \includegraphics[width = 0.93\textwidth]{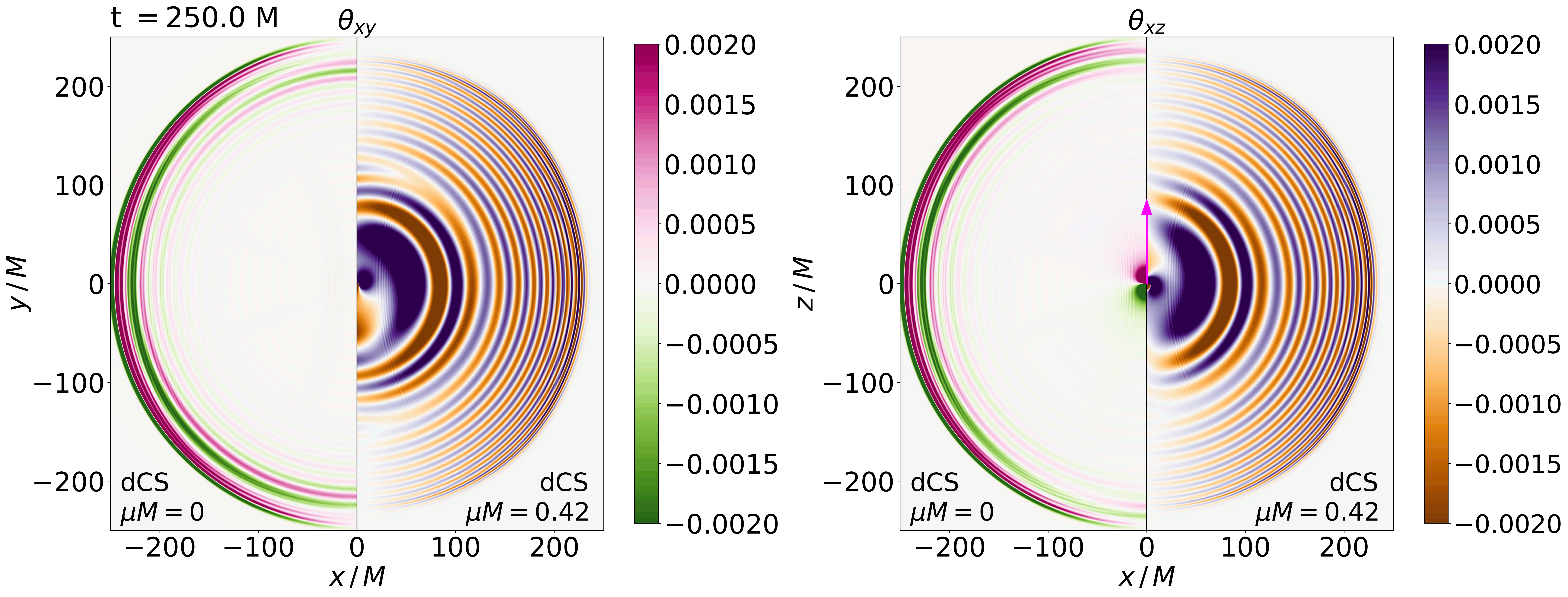}
     \caption{2D snapshots of the \dCS field $\Theta$, initialized as a $l=m=1$ Gaussian and evolved around a \bh with $a/M=0.99$, at $t=250M$.
     We show the field in the equatorial plane (left panel) and in the $xz$ plane (right panel),
     where the magenta arrow indicates the \bh{'s} rotation axis.
     We compare the evolution of a massless \dCS field ($\mu M=0.0$, left halves) and that of a massive \dCS field with $\mu M=0.42$ (right halves).
     Left panel: the massless field (left half) decays as the $l=m=1$ perturbation is propagating off the grid. The massive field (right half) develops an oscillating scalar cloud.
     Right panel: the massless field (left half) develops the \dCS hair, i.e., a $l=1$, $m=0$ dipolar configuration. The massive field (right half), exhibits a suppressed dipolar configuration close to the \bh that is smoothly connected to an oscillating cloud at large distances.}
    \label{fig:xyxz_t250_dCSvsmdCS_out}
\end{figure*}

\begin{figure*}[]
    \centering
    \includegraphics[width = 0.9\textwidth]{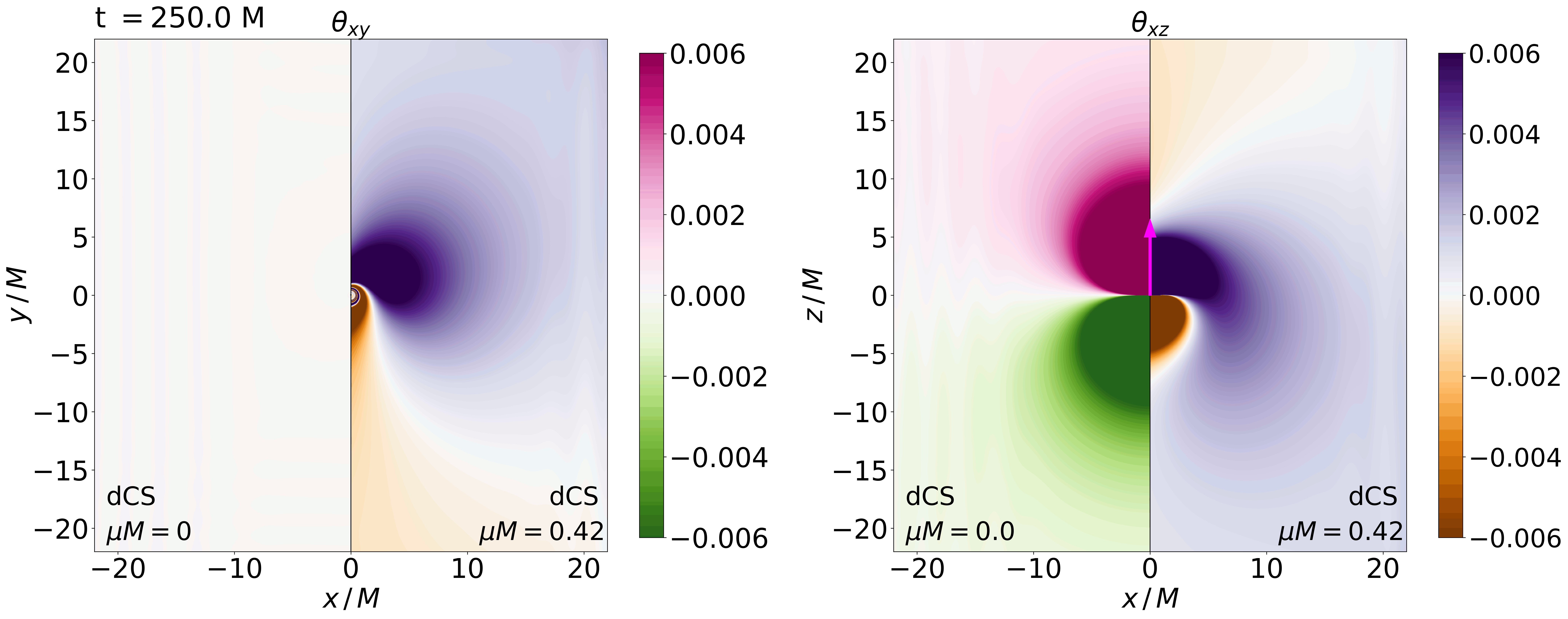}
     \caption{Same as Fig.~\ref{fig:xyxz_t250_dCSvsmdCS_out} but zoomed in near the \bh{.}
     Left panel: \dCS field in the $xy$ plane. The massless field (left half) has decayed, while the field with $\mu M=0.42$ (right half) has formed an oscillating scalar cloud.
     Right panel: \dCS field in the $xz$ plane. The massless field (left half) develops the axi-symmetric\dCS hair.
     The massive \dCS field develops a similar dipolar structure with a smaller amplitude that smoothly transitions to the oscillating cloud induced by the mass term.}
    \label{fig:xyxz_t250_dCSvsmdCS_in}
\end{figure*}

This behaviour becomes even clearer in the 2D snapshots in
Figs.~\ref{fig:xyxz_t250_dCSvsmdCS_out}
and~\ref{fig:xyxz_t250_dCSvsmdCS_in}
where we display the amplitude of the scalar field
at $t=250M$
on the entire numerical domain and close to the \bh{,}
respectively.
Specifically, we display $\Theta$,
evolved around a \bh with spin $a/M=0.99$,
in the equatorial plane (left panels) and the $xz$ plane (right panels).
We compare the massive \dCS field with $\mu M=0.42$ (right halves of each panel)
against the massless \dCS field (left halves of each panel).
In the massless case,
the initial $l=m=1$ perturbation propagates outwards
so that the field vanishes in the equatorial plane
(see left half of left panel)
and only the $l=1$, $m=0$ \dCS hair that is sourced by the Pontryagin density remains (see left half of right panel).
This is to be expected because the Pontryagin density vanishes in the equatorial plane due to the axisymmetry of the background.
The massive case, displayed in the right halves of Figs.~\ref{fig:xyxz_t250_dCSvsmdCS_out}
and~\ref{fig:xyxz_t250_dCSvsmdCS_in},
exhibits two distinct features:
(i) it also has a dipolar structure close to the \bh{,}
but with a smaller magnitude than in the massless case (see right panel in Fig.~\ref{fig:xyxz_t250_dCSvsmdCS_in}), and
(ii) it has an oscillatory pattern at large distances,
both in the equatorial and in the $xz$ plane,
due to the mass term.

For the results presented thus far,
we only initialized the $l=m=1$ multipole of the field $\Theta$.
Hence, our findings refer to the growth of the
$l=1$, $m=0$ scalar mode.
How robust are our results against different types of initial data?
What would be the fate of the \dCS hair if it was already present in the initial data?
And, reversely, does the formation of the oscillating \dCS dipole require the presence of the massive $l=m=1$ mode?

\begin{figure*}
    \centering
    \includegraphics[width = 0.9\textwidth]{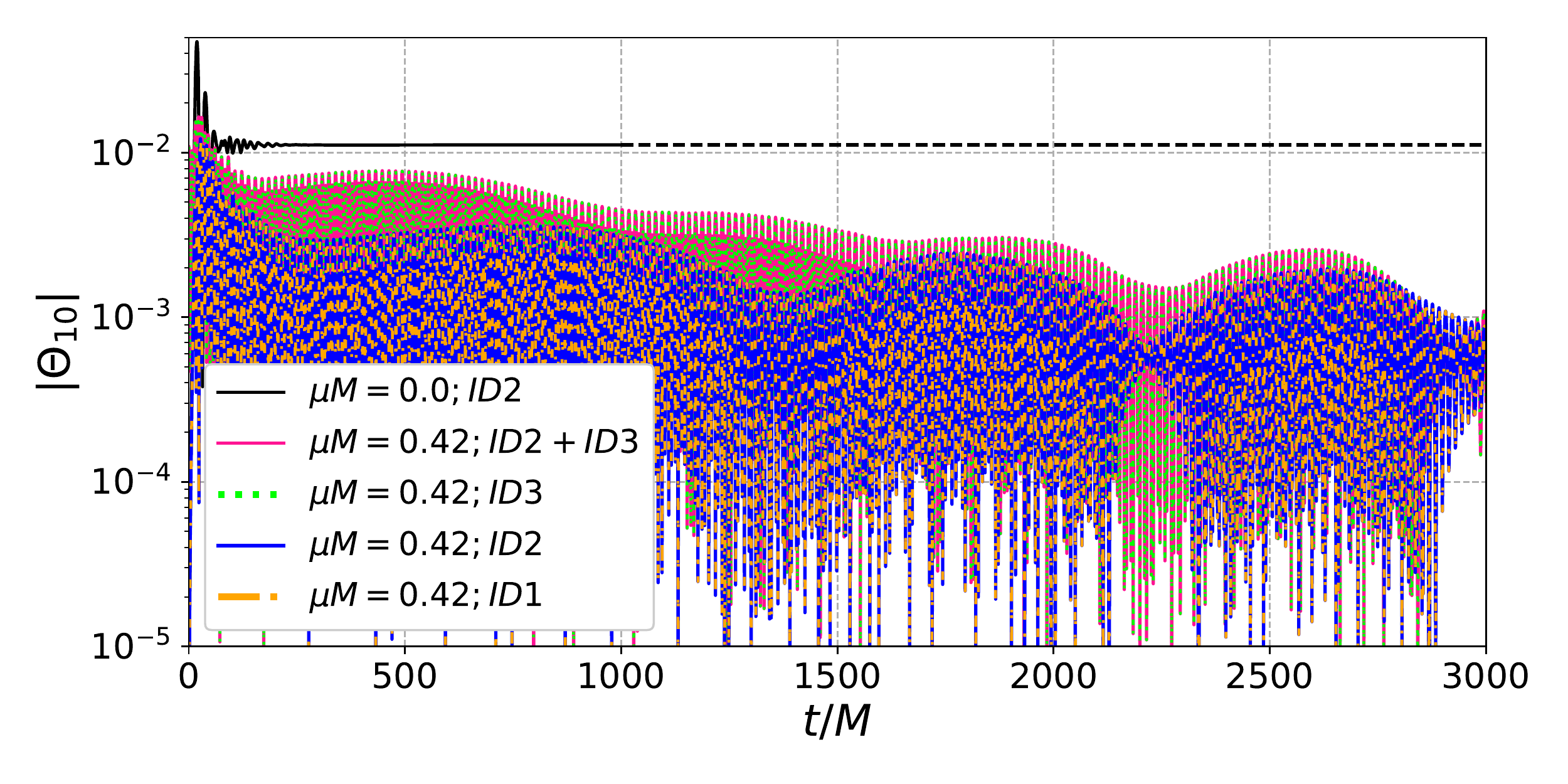}
    \caption{Evolution of the $l = 1, m= 0$ mode of a field with $\mu M=0.42$ evolving in the background of a \bh with $a/M=0.99$ in \dCS gravity ($\aCSh~= 1$).
    The field is initialized as an $l=m=1$ Gaussian, ID2 with $\Sigma_{11}$ in Eq.~\eqref{eq:ID2} (solid blue line), or as an $l=m=1$ Gaussian plus the \dCS hair, ID2+ID3 (solid pink).
    We compare the former with a simulation where the field is initially zero, ID1, (orange dash-dotted line) and the latter with a simulation where the field is initialized as the \dCS hair, ID3, indicating $\Theta_{11} = 0$ (dotted green).
    We find that the dCS, $\Theta_{10}$, multipole does not differ between these two comparisons, indicating that our results for the growth and evolution of  $\Theta_{10}$ are independent of our choice of initial data for the massive, $\Theta_{11}$, multipole.
    As an additional comparison, we show the evolution of a massless field with $l=m=1$ Gaussian initial data (solid black line).
    We ran the massless simulation until $t=1000$M and indicate its extrapolated value (dashed black line).
    }
    \label{fig:ID2_vs_ID2_ID3}
\end{figure*}

We present answers to these questions in Fig.~\ref{fig:ID2_vs_ID2_ID3},
where we compare the evolution of $|\Theta_{10}|$
of a field with $\mu M=0.42$ around a \bh with $a/M=0.99$
for different initial data:
\begin{enumerate*}[label={(\roman*)}]
\item a Gaussian shell with $\Sigma_{11}$ in Eq.~\eqref{eq:ID2},
and $\Theta_{10}(t=0)=0$ (solid blue line),
\item a $\Sigma_{11}$ Gaussian shell and $\Theta_{10}$ given by the \dCS hair in Eq.~\eqref{eq:ID3} (solid pink line),
\item $\Theta_{11}(t=0)=0$ and $\Theta_{10}$ given by the \dCS hair in Eq.~\eqref{eq:ID3} (dotted green line),
and \item $\Theta(t=0)=0$ (dash-dotted orange line).
\end{enumerate*}
For comparison we also show the evolution of a massless field with ID2 ($\Sigma_{11}$ in Eq.~\eqref{eq:ID2}) and $\Theta_{10}=0$ (solid black line).
The initially hairy, massive field (solid pink line) decays and approaches the magnitude of the initially hairless, massive field (solid blue line).
That is, the transformation of the time-independent hairy solution in massless \dCS gravity
into an oscillating dipole with a lower amplitude appears to be a robust feature in massive \dCS gravity
and independent of the initial data choice for the $l = 1, m =0$ multipole.
Similarly, we do not find any difference in the evolution of $\Theta_{10}$ for different initial data of the $l=m=1$ mode. Compare, e.g., the field initialized as zero (dash-dotted orange line) with the one initialized as a $\Sigma_{11}$ Gaussian shell (solid blue line).

In summary, we investigated the effect of a mass term on the formation and evolution of the \dCS hair.
We have identified
a new, oscillating hairy solution in massive \dCS gravity
consisting of an axi-symmetric
dipole near the \bh
that smoothly transitions to an oscillating dipole
at large distances.
This transition and the oscillation frequency are determined by the field's mass parameter.
We verified that the formation of an oscillating dipole in massive \dCS gravity is a robust feature
independent of the initial data.

\subsection{Characteristic spectra in massive \dCS gravity}\label{subsec:mdCS_spec}
In this section we determine the frequency spectra of the
 $l=1, m=0,1$ multipoles of the
\dCS axion.
To do so, we perform a
spectral analysis of our data.

First, we obtain the time series data of the scalar field's multipoles, $\Theta_{lm}(t,\rex)$, by
interpolating the field $\Theta$ onto spheres of constant extraction radii $\rex$ and projecting it onto $s=0$ spherical harmonics.
We then apply a window function (typically, a Blackman-Harris window~\cite{Harris:1978})
to the time series data to mitigate the effects of the initial transient,
and compute the Fourier amplitude
\begin{equation}
    \tilde{\Theta}_{lm}(\omega_i) = \frac{1}{N}\sum_{j=0}^{N-1}\Theta_{lm}(t_j) e^{-i\omega_i t_j} \,,
\end{equation}
via a Fast Fourier Transform algorithm.
Finally, we compute the power spectrum,
$P_{lm}(\omega)=|\tilde{\Theta}_{lm}(\omega)|^2$,
of the $(l,\,m)$-mode.
A monochromatic signal,
$\Theta_{lm}\sim \exp\left(-i\bar{\omega}t+\bar{\nu} t\right)$,
with characteristic frequency $\bar{\omega}$ and growth or decay rate $\bar{\nu}$,
corresponds to a power spectrum
consisting of a peak described by a three-parameter Lorentzian (or Breit-Wigner) function
\begin{equation}\label{eq:Lorentzian}
\Lambda(\omega;\bar{I},\bar{\omega},\bar{\nu})=\bar{I}\frac{\bar{\nu}^2}{\left((\omega-\bar{\omega})^2+\bar{\nu}^2\right)}
\,.
\end{equation}
The peak height is set by $\bar{I}$, while its position and width are determined by $\bar{\omega}$ and $\bar{\nu}$, respectively.

\begin{figure}[h!]
    \centering
    \includegraphics[width = \columnwidth]{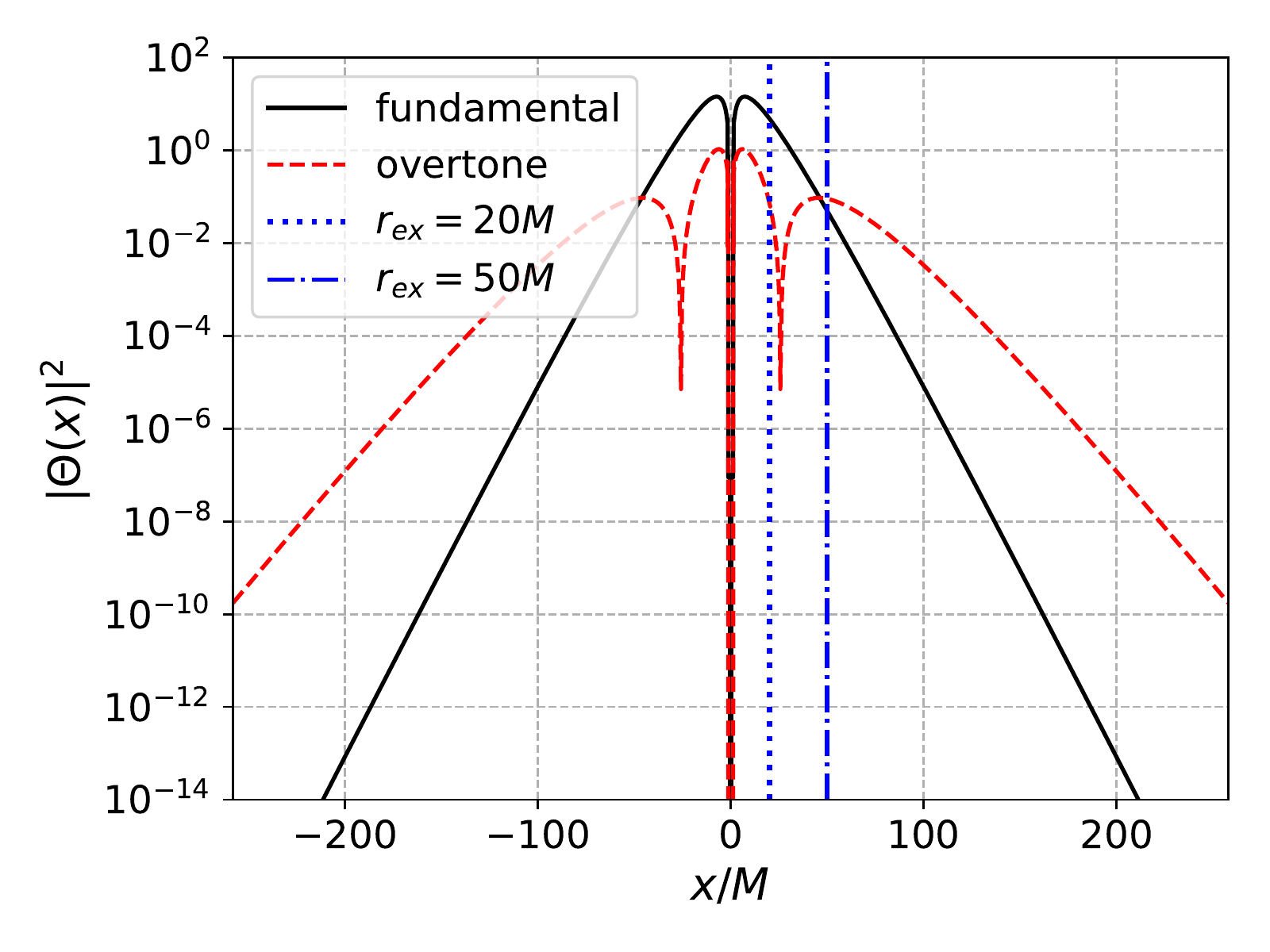}
    \caption{Initial profiles of massive scalar field $\Theta$ along the x-axis with mass $\mu M = 0.42$ evolving in the background of a BH with spin $a/M = 0.99$ in \dCS gravity ($\aCSh = 1$).
    We initialize the scalar field as a \QBS, ID4 in Sec.~\ref{subsec:IDSF},
    with fundamental frequency $M \bar{\omega} = 0.409$ (solid black) and overtone frequency $M\bar{\omega} = 0.415$ (dashed red).
    For the spectal analysis, we extract the multipoles $\Theta_{10}$ and $\Theta_{11}$ at
    $\rex=20$M (dotted blue) and $\rex=50$M (dash-dotted blue).
    } \label{fig:fund_and_overt}
\end{figure}

Ideally, we would like the time series extracted from our numerical simulations to be monochromatic signals with characteristic frequency $\bar{\omega}=\omega_{lmn}$, where $l$ and $m$ are the harmonic mode indices and $n$ is the overtone number.
In practise, however, the extracted signals are projected only into $(lm)$ multipoles, so they still are a superposition of the fundamental ($n=0$) and overtone modes.
Depending on their relative amplitude, the superposition of fundamental and overtone modes can give rise to strong modulations of the signal and beating patterns~\cite{Witek:2012tr}.
To overcome this complication we fix the extraction radius $\rex$ to a location where the overtone mode nearly vanishes.
In  Fig.~\ref{fig:fund_and_overt} we present the profiles of the fundamental and first overtone modes
of the initial scalar field, determined by a
\QBS
with $l=m=1$ and mass parameter $\mu M=0.42$ around a \bh with spin $a/M=0.99$.
In this case, the overtone mode has a node at $\rex\simeq20$M.
Thus, the signal at this location is dominated by the fundamental mode and, hence, nearly monochromatic.
Therefore, we choose the waveform extracted at $\rex=20$M
for the spectral analysis.
To measure the first overtone frequencies we additionally extract
data at $\rex\simeq 50 M$, where the fundamental mode is subdominant; c.f. Fig.~\ref{fig:fund_and_overt}.

We measure the three parameters, $(\bar{I},\bar{\omega},\bar{\nu})$,
via a nonlinear regression
using Eq.~\eqref{eq:Lorentzian} as fitting function.
This method produces reliable estimates,
within $0.5\%$,
of the characteristic frequency, $\omega_{lmn}$, but is greatly limited in correctly measuring the growth rate.
In particular, for long-lived modes which grow or decay
on timescales larger than what can be evolved in $3+1$ simulations
(with reasonable computational resources),
the width of the peaks is dominated by windowing effects and insufficient frequency resolution.
Longer time series and different techniques,
such as those employed in~\cite{Dolan:2012yt,Dima:2020rzg}
are required for more accurate estimates of the growth rates.

\begin{figure}[h!]
  \hspace{0.5em}\centering\includegraphics[width = 0.455 \textwidth]{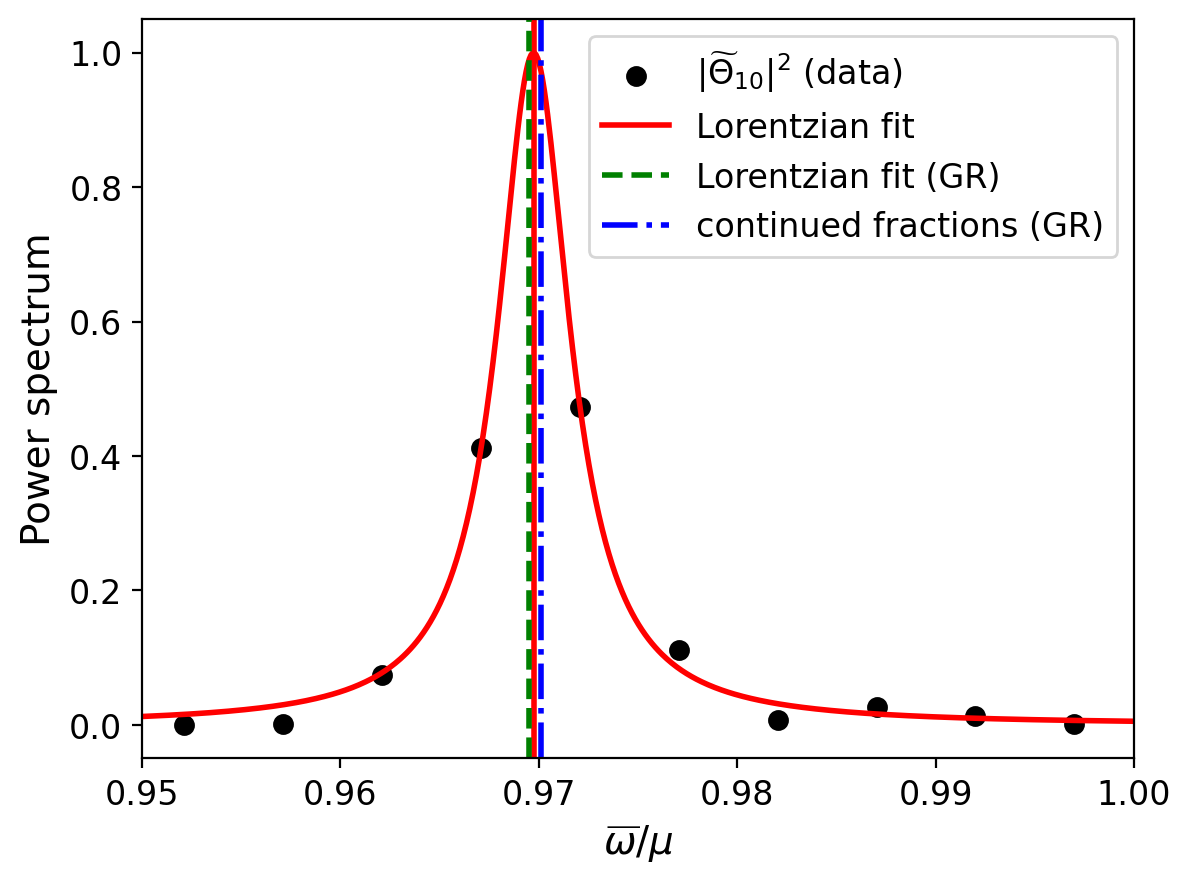}
    \hfill
    \centering\includegraphics[width = 0.44 \textwidth]{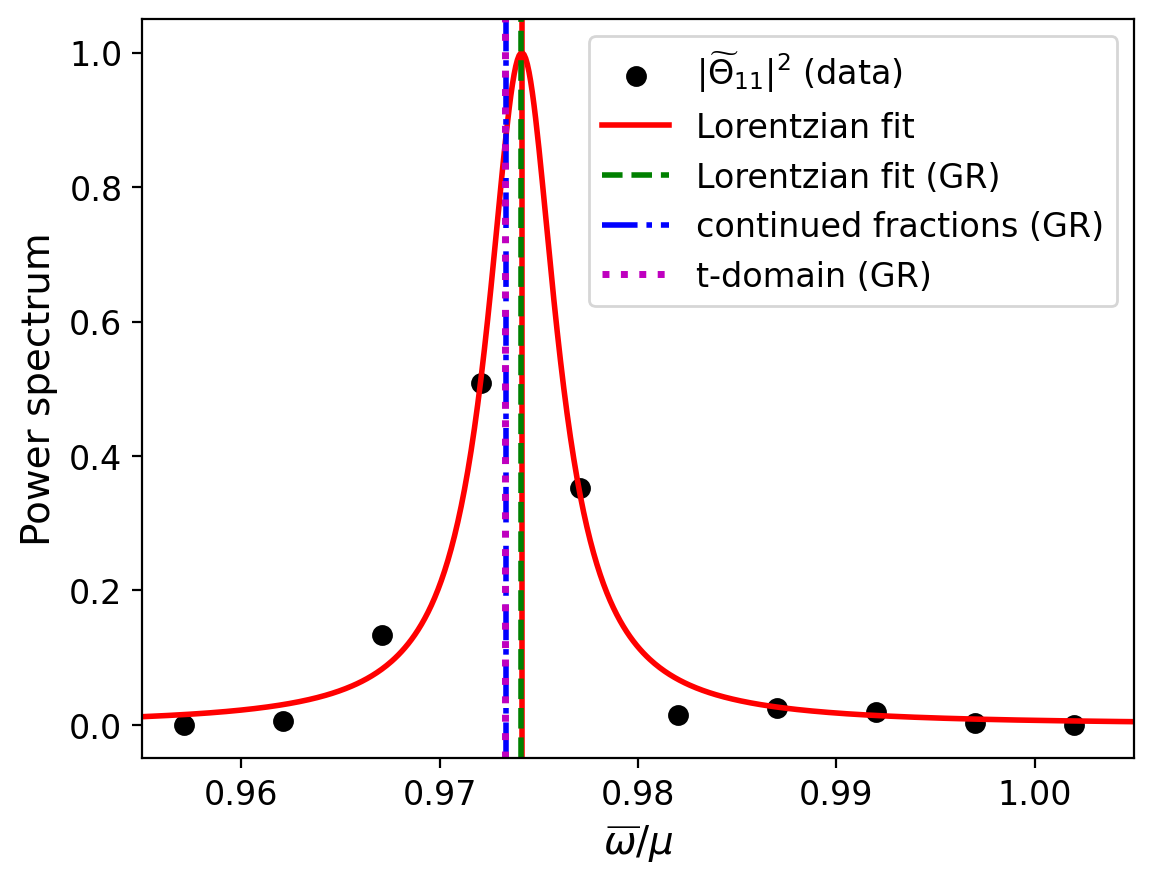}
    \caption{Power spectra reconstructed from simulations with Gaussian initial data (ID2 in Sec.~\ref{subsec:IDSF})
    and parameters $a/M=0.99$, $\mu M=0.42$,
    normalized by the maximum value.
    We indicate the numerical data (black dots), extracted at $\rex=20 M$,
    and the Lorentzian fit to our data points (solid red line) obtained in massive \dCS gravity.
    For comparison we also show the peak of a Lorentzian fit for the equivalent \GR simulation (dashed green line) and the frequency obtained with the continued fraction method in \GR~\cite{Dolan:2007mj} (dot-dashed blue line).
    Top panel:
    data and fits for the $(l,\,m,\,n)=(1,0,0)$ multipole.
    The fit of the data gives a peak frequency of $\bar{\omega}/\mu=0.970\pm0.005$.
    Bottom panel:
    data and fits for the $(l,\,m,\,n)=(1,1,0)$ multipole.
    The fit of the data gives a peak frequency of $\bar{\omega}/\mu=0.974\pm0.005$.
\label{fig:PWS_Gauss}
}
\end{figure}

\begin{figure}[!h]
    \hspace{0.5em}\centering\includegraphics[width = 0.465 \textwidth]{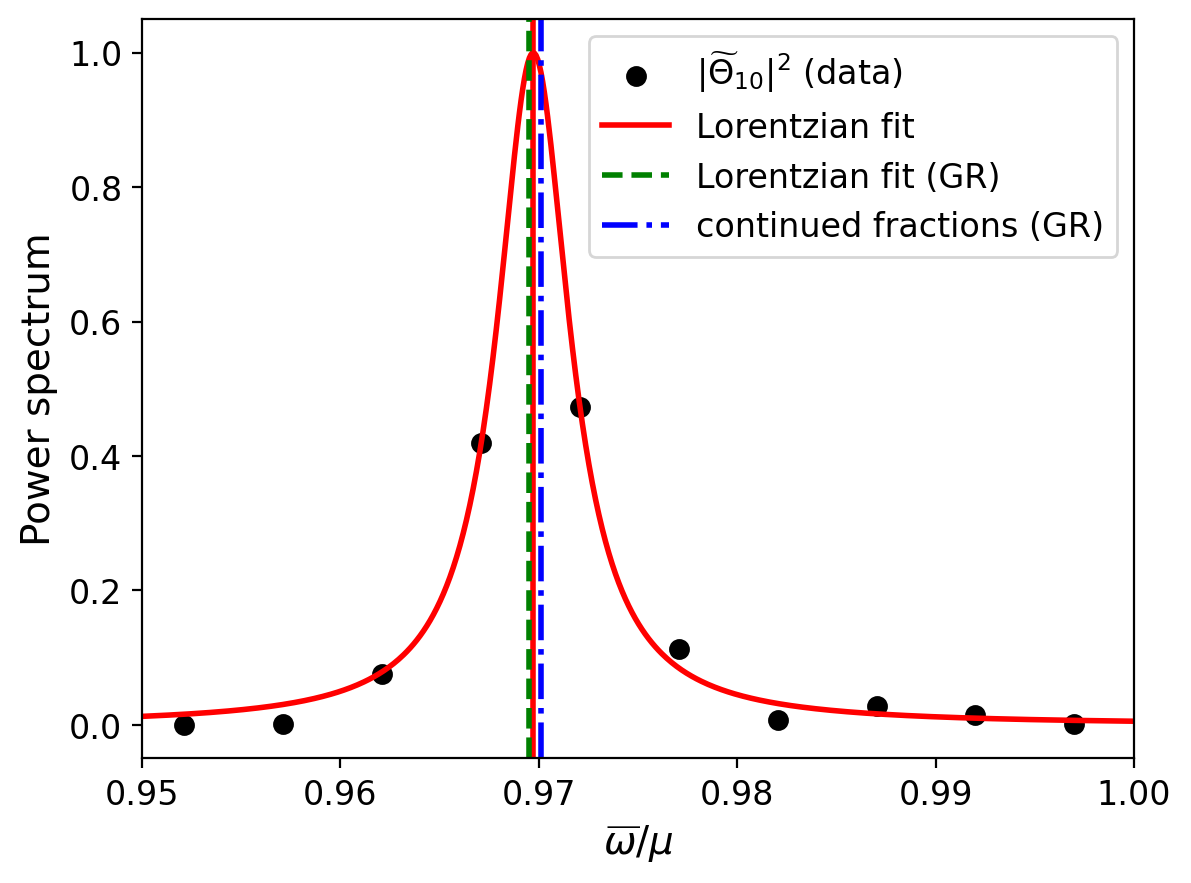}
    \hfill
    \centering\includegraphics[width = 0.45 \textwidth]{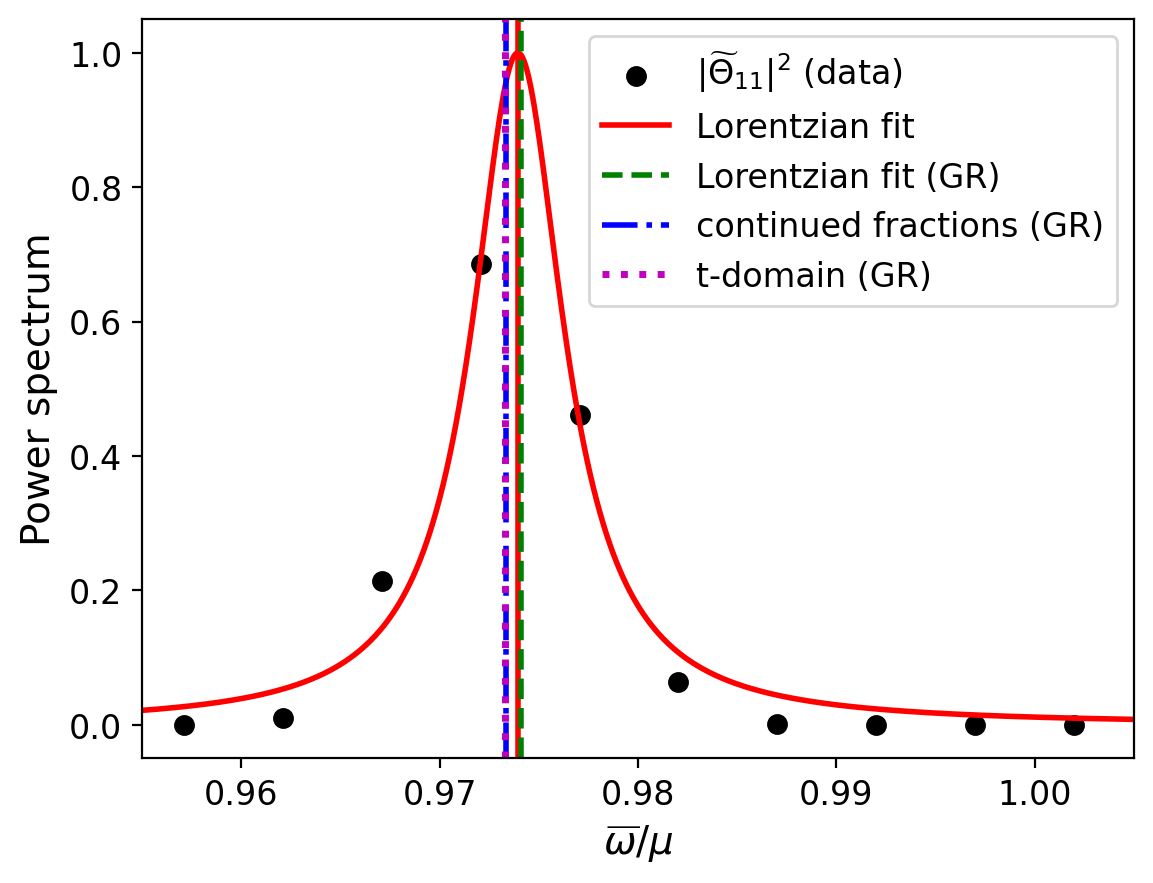}
    \caption{Same as Fig.~\ref{fig:PWS_Gauss} but for \QBS initial data (ID4 in Sec.~\ref{subsec:IDSF})
    Top panel:
    data and fits for the $(l,\,m,\,n)=(1,0,0)$ multipole, and the frequency estimated from the fit is $\bar{\omega}/\mu=0.969\pm0.005$.
    Bottom panel:
    data and fits for the $(l,\,m,\,n)=(1,1,0)$ multipole, and the frequency estimated from the fit is $\bar{\omega}/\mu=0.974\pm0.005$.}
    \label{fig:PWS_BS}
\end{figure}

\begin{figure}[!h]
    \hspace{-0.5em}\centering\includegraphics[width = 0.44 \textwidth]{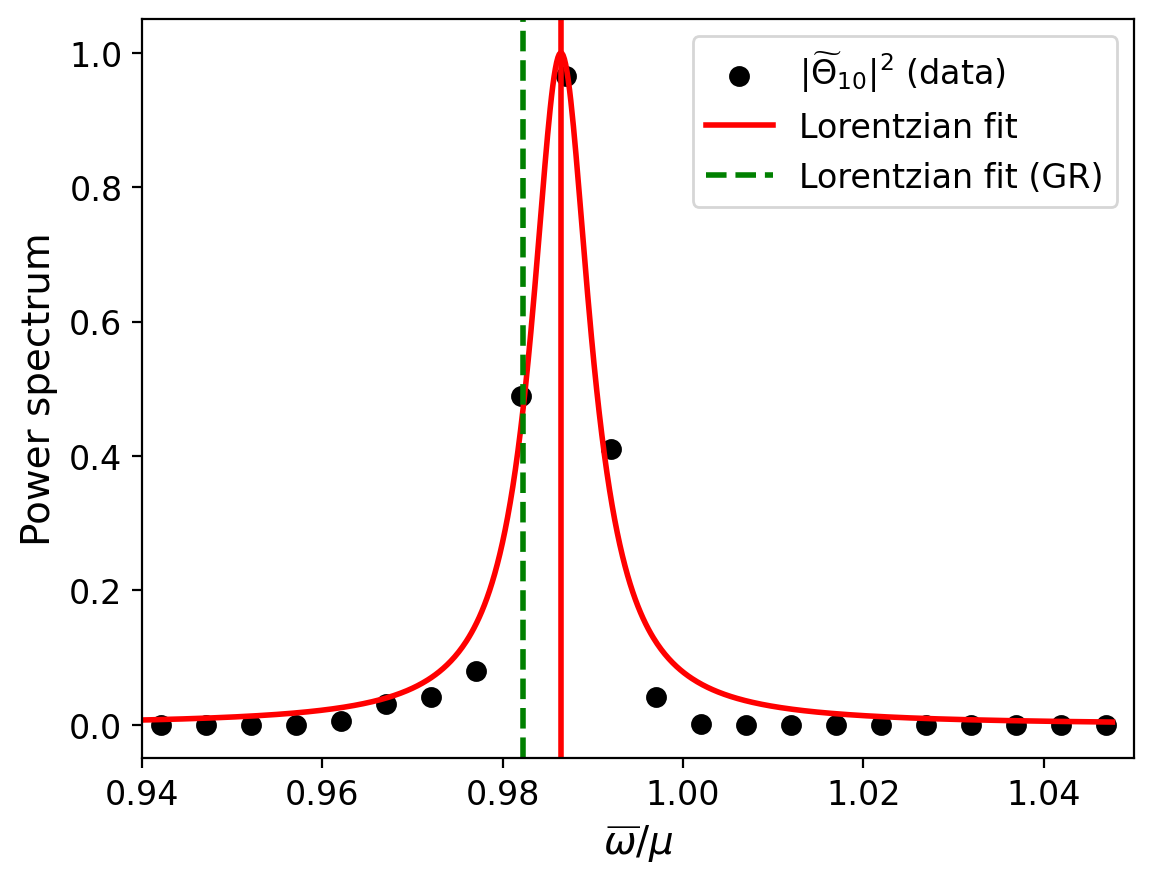}
    \hfill
    \centering\includegraphics[width = 0.46 \textwidth]{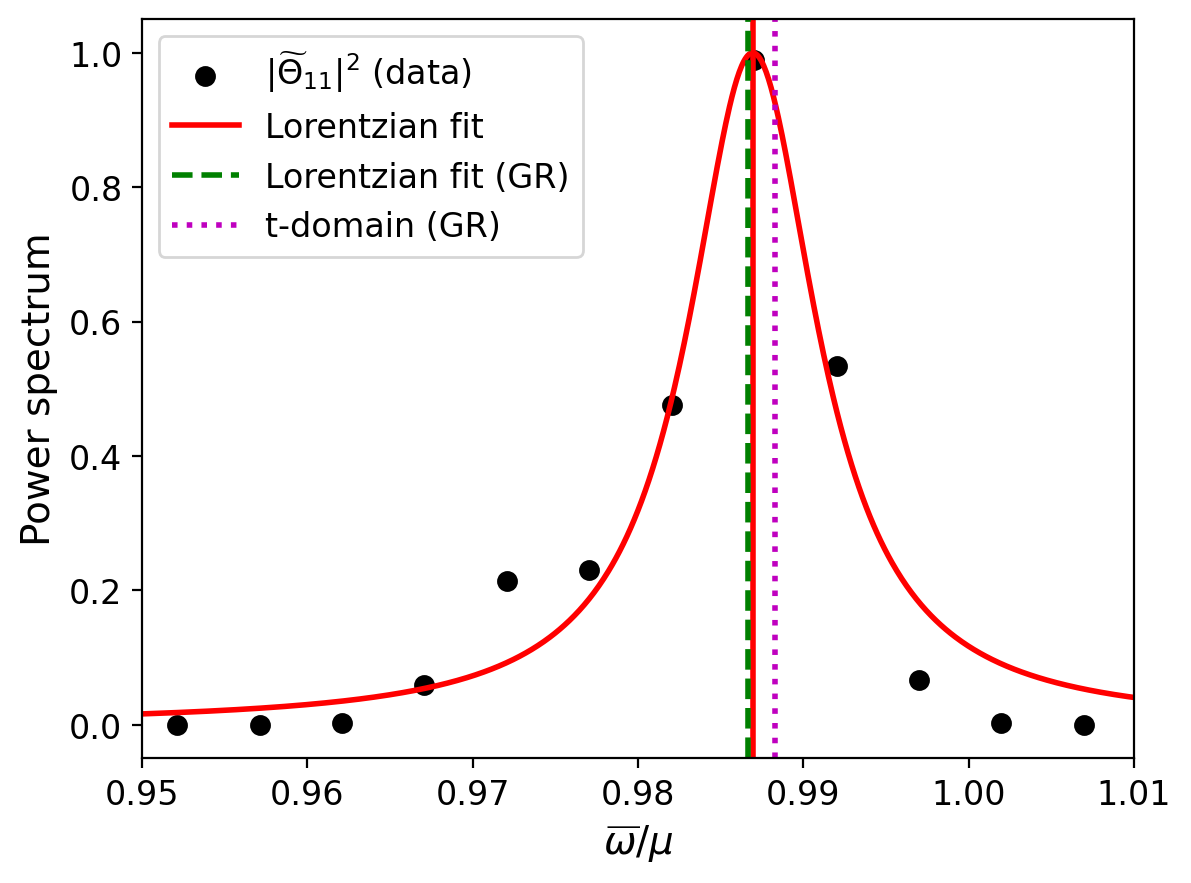}
    \caption{Same as Fig.~\ref{fig:PWS_Gauss} but for data extracted at $\rex=50M$,
    where the first overtone dominates over the fundamental $l=1$ mode.
    Top panel:
    data and fits for the $(l,\,m,\,n)=(1,0,1)$ multipole,
    and the frequency estimated from the fit is $\bar{\omega}/\mu=0.986\pm0.005$.
    Bottom panel:
    data and fits for the $(l,\,m,\,n)=(1,1,1)$ multipole,
    and the frequency estimated from the fit is $\bar{\omega}/\mu=0.987\pm0.005$.
    The power spectrum also contains the fundamental
    mode (see shoulder to the left of the fitted peak), but it is subdominant with respect to the $n=1$ overtone.}
    \label{fig:PWS_Overtone}
\end{figure}

We present the results of the spectral analysis
in Figs.~\ref{fig:PWS_Gauss},~\ref{fig:PWS_BS} and~\ref{fig:PWS_Overtone}.
We concentrate on the simulations with parameters
$a/M=0.99$ and $\mu M=0.42$.
We compute the spectra for the simulations with Gaussian initial data
with $\Sigma_{lm}=\Sigma_{11}$
(ID2 in Sec.~\ref{subsec:IDSF}),
and for \QBS initial data
(ID4 in Sec.~\ref{subsec:IDSF})
shown in Fig.~\ref{fig:fund_and_overt}.

\clearpage
\pagebreak

Fig.~\ref{fig:PWS_Gauss} presents the power spectra of
the $l=1,m=0,n=0$ (left panel) and of the $l=m=1,n=0$ (right panel) multipoles
for the simulation with Gaussian initial data,
extracted at $\rex=20$M, after an evolution time of $t=3000 $M.
We measure the characteristic frequency of the $l=1,m=0,n=0$ mode to be $\bar{\omega}/\mu\simeq 0.970 \pm 0.005$,
and
that of the $l=m=1,n=0$ mode to be $\bar{\omega}/\mu\simeq 0.974 \pm 0.005$.
We took the frequency resolution of our discrete Fourier series as a conservative estimator of the uncertainty; see App.~\ref{App:FreqErrEst} for a detailed  discussion.

We compare these
estimates, obtained from the peak in the Lorentzian fit (solid red line),
against the \QBS frequency of a massive scalar field in \GR{.}
We obtain the latter in three different ways:
\begin{enumerate*}[label={(\roman*)}]
\item by fitting the data from a 3+1 simulation in which we turn off the dCS coupling, $\aCSh=0$, (dashed green line);
\item by computing the frequency via the continued fraction method~\cite{Dolan:2007mj} (dot-dashed blue line);
\item
by fitting data from an effective, 1+1 evolution in the time domain~\cite{Dolan:2012yt} for the $l=m=1$ mode (right panels; purple dotted line).
\end{enumerate*}
For both harmonic modes, we see no statistical evidence of a deviation from the characteristic frequencies of a minimally-coupled massive scalar field around a Kerr \bh (i.e. $\aCSh=0$).
This conclusion is robust
against different initial data:
we obtain compatible frequency measurements for
\QBS initial data,
as can be see from Fig.~\ref{fig:PWS_BS}.

We also extracted signals at $\rex=50M$, where the $n=1$ overtone becomes larger than the fundamental mode $n=0$. In Fig.~\ref{fig:PWS_Overtone} we show the peaks corresponding to modes $\left(l,m,n\right)=\left(1,0,1\right)$ and $\left(l,m,n\right)=\left(1,1,1\right)$, respectively.
Concerning the scalar's $l=1,m=0$ multipole,
we observe a larger discrepancy with respect to the reference measure with $\aCSh=0$ that is, nonetheless, compatible with a statistical fluctuation.
We find that results are overall consistent with \GR,
within $0.5\%$.

To summarize, we observe that the spectrum of characteristic oscillations of the massive \dCS field is determined
by the mass of the field.
Thus, it appears indistinguishable from the \QBS spectrum of a massive scalar field
around a Kerr \bh in \GR{}
at sufficiently large distances.
This conclusion is consistent with the results
of Macedo~\cite{Macedo:2018txb}
for \QBS{s} of a massive \dCS field around nonrotating
\bh{s.}
Because the 3+1 time domain approach does not allow for sufficiently long evolutions
(at a resonable computational cost)
to accurately measure to growth or decay rate of the scalar modes, we postpone the computation of this quantity
in massive \dCS gravity
to future work using different techniques.

\section{Summary and conclusions}~\label{sec:Conclusion}
In this paper we study
the phenomenology of a scalar field in massive \dCS gravity, and its evolution around rotating \bh{s}.
This line of research broadens
our understanding  on how \bh observations may be used to
probe for physics beyond the standard model.

Here, we focus on ``new physics'' in the shape of fundamental scalar fields that can leave
potentially observable
footprints in the phenomenology of \bh{s}.
In particular,
working in the decoupling approximation we find new, oscillating scalar
configurations
around rotating \bh{s} that may give rise to new \bh solutions if the backreaction of the scalar onto the metric is taken into account.
Our results have a twofold interpretation:
On the one hand, we may interpret the scalar field as new solutions to \dCS modified gravity if a mass-term is generated by nonperturbative effects.
On the other hand,
we may interpret the scalar field as an axionlike particle
-- a popular dark matter candidate --
sensitive to parity violation in the gravity sector.
Thus, our results aid ongoing efforts to
devise new observational tests of gravity
or probes for ultralight particles that
pose
a class of dark matter models.

This project has been guided by three questions stated in the introduction, Sec.~\ref{sec:intro},
and we use the same structure to summarize
our results here.

Massive scalar fields
can form \QBS{s},
or scalar clouds,
around rotating \bh{s} in \GR
that are predominantly
determined by the
$\Theta_{11}$
multipole.
{\textbf{How does the nonminimal coupling to curvature affect
such a
massive scalar cloud?}}
To address this question, we
analyze the evolution
of the $\Theta_{11}$
mode in massive \dCS gravity
for a range of mass parameters and \bh spins.
We observe that the \QBS frequencies and,
when excited,
their massive power-law tails
are consistent with those of massive scalar fields in \GR{.}
In addition, we compare the structure of the scalar cloud in the equatorial plane
in massive \dCS gravity against that found in \GR{.}
We observe no difference between the two,
so we conclude that,
to leading order,
the coupling to the Pontryagin density has
essentially no effect on the
$l=m=1$ \QBS{.}
Thus, ultralight scalars
representing, e.g., wavelike dark matter candidates
appear insensitive to this type of parity-violating corrections to \GR
far from \bh{s}.
This may come as no surprise because the Pontryagin density primarily sources the $l=1,m=0$ dipole of the \dCS axion.

This leads us to the second question:
{\textbf{How does the mass term affect the \dCS hair?}}
To address this question we compare the evolution of the $\Theta_{10}$ mode in massless and in massive \dCS gravity.
In both cases, the Pontryagin density causes
the formation of a (quasi)stationary
scalar dipole close to the \bh{,}
that is absent in \GR{.}
We observe that the mass term suppresses the
amplitude of the \dCS hair.
What is the origin of this suppression?
We can exclude mode mixing between the $l=1,m=0$ and $l=m=1$ multipoles
because of the symmetries of the background spacetime.
In particular, modes with different azimuthal numbers $m$ decouple around the Kerr metric.
It is also not due to the Pontryagin density, which inherits the axisymmetry of the background spacetime and only sources $m=0$ multipoles.
Instead, we interpret the reduction of the axion's amplitude as a Yukawa suppression
due to the field's mass.

\begin{figure}
    \centering
    \includegraphics[width = 0.9\columnwidth]{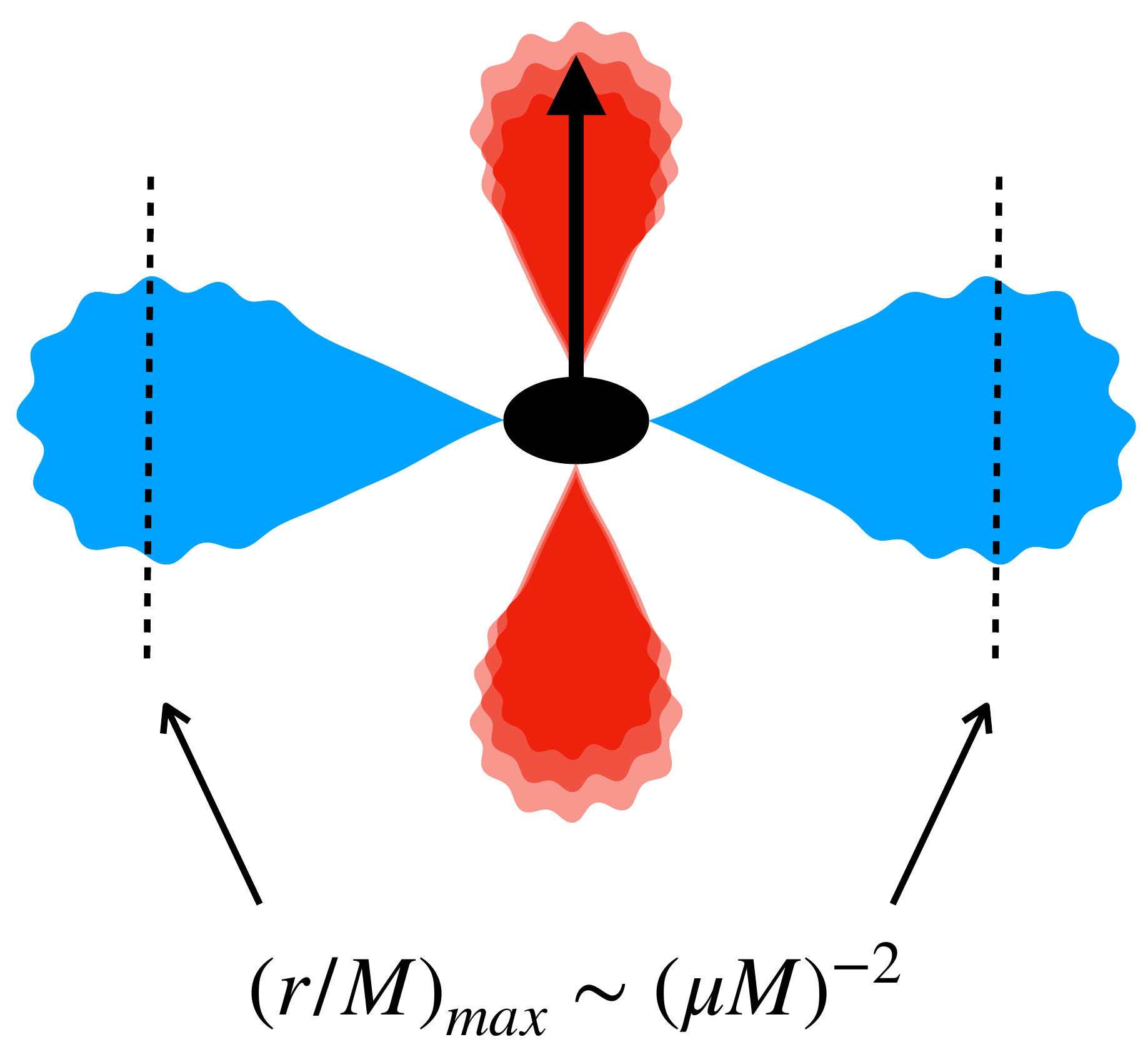}
    \caption{
     Updated sketch (see Fig.~\ref{fig:sketch1}) of the oscillating massive dCS hair sourced by a rotating BH.
     The blue clouds correspond to the \QBS{s} in the equatorial plane and  $(r/M)_{max} \sim (\mu M)^{-2}$ indicates their maximum.
     The red gradient along the axis of rotation indicates the oscillatory pattern of the axion that is due to the formation of a \QBS peaking roughly at $r\simeq r_{max}$.}    \label{fig:sketch2}
\end{figure}

We also observe that the mass term
imprints an oscillatory pattern on the \dCS hair,
$\Theta_{10}$,
far away from the \bh{}
as illustrated in Fig.~\ref{fig:sketch2}.
The transition from the
quasistationary to an oscillating
configuration occurs
at $r/M\sim (\mu M)^{-2}$.
The oscillation frequency of the massive \dCS hair is determined by its mass parameter.
Consequently, \bh hair in \dCS gravity would be sensitive to the presence of a mass term in the field's potential.

This leads us to the last question:
{\textbf{What is the scalar's characteristic
frequency spectrum in massive \dCS gravity?}}
We analyze the frequency spectrum of the $l=1,m=0$ and the $l=m=1$
scalar multipoles
in massive \dCS gravity.
We find that both are determined
essentially
by the scalar's mass parameter.

In conclusion,
we find that rotating \bh{s} in massive \dCS gravity give rise to new configurations of the \dCS axion
that connects an axi-symmetric dipole close to the \bh{} with an oscillating axion ``cloud'' at large distances.
We sketch this new configuration in Fig.~\ref{fig:sketch2}.

Based on our results, we speculate that
massive \dCS gravity admits new \bh solutions with an oscillating steady-state axion hair at the onset of superradiance, similar to new hairy solutions in \GR{}~\cite{Herdeiro:2014goa}.
Such a study would require to solve the massive \dCS equations in full or perturbatively to higher order in the metric, and it is beyond the scope of the present paper.

Furthermore, while we observe the onset  of massive \QBS{s} in \dCS gravity, they appear to evolve on similar timescales as scalar \QBS{s} in \GR{.}
This makes numerical simulations in $3+1$ dimensions not well suited to conduct a detailed analysis of their evolution.
Thus, we leave the hunt for superradiant instabilities in massive \dCS gravity to future work
using different techniques.

In conclusion, this work is the first in a series
to explore the properties of \bh{s} in the presence of axionlike particles, representing a large class of dark matter candidates, when parity is violated in the gravity sector.

\section*{Acknowledgements}
We thank  Stephon Alexander,
Noora Ghadiri,
Leah Jenks, Hector O. Silva, Leo Stein and Nicolas Yunes for insightful discussions and comments.
We are indebted to Cheng-Hsin Cheng for providing useful technical insights
for the visualization of our data.

The authors acknowledge support provided by the National Science Foundation under NSF Awards No. OAC-2004879 and No. PHY-2110416.
We acknowledge the Texas Advanced Computing Center (TACC) at the University of Texas at Austin for providing HPC resources on Frontera via allocations PHY22018 and PHY22041.
This work used the Extreme Science and Engineering Discovery Environment (XSEDE) system Expanse through the allocation TG-PHY210114, which was supported by NSF grants No. ACI-1548562 and No. PHY-210074.
This research used resources provided by the Delta research computing project, which is supported by the NSF Award No. OCI-2005572 and the State of Illinois.

We thank KITP  for its hospitality during  the workshop ``High-Precision Gravitational Waves'' which was supported in part by NSF Award No. PHY-1748958.
We acknowledge support by NSF Award No. NSF-1759835 for the ``New frontiers in Strong Gravity'' workshop where part of this work has been completed.
We thank Steven Brandt for supporting travel to the Northamerican Einstein Toolkit workshop via the NSF Award No. OAC-1550551
where part of this work was presented.

This work used the open-source softwares
\textsc{xTensor}~\cite{xact_url,Brizuela:2008ra},
the \ETK~\cite{Loffler:2011ay,EinsteinToolkit:2022_11},
\Canuda~\cite{witek_helvi_2023_7791842},
\PostCactus~\cite{2021ascl.soft07017K}.
The \CanudadCS{} code developed to conduct the simulations in this work is open source and available in a git repository~\cite{CanudadCS_repo}.
A YouTube playlist with 2D animations rendered from data produced with our simulations is available at~\cite{YoutubeLinkmdCS}.

\clearpage
\appendix
\section{Harmonic decomposition of the Pontryagin density}~\label{app:SphHamDecPont}

In this appendix we complement our numerical results
by seeking analytic insight into the effects of the nonminimal coupling to gravity onto the $(lm)$ multipoles of the massive \dCS field.
Therefore, we project the Pontryagin density
(that sources the \dCS field)
onto a basis of spherical harmonic functions,  $Y_{lm}(\theta,\phi)$.
Working in \BL coordinates, $(t,\rBL,\theta,\phi)$,
the projection is given by
\begin{align}\label{appeq:SphDecPont}
\CS  =  & \sum_{l,m} \mathcal{P}_{lm}(\rBL) Y_{lm}(\theta,\phi)
\,,\\
\mathcal{P}_{lm} := & \int Y^*_{lm}(\theta,\phi) (\CS)d(\cos\theta)d\phi
\,,
\end{align}
where the spherical harmonics are normalized
such that
\begin{align}\label{appeq:SphHarmNorm}
\int Y^*_{lm}(\theta,\phi) Y_{kn}(\theta,\phi)d(\cos\theta)d\phi
= \delta_{lk}\delta_{mn}
\,.
\end{align}
In the current project,
we concentrate on Kerr \bh{s} as our background spacetime.
Therefore, we focus on writing the Pontryagin density as a superposition of spherical harmonics
evaluated on a Kerr background.
In \BL coordinates,
the Kerr metric is given by Eq.~\eqref{eq:KerrBL}
and the Pontryagin density reads
\begin{align}\label{appeq:PontKerr}
\CS  = & 96M^2\frac{3 \rBL^5 a \cos\theta - 10 \rBL^3 a^3\cos^3\theta + 3 \rBL a^5 \cos^5\theta}{\Sigma^6}
\,,
\end{align}
where the metric function $\Sigma$ is given in Eq.~\eqref{eq:KerrMetricFunction}.
Note, that the Pontryagin density inherits the axisymmetry of the background spacetime and,
therefore, it has no dependence on the azimuthal angle $\phi$.
Consequently, the only nonvanishing spherical harmonic components, $\mathcal{P}_{lm}$,
are those with $m=0$.
Moreover, the parity symmetry properties of the Pontryagin density imply that it is composed of a combination of odd $l$
modes.
Taking the above considerations into account,
we can decompose the Pontryagin density as
\begin{align}\label{appeq:SphDecPont}
\CS  =  & \sum_{j=0}^{\infty} p_{j}(\rBL) Y_{(2j+1)0}(\theta,\phi)
\,,
\end{align}
with the projection coefficients
\begin{align}
p_j := & 2\pi\int_{-1}^{+1} Y^*_{(2j+1),0} (\CS) d(\cos\theta)
\,.
\end{align}
As a result, one can conclude that the nonminimal coupling to the Pontryagin density behaves as a source term only for the scalar field multipoles with $m=0$ and odd $l$,
and it does not introduce any new mode-mixing
at the level of the scalar field equation.

\section{Error estimates}
In this appendix we
quantify
the discretization error of our numerical simulations
through a convergence analysis
and the statistical error of the frequency estimates.

\subsection{Convergence tests}\label{App:ConvTest}
To assess the numerical discretization error, we perform a convergence analysis of the run a099\_SF11\_mu042 in Table~\ref{tab:simulations}.
This run is representative of the most demanding setups as it evolves
a scalar field with mass parameter $\mu M = 0.42$ in the background of a Kerr \bh with spin $a/M = 0.99$.
The grid setup is identical to the simulations outlined in Sec.~\ref{Simulations},
but we vary the grid spacing $\dif x_{h}/M=0.9$ (high resolution),
$\dif x_{m}/M = 1.0$ (medium resolution; standard choice for our simulations)
and $\dif x_{l}/M = 1.1$ (low resolution);
see  Table~\ref{Table:conv}.
In Table~\ref{Table:conv} we also list the resolution $h=\dif x/ 2^6$ on the innermost refinement level that encompasses the \bh{.}

We compare the
$l=1, m=0$ and $l=m=1$ multipoles of the massive \dCS scalar
extracted at $\rex=10M$
for the high, medium, and low resolution runs.
We observe that all runs perfectly align, i.e., they are consistent across the different resolutions.

\begin{figure*}[t]
    \centering
    \includegraphics[width = 0.48 \textwidth]{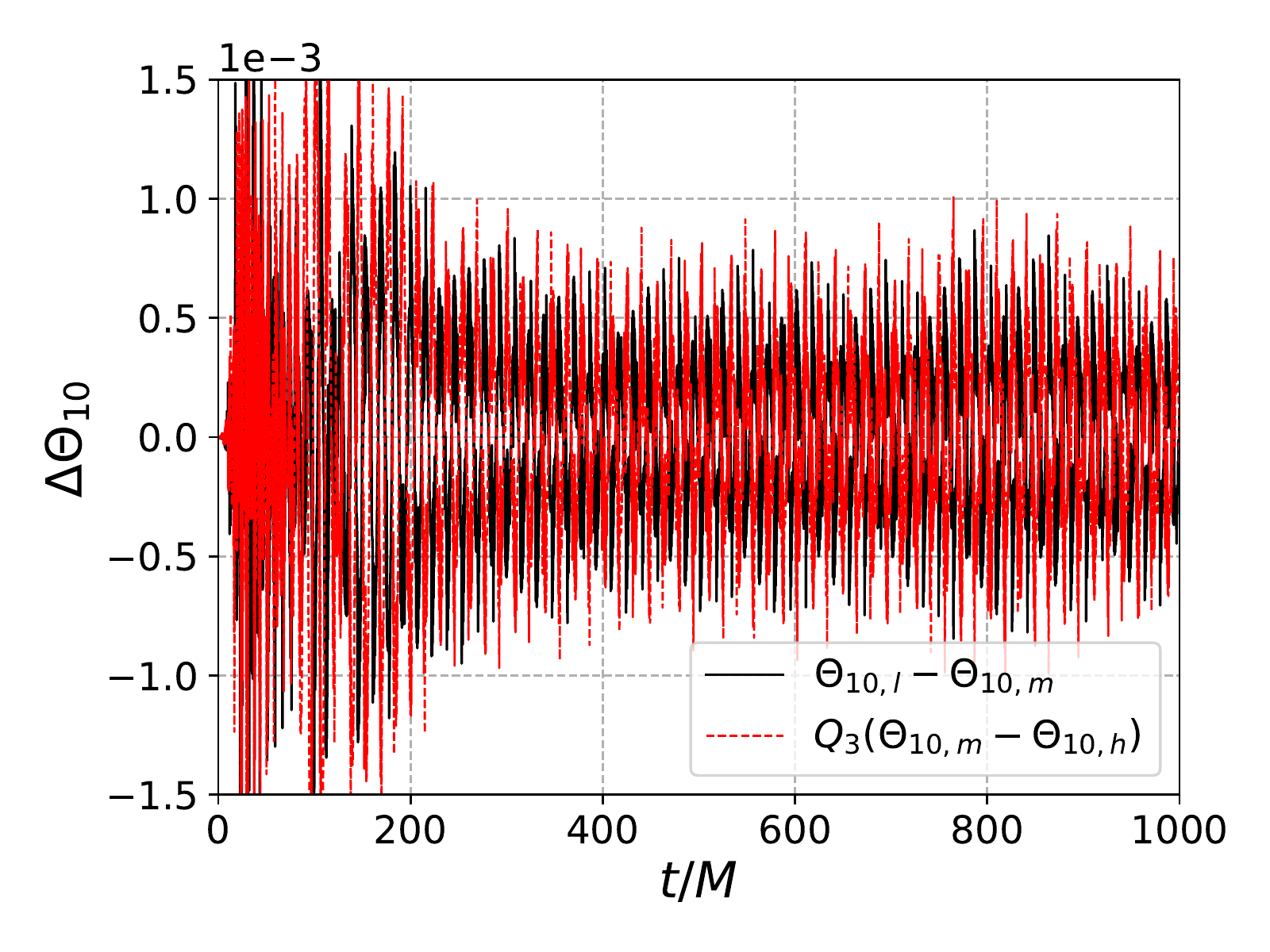}
    \includegraphics[width = 0.48 \textwidth]{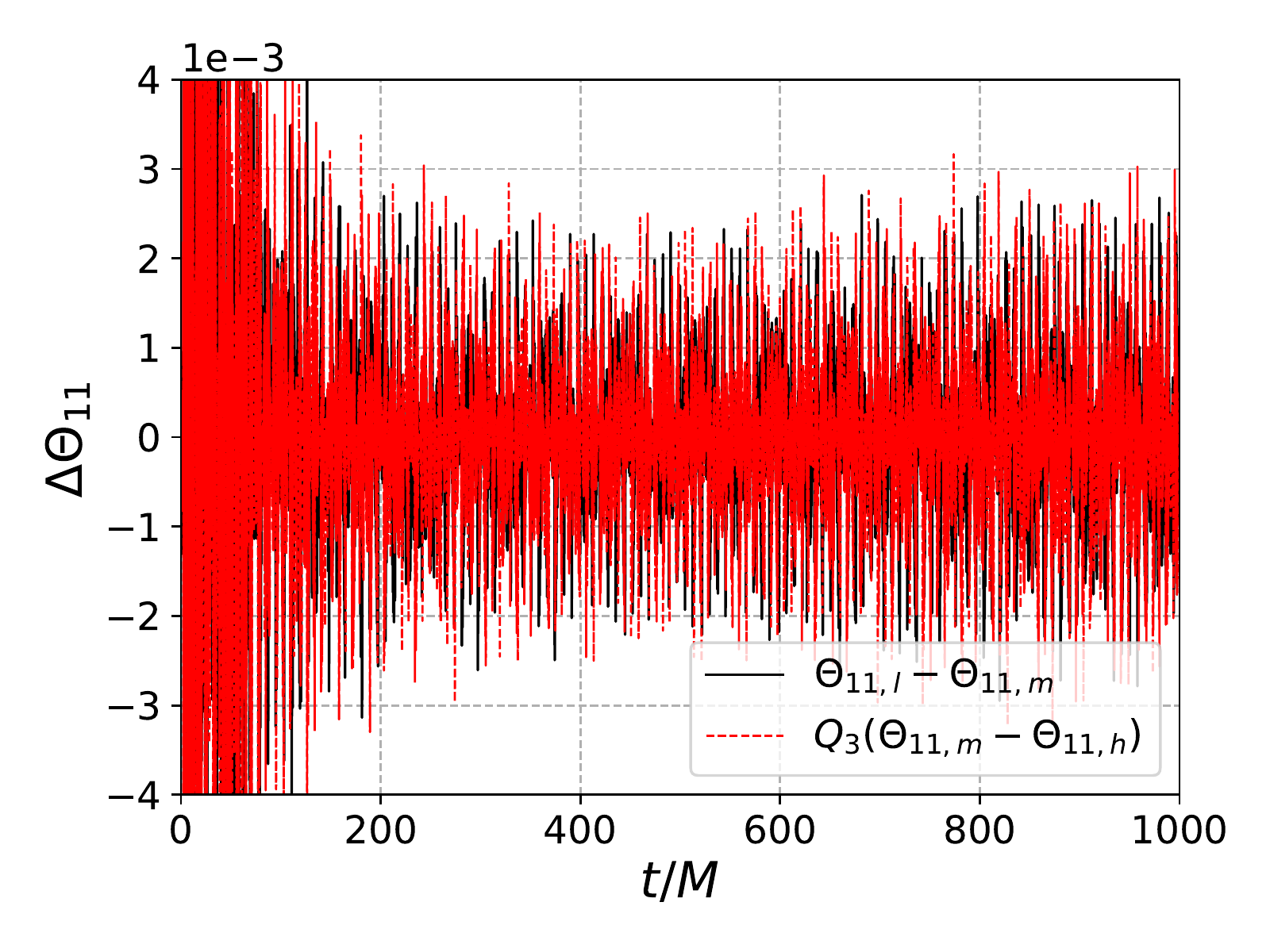}
    \caption{Convergence plot of the
    $l = 1, m = 0$ (left) and $l = m =1$ (right) multipoles of the massive \dCS field
    (with $\mu M=0.42$ around a \bh with $a/M=0.99$).
    We compare the difference between the low and medium resolution run (solid black line),
    and the medium and high resolution run (dashed red line).
    The latter is rescaled by the factor $Q_3 = 1.221$ indicating third order convergence.}
    \label{Fig:ConvTest}
\end{figure*}

In Fig.~\ref{Fig:ConvTest} we show the convergence plot for the $l=1,m=0$ (left) and $l=m=1$ (right) multipoles of the massive \dCS field.
Specifically, we show the difference between the low and medium resolution waveform (solid black line),
and the medium and high resolution run (dashed red line).
The latter difference has been rescaled by the factor $Q_{3}=1.221$ indicating third order convergence.
This is consistent with our numerical code that employs a fourth order finite difference and time integration scheme, supplemented with second order interpolation at refinement boundaries.

From the convergence analysis we can now estimate the numerical error.
In particular,
for the $l=1,m=0$ multipole we find
$\Delta \Theta_{10}/\Theta_{10,h}\lesssim 12\%$ at late times $t\sim1,000M$.
For the $l=m=1$ multipole we find
$\Delta \Theta_{11}/\Theta_{11,h}\lesssim 0.15\%$ at late times $t\sim1,000M$.

\begin{table}[t]
    \centering
    \begin{tabular}{|c|c|c|}
        \hline
         Run &
         $\dif x/M$ &
         $h/M$  \\
         \hline
    a099\_SF11\_mu042\_low & ~$1.1$~ & $1.719\times 10^{-2}$ \\
     a099\_SF11\_mu042\_med & $1.0$ & $1.563\times 10^{-2}$ \\          a099\_SF11\_mu042\_high & $0.9$ & $1.406\times 10^{-2}$\\
    \hline
    \end{tabular}
    \caption{List of runs for the convergence tests evolving the massive \dCS field  with $\mu M = 0.42$
    around a \bh with spin $a/M = 0.99$.
    The field initially has a Gaussian profile with $\Sigma_{11}$ (ID2).
    We denote the resolution $\dif x/M$ on the outermost refinement level and the resolution $h=\dif x / 2^{n-1}$ (with $n=7$) on the innermost refinement level.}
    \label{Table:conv}
\end{table}

\subsection{Frequency error estimate}~\label{App:FreqErrEst}

To obtain the frequency estimates we perform a Fourier transformation of the numerical time series data that lasts for about $T\sim 3 \times 10^3 M$ and has a finite resolution.
This gives rise to an error due to the frequency resolution that we estimate to
$\sigma_{\hat{\omega}}=\Delta\hat{\omega}=\Delta\omega/\mu=2\pi/(T\mu)\simeq 0.005$ (or equivalently, $\sim 0.5\%$).

Additionally, we estimate the statistical error.
Therefore, we perform a more sophisticated analysis by
propagating the numerical error of the time series data
through the Discrete Fourier Transform and the Breit-Wigner fitting procedure.
To estimate the uncertainty in the extracted multipole $\Theta_{lm}(t;r=r_{\rm ext})$
we use the conservative estimator given by the maximum amplitude error
(excluding the initial transient)
taken from the convergence tests,
$\sigma_{\Theta}={\rm max}\lbrace\Delta\Theta_{lm}\rbrace\lesssim 3\times 10^{-3}$; see Appendix~\ref{App:ConvTest}.
Taken this conservative error in the multipole amplitude, assuming that it remains approximately constant in time and treating it as uncorrelated,
the propogated error in the frequency power spectrum is given by
\begin{equation}
    \sigma_{\rm \tilde{P}}=\frac{2\sqrt{\tilde{P}}}{\sqrt{N}}\sigma_{\Theta}
\,.
\end{equation}
Applying a nonlinear least square regression algorithm
yields the
statistical error on the fitted frequency.
It is $\sigma_{\hat{\omega}}\simeq 2\times10^{-4}$ ($0.02\%$) for the $l=m=1$ multipole frequency, and $\sigma_{\hat{\omega}}\simeq 4\times10^{-4}$ ($0.04\%$) for the $l=1$, $m=0$ multipole.
Taking both contributions together, we estimate the error in the frequency to be
$\sigma_{\hat{\omega}}\lesssim 5.4\times10^{-3}$.

\section{General \dCS gravity at decoupling}\label{appsec:GeneralDCS}
For completeness,
we derive the equations of motion
and the evolution equations (at decoupling)
of \dCS gravity with a general coupling function $f(\Theta)$ between the \dCS axion and the Pontryagin density.
We vary the action, given in Eq.~\eqref{eq:ActiondCS},
with respect to the metric and the pseudoscalar field and obtain
\begin{subequations}
\label{appeq:eoms}
\begin{align}
\label{appeq:dCSKG}
&\Box\Theta - \dVth + \frac{\aCS}{4}\dfth \CS = 0
\,,\\
\label{appeq:dCSTenEoM}
&G_{ab} - \frac{1}{2} T^{\Theta}_{ab} + \aCS ~\CCS_{ab} = 0
\,,
\end{align}
\end{subequations}
where $G_{ab}=R_{ab}-\frac{1}{2}g_{ab}R$ is the Einstein tensor,
and $T^\Theta_{ab}$ is the canonical scalar field
energy-momentum tensor
\begin{equation}\label{appeq:TmnSF}
T^\Theta_{ab}=\nabla_a\Theta\nabla_b\Theta-g_{ab}
\left( \frac{1}{2} (\nabla\Theta)^2+\Vth\right)
\,.
\end{equation}
The C-tensor captures the modification of Einstein's equations and it is given by
\begin{equation}\label{appeq:ctensor}
    \CCS^{ab}\equiv \E_c\,\tensor{\epsilon}{^c^d^e^(^a}\nabla_e\tensor{R}{^b^)_d} + \F_{cd}\,^*R^{d(ab)c}
\end{equation}
where the auxiliary tensors $\E_a$ and $\F_{ab}$ are defined as
\begin{subequations}
\begin{align}
    \E_a \equiv& \dfth\nabla_a\Theta
    \,,\\
    \F_{ab} \equiv& \dfth\nabla_a\nabla_b\Theta+\ddfth\nabla_a\Theta\nabla_b\Theta
    \,.
\end{align}
\end{subequations}
As described in Sec.~\ref{ssec:ActionAndEoMs}, we work in the decoupling approximation around
vacuum \GR{.}
That is, we study the evolution of the massive \dCS field in a background spacetime that is determined by Einstein's equations in vacuum.
With these assumptions the Ricci tensor and its derivative vanish, $R_{ab}=0=\nabla^eR_{ab}$.
Consequently the first term in Eq.~\eqref{appeq:ctensor} vanishes.
Then we can write the C-tensor in terms of the dual Weyl tensor as,
\begin{equation}
\CCS^{ab}_{\text{vac;GR}}= \F_{cd}\,^{*}W^{d(ab)c}
\,.
\end{equation}
Following the same steps as in Section \ref{ssec:ActionAndEoMs}, we apply the decoupling approximation to the field equations \eqref{appeq:eoms}, rescale $\Theta \rightarrow (\aCS / M^2) \Theta$
and perform the $3+1$ decomposition
to find
\begin{subequations}
\label{appeq:dtThetaKTheta_Gen}
\begin{align}
\label{appeq:dtThetaADM_Gen}
\dif_{\rm t} \Theta       = & -   \alpha \KTheta
\, \\
\label{appeq:dtKThetaADM_Gen}
\dif_{t} \KTheta =  & - \alpha  D^{i} D_{i}\Theta - D^{i}\alpha D_{i} \Theta
        \\
        & + \alpha\left( K \KTheta + V'(\Theta) - \frac{\aCSh M^2}{4}\dfth\CS \right)
\,\nonumber
\end{align}
\end{subequations}
where
$\dif_{\rm t} = \left(\p_{t} - \Lie_{\beta}\right)$ with $\Lie_{\beta}$ being the Lie derivative along the shift vector,
and we introduce the dimensionless parameter $\aCSh$ that allows us to switch
the coupling to the Pontryagin density on ($\aCSh=1$) and off ($\aCSh=0$).
This enables us to compare against the evolution of a massive scalar field in \GR{.}

In the decoupling limit,
the Pontryagin density is evaluated on the \GR background.
It is convenient to write the Pontryagin density in terms of the gravito-electric, $E_{ij}$, and gravito-magnetic, $B_{ij}$, components of the Weyl tensor
given by Eqs.~\eqref{eq:EijInGR} and \eqref{eq:BijInGR}.
Then, the Pontryagin density is
\begin{align}
    \CS=\,^*W_{abcd}W^{bacd} = -16E^{ij}B_{ij}
    \,.
\end{align}

To estimate the effect that the \dCS field and its nonminimal coupling would have onto the metric,
if back-reacted onto the spacetime,
we introduce the effective energy-momentum tensor
\begin{equation}\label{appeq:Tmneff}
    T^{\rm eff}_{ab}:=T^{\rm \Theta}_{ab}-2\aCSh M^2 \CCS^{\textrm{vac;GR}}_{ab}
    \,.
\end{equation}
In $3+1$ form, the energy-momentum tensor can be decomposed into
the energy density, $\rho^{\rm eff}:=n^an^bT_{ab}^{\rm eff}$,
the energy flux, $j_i^{\rm eff}:=-\gamma_i^an^bT_{ab}^{\rm eff}$,
and the spatial stress tensor, $S^{\rm eff}_{ij}:=\gamma_i^a\gamma_j^bT_{ab}^{\rm eff}$.
They are given by
\begin{subequations}
\label{appeq:SplitTmunuVac}
\begin{align}
\label{appeq:RhoEff}
   \rho^{\rm eff} = & \frac{1}{2}\left( K_{\Theta}^2 + 2 \Vth + D_i\Theta D^i\Theta \right) + 2\aCSh M^2 \left(B^{ij} \mathcal{F}_{ij}\right)
\,,\\
\label{eq:Ji}
   j_i^{\rm eff} = & K_{\Theta} D_i\Theta + 2\aCSh M^2 (B_{ij}\F^j - \epsilon_{ijk}E^{jl}\mathcal{F}_{l}{}^k)
\,,\\
\label{eq:Sij}
   S_{ij}^{\rm eff} = & D_i\Theta D_j\Theta + \frac{1}{2}\gamma_{ij}(K_\Theta^2 - D_k\Theta D^k\Theta - 2 \Vth)
   \nonumber \\
   & + 2\aCSh M^2 \left( 2\epsilon_{(i|kl}E_{|j)}{}^l\F^k - 2B_{(i}{}^k\F_{j)k} + \gamma_{ij} B^{kl}\mathcal{F}_{kl} \right.
   \nonumber\\
   & \left. + B_{ij} (\F_{nn} + {\rm tr}\F) \right)
   \,.
\end{align}
\end{subequations}
Here, we introduced the decomposition of the auxiliary tensor as
$\F_{nn}=\F_{ab}n^a n^b$,
$\F_{i}=-\gamma^{a}{}_{i}n^b\F_{ab}$
and
$\F_{ij}=\gamma^{a}{}_{i}\gamma^{b}{}_{j}\F_{ab}$,
and its trace ${\rm tr}\F = \gamma^{ij}\F_{ij}$.
They are explicitly given by
\begin{subequations}\label{eq:AuxF}
\begin{align}
\F_{nn} = & \left( D_kD^k\Theta - K K_\Theta - \dVth \right) \dfth \nonumber \\
        & + \frac{\aCSh M^2}{4}\CS \dfth^2 + K_\Theta^2 \ddfth
    \,,\\
\F_i = & (D_iK_\Theta - K_{ij}D^j\Theta) \dfth + K_\Theta D_i \ddfth
    \,,\\
\F_{ij} = & (D_iD_j\Theta - K_{ij}K_{\Theta})\dfth + D_i\Theta D_j \Theta \ddfth
    \,.
\end{align}
\end{subequations}
In the expression for $\F_{nn}$
we substituted the time derivate of $\KTheta$ with its evolution equation,
Eq.~\eqref{appeq:dtKThetaADM_Gen}.

For completeness,
we also provide the \BSSN formulation of the \dCS scalar's evolution equation and the energy density, flux and stress tensor
to be used in an implementation of the general \dCS equations.
In fact, these are the equations implemented in
\CanudadCS{.}
The \BSSN variables are given in Eqs.~\eqref{eq:BSSNvars}.
Note that, in the following, we introduce the tilde over variables to indicate that they are expressed in terms of the \BSSN variables.
The evolution equations, Eqs.~\eqref{appeq:dtThetaKTheta_Gen},
become
\begin{subequations}
\label{eq:dtdCSKG_decoupBSSN_Gen}
\begin{align}
\label{eq:dtThetaBSSN}
\dif_{\rm t} \Theta   = & -\alpha \KTheta
\,,\\
\label{eq:dtKThetaBSSN}
\dif_{\rm t} \KTheta  = & - W^2 \tD^{i}\alpha \tD_{i} \Theta - \alpha  \left(W^2 \tD^{i} \tD_{i}\Theta - W \tD^i \Theta \tD_i W \right. \nonumber \\
& \left. - K \KTheta - \dVth + \frac{\aCSh M^2}{4}\CS \dfth\right)
\,,
\end{align}
\end{subequations}
and indices are raised with the conformal metric $\tilde{\gamma}_{ij}$.
In \BSSN variables, the Pontryagin density $\CS$ is given by
\begin{align}
\CS = & - 16 \tgam^{ia}\tgam^{jb}\tE_{ab} \tB_{ij}
\end{align}
with
\begin{subequations}
\label{eq:tEtBinBSSN}
    \begin{align}
    \label{eq:tE_BSSN}
    \tE_{ij} = & W^2 E_{ij}
    \\ = &
        W^2R^{\rm tf}_{ij} + \frac{1}{3} \tA_{ij} K - \tA_{i}{}^{k}\tA_{jk} + \frac{1}{3}\tgam_{ij} \tA_{kl}\tA^{kl}
    \,,\nonumber\\
    \label{eq:tB_BSSN}
    \tB_{ij} = & W^2 B_{ij}
    \\ = &
        -W\teps_{(i|}{}^{kl}\tD_{l}\tA_{|j)k} -\teps_{(i|}{}^{kl}\tA_{|j)l}\tD_{k}W
    \,.\nonumber
    \end{align}
\end{subequations}
Note that we insert the BSSN variables into our expressions for $B_{ij}$ and $E_{ij}$, and then rescale them by $W^2$ for convenience.

Similarly, we substitute the \BSSN variables into the
energy density, flux and spatial stress tensor~\eqref{appeq:SplitTmunuVac}
to find
\begin{subequations}
\label{appeq:SplitTmunuBSSN}
\begin{align}
\label{appeq:RhoEffBSSN}
  \rho^{\rm eff} = & \frac{1}{2}\left(K_{\Theta}^2+2\Vth+ W^2(\tD_i\Theta)(\tD^i\Theta)\right)
  \nonumber\\ &
  + 2\aCSh M^2 (\tB^{ij}\tF_{ij})
\,,\\
\label{appeq:JiEffBSSN}
   j_i^{\rm eff} = & K_{\Theta} \tD_i\Theta + 2\aCSh M^2 \left(\tB_{ij}{\tilde{\F}^j} - \frac{1}{W}\teps_{ijk}\tE^{jl}\tF_{l}{}^k\right)
\,,\\
\label{appeq:SijEFFBSSN}
   S_{ij}^{\rm eff} = & \tD_i\Theta \tD_j\Theta + \frac{1}{2W^2}\tgam_{ij}(K_\Theta^2 - W^2\tD_k\Theta \tD^k\Theta - 2 \Vth)
   \nonumber \\
   & + \frac{2\aCSh M^2}{W^2} \left( 2W\teps_{(i|kl}\tE_{|j)}{}^l{\tilde{\F}^k} - 2\tB_{(i}{}^k\tF_{j)k}
   \right.\nonumber\\
   & \left.
   + \tgam_{ij} \tB^{kl}\tF_{kl}
   + \tB_{ij} (\F_{nn} + {\rm tr}\F) \right)
\,.
\end{align}
\end{subequations}

The auxiliary tensors become
\begin{subequations}\label{eq:AuxFBSSN}
\begin{align}
\F_{nn} = & \dfth
\Bigl(-K K_\Theta + W^2\tD_k\tD^k\Theta - W\tD_k\Theta \tD^k W
\nonumber\\
&  - \dVth\Bigr)
        +\frac{\aCSh M^2}{4}\CS \dfth^2 +{K_\Theta}^2\ddfth
    \,,\\
\F_i = & (\tD_i K_\Theta) \dfth - (\tA_{ij}+\frac{1}{3}K\tgam_{ij})(\tD^j\Theta)\dfth \nonumber \\ &
    + K_\Theta(\tD_i\Theta)\ddfth
    \,,\\
\tF_{ij} = & W^2 \mathcal{F}_{ij} \nonumber \\
          =  & -(\tA_{ij}+\frac{1}{3}K\tgam_{ij})K_\Theta\dfth + W^2(D_iD_j\Theta)\dfth \nonumber \\
          & + W^2(\tD_i\Theta)(\tD_j\Theta)\ddfth
    \,,
\end{align}
\end{subequations}
where we have rescaled $\F_{ij}$ by $W^2$ for convenience
and introduced the rescaled quantity $\F^i=W^2\tilde{\F}^i$
. Finally, the trace of the spatial projection is given by ${\rm tr}\F = \gamma^{ij}\F_{ij} =\tgam^{ij}\tF_{ij}$
and
\begin{eqnarray}
    D_iD_j\Theta &=&
        \frac{\tD_k\Theta}{W}\left({\tgam^k}_i(\tD_j W) + {\tgam^k}_j(\tD_i W) - \tgam_{ij} (\tD^kW) \right)
        \nonumber \\ & &
        + \tD_i\tD_j\Theta
        \,.
\end{eqnarray}

\section{Snapshots of numerical simulations}~\label{app:snapshots}
\begin{figure*}[h!]
    \centering
  \includegraphics[width = 0.95\textwidth]{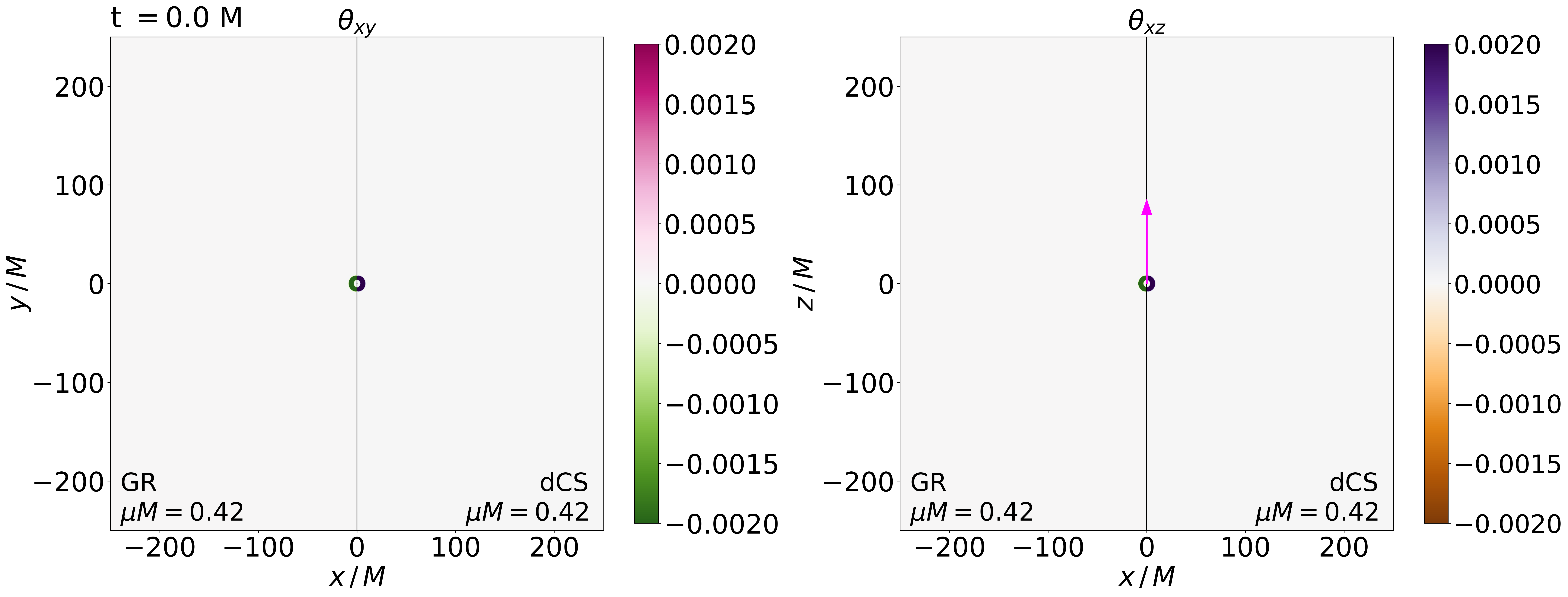}
  \\
  \includegraphics[width = 0.95\textwidth]{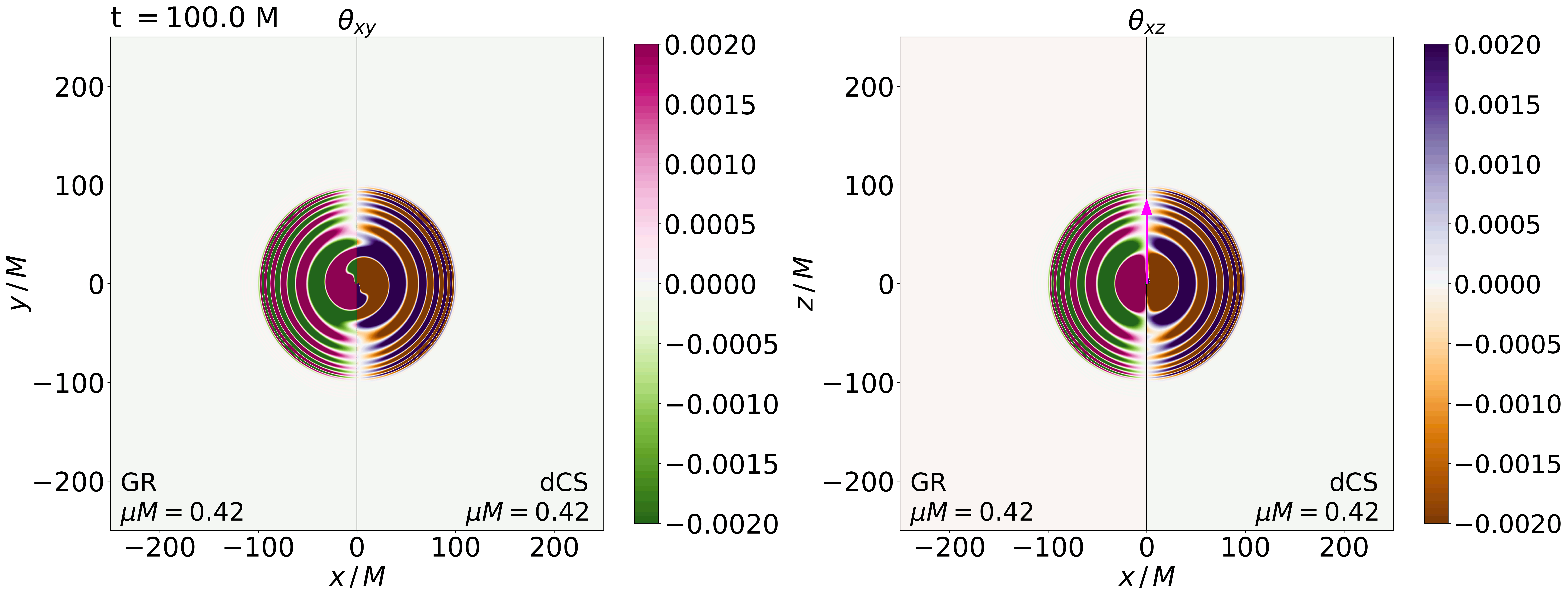}
  \\
  \includegraphics[width = 0.95\textwidth]{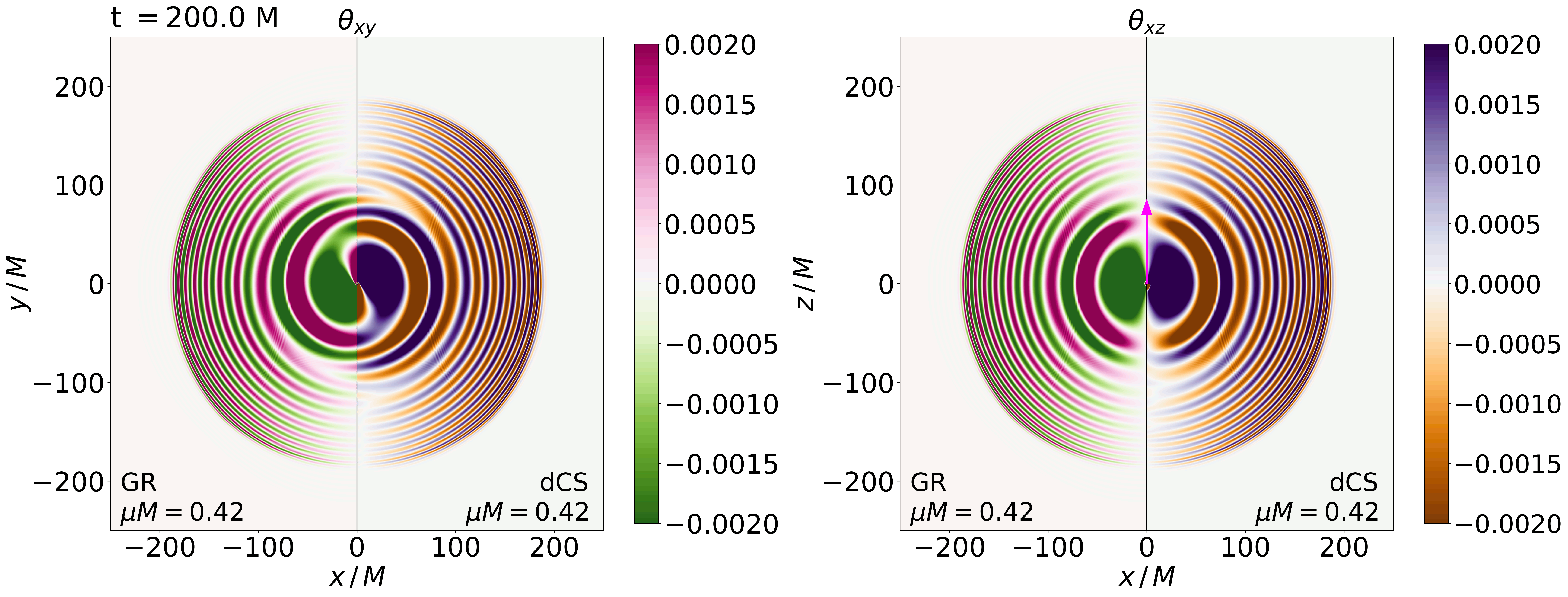}
    \caption{\textbf{\GR vs massive \dCS}. Snapshots of massive scalar field $\Theta$ with $\mu M = 0.42$ coupled to a BH with $a/M = 0.99$ in the $xy$ plane (left) and $xz$ plane (right) where z is the axis of rotation.}\label{fig:snaps_GRvsmdCS}
\end{figure*}

\begin{figure*}[h!]
    \centering
   \includegraphics[width = 0.95\textwidth]{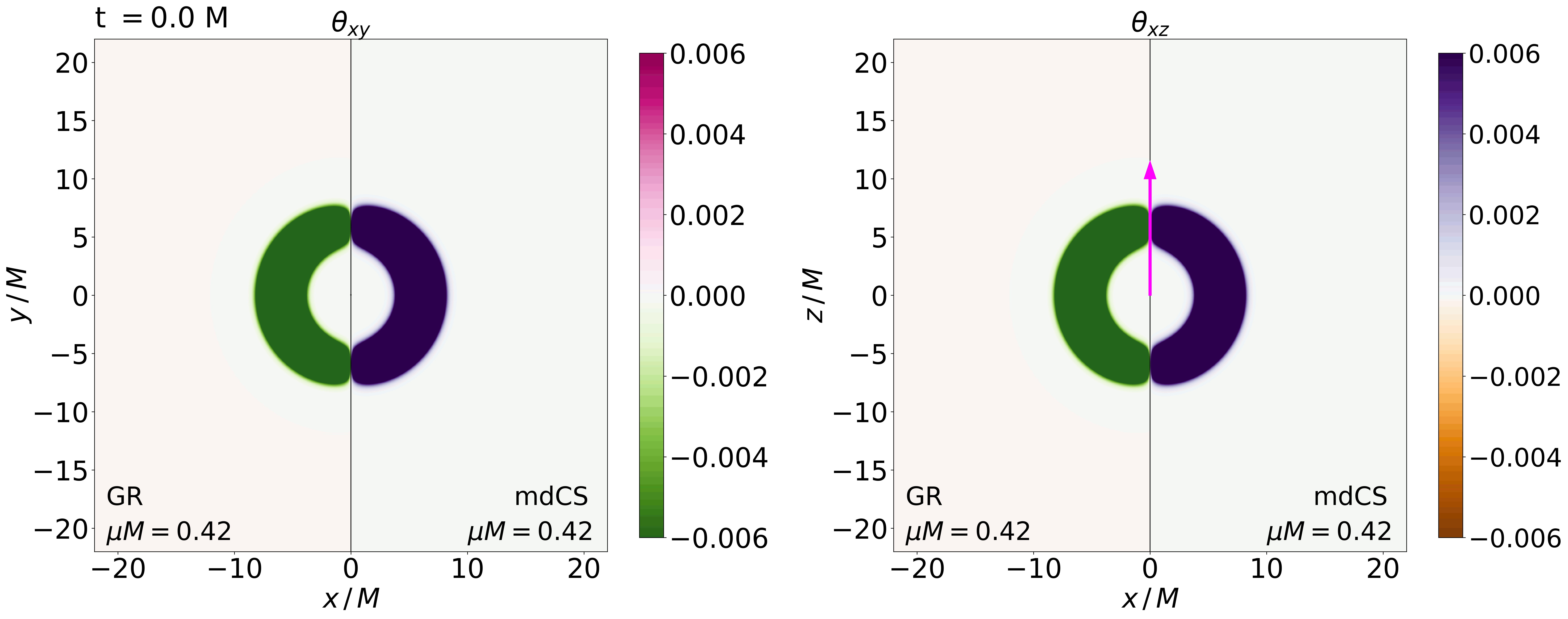}
   \\
   \includegraphics[width = 0.95\textwidth]{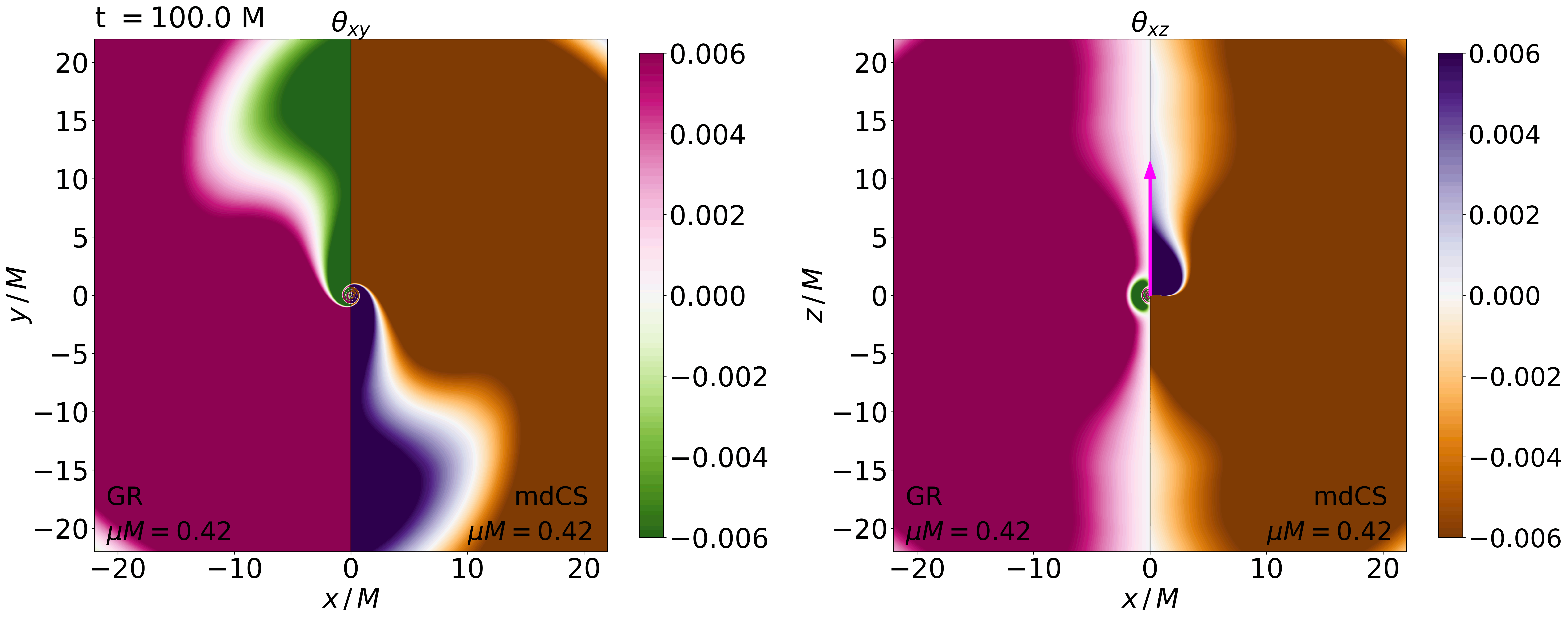}
   \\
   \includegraphics[width = 0.95\textwidth]{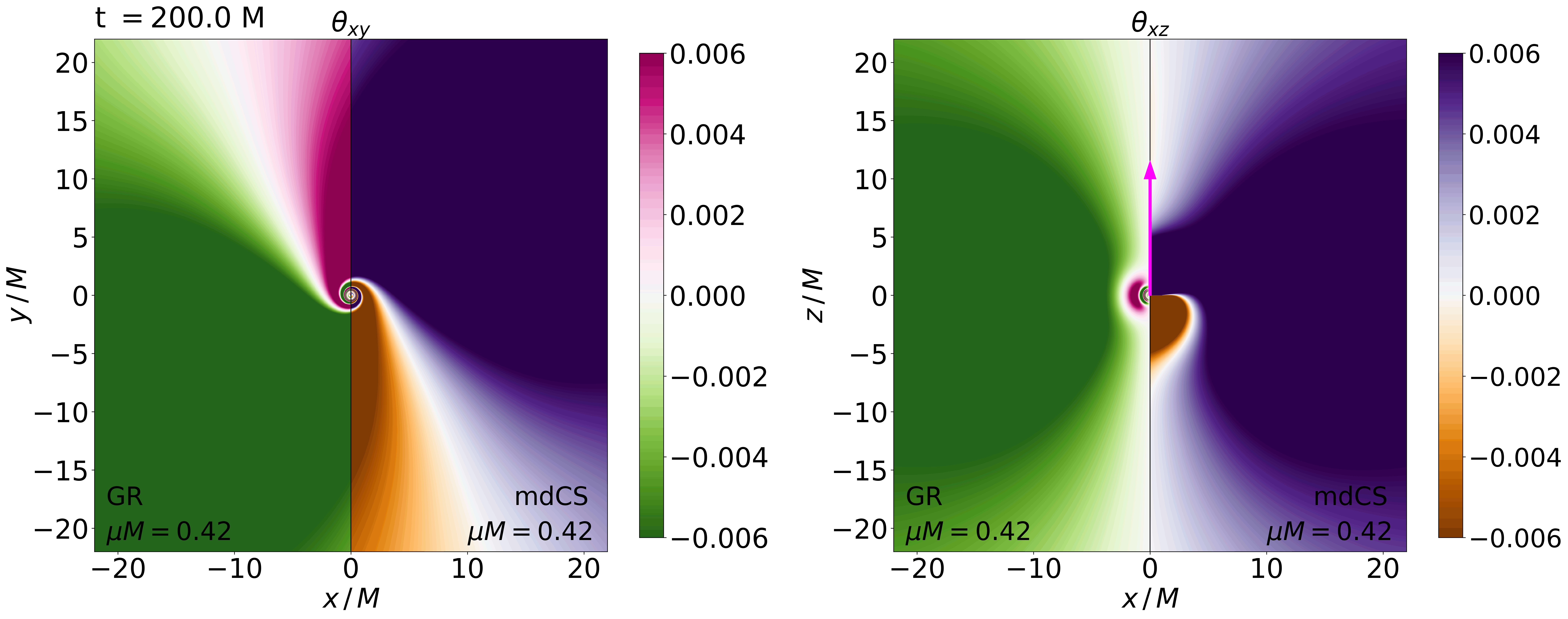}
    \caption{\textbf{\GR vs massive \dCS \,-- closeup}. Snapshots of massive scalar field $\Theta$ with $\mu M = 0.42$ coupled to a BH with $a/M = 0.99$ in the $xy$ plane (left) and $xz$ plane (right) where z is the axis of rotation.
    }\label{fig:snaps_GRvsmdCS_closeup}
\end{figure*}

\begin{figure*}[h!]
    \centering
   \includegraphics[width = 0.95\textwidth]{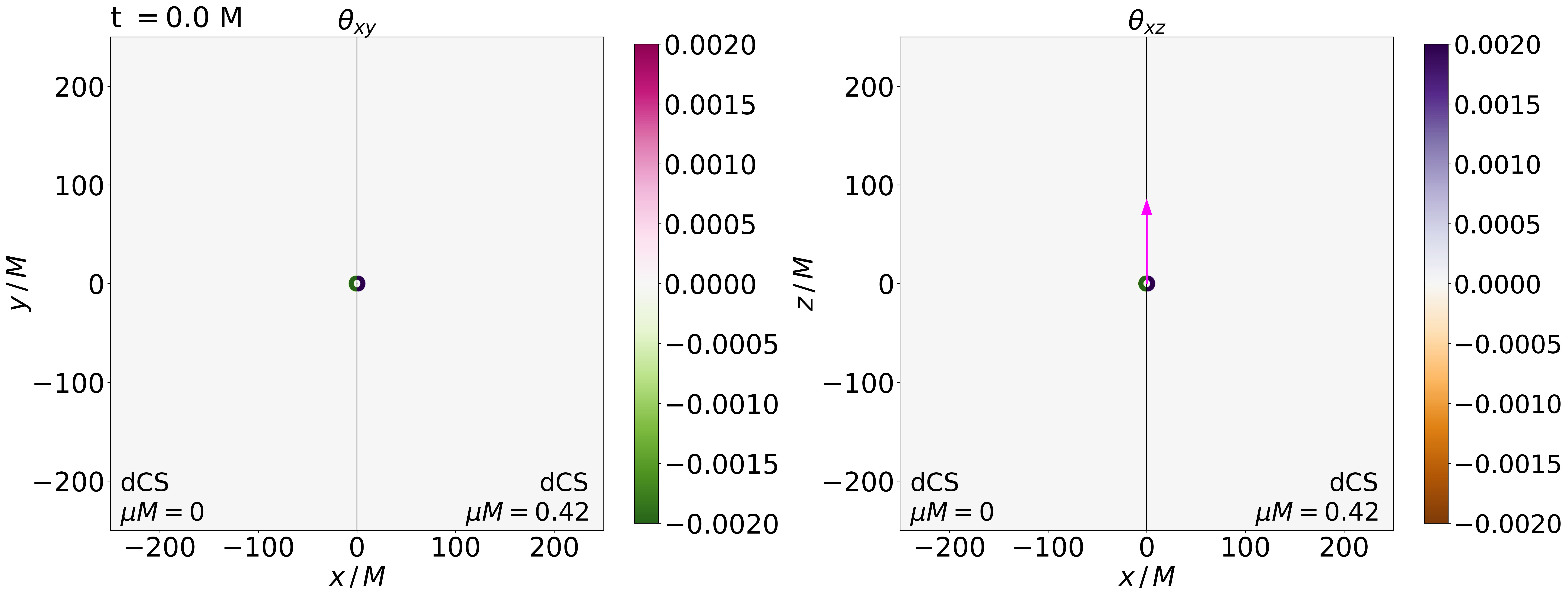}
   \\
   \includegraphics[width = 0.95\textwidth]{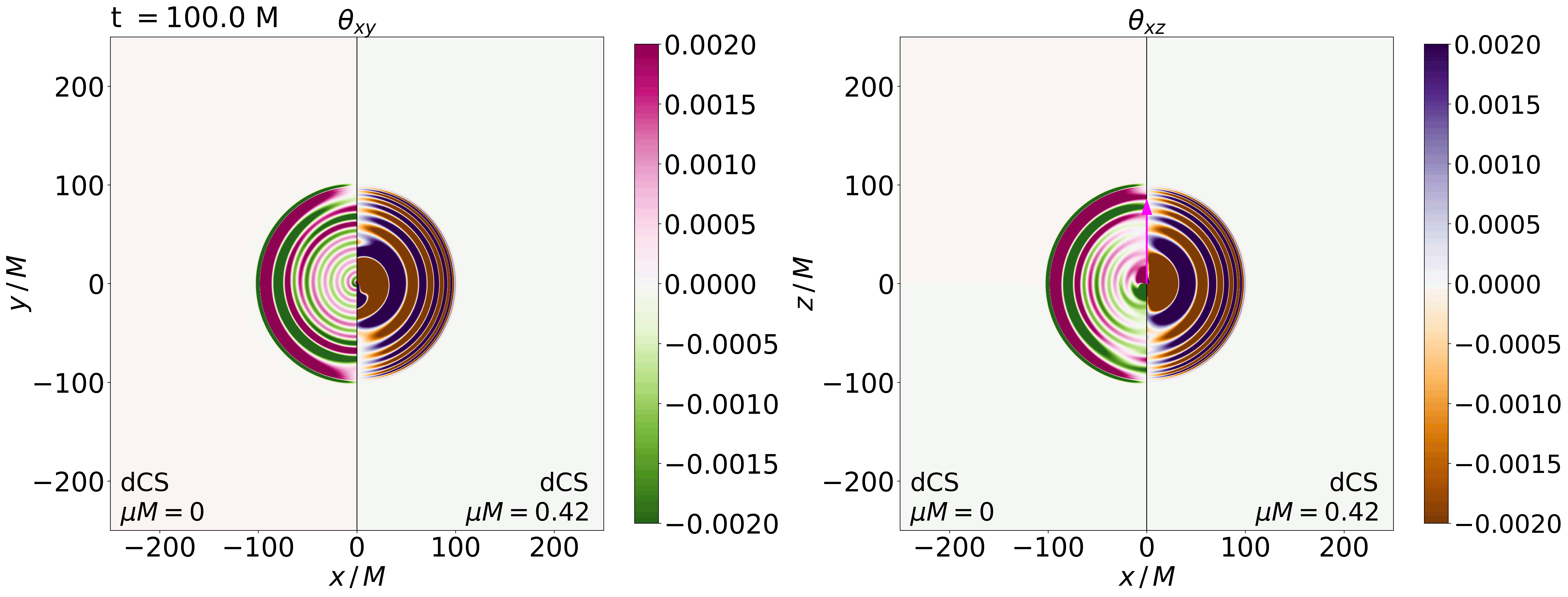}
   \\
   \includegraphics[width = 0.95\textwidth]{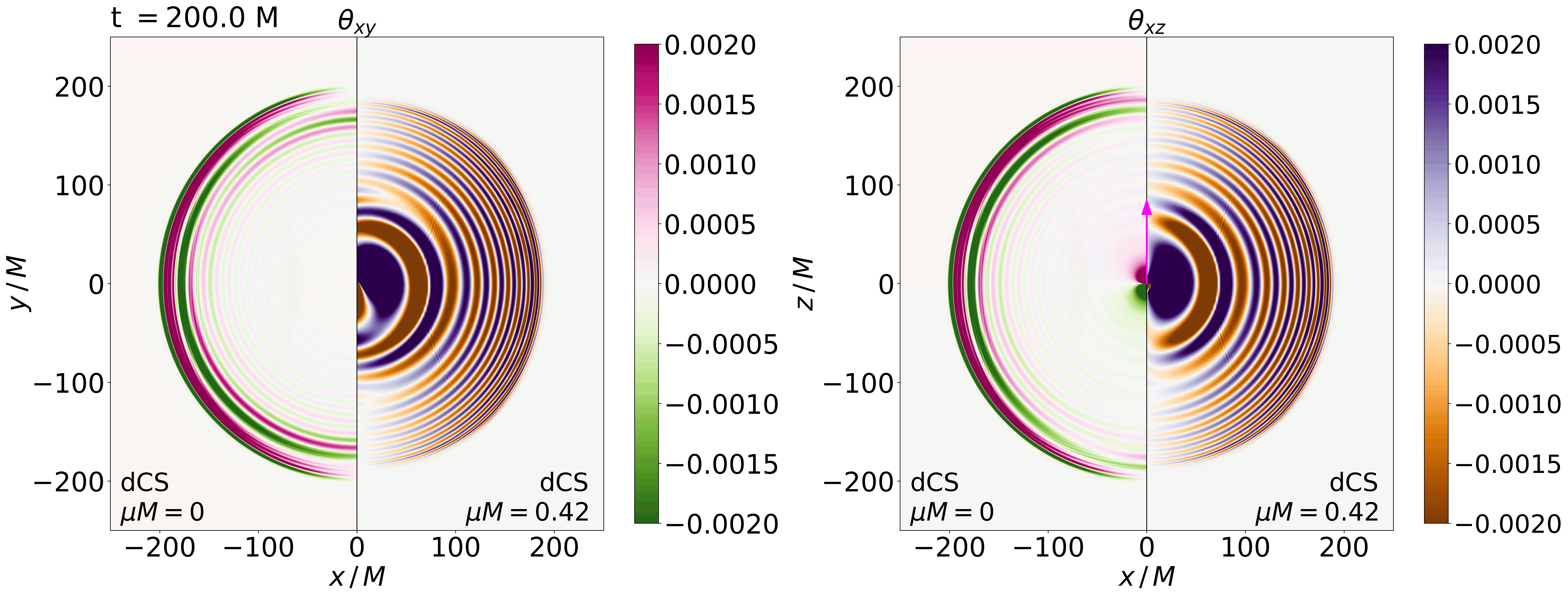}
    \caption{\textbf{\dCS vs massive \dCS}. Snapshots of massive scalar field $\Theta$ with $\mu M = 0.42$ coupled to a BH with $a/M = 0.99$ in the $xy$ plane (left) and $xz$ plane (right) where z is the axis of rotation.}\label{fig:snaps_dCSvsmdCS}
\end{figure*}

\begin{figure*}[h!]
    \centering
    \includegraphics[width = 0.9\textwidth]{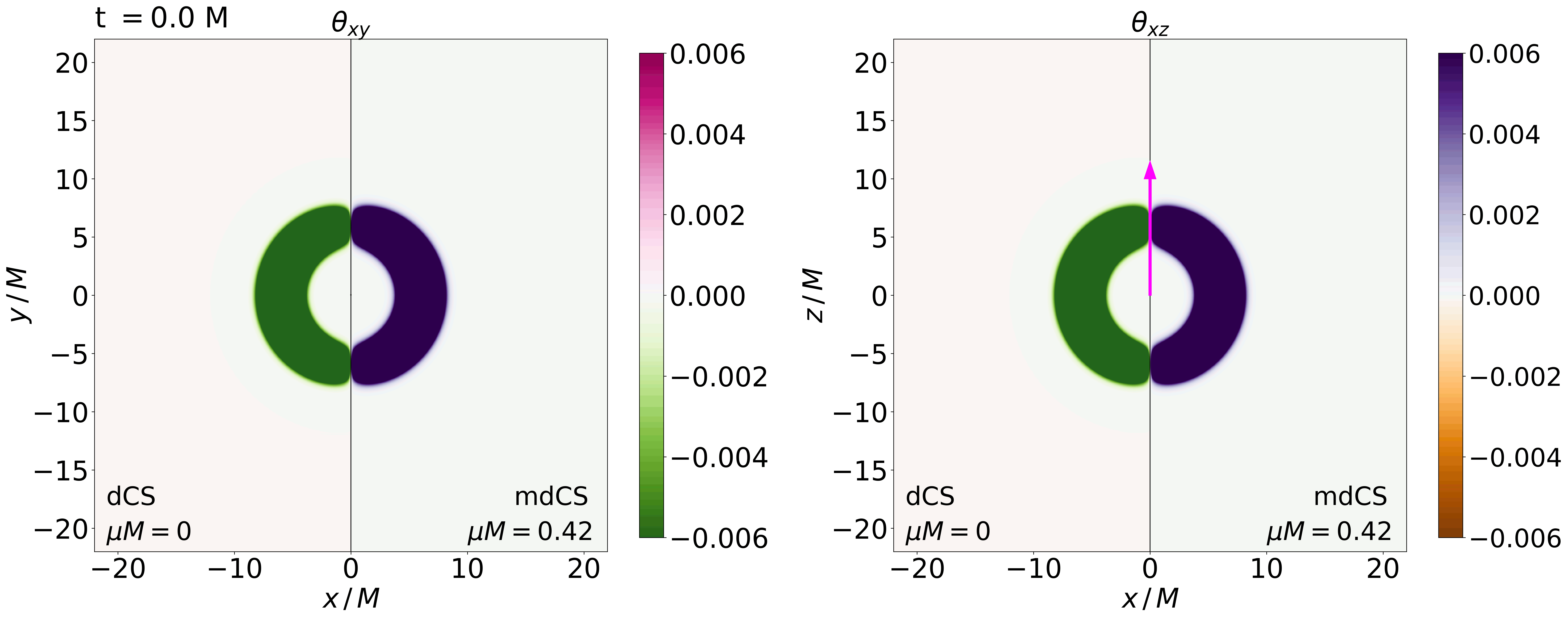}
    \\
    \includegraphics[width = 0.9\textwidth]{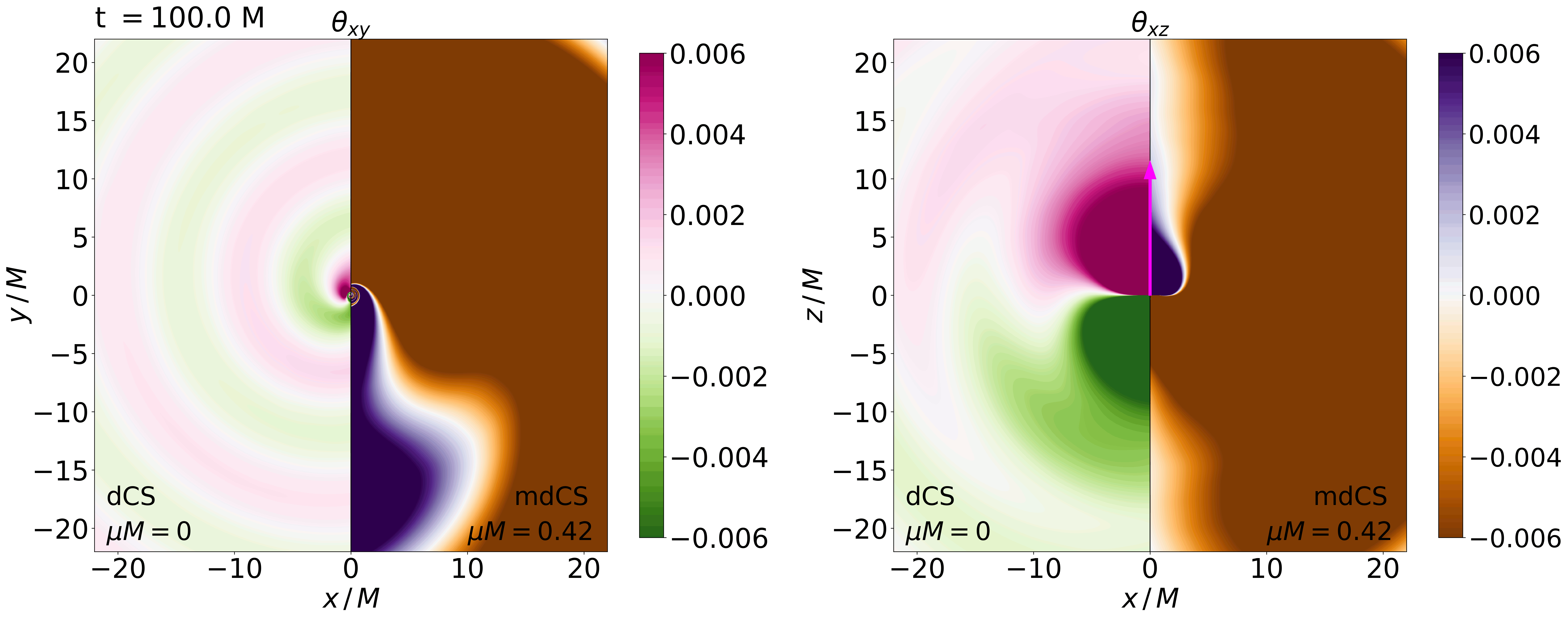}
    \\
    \includegraphics[width = 0.9\textwidth]{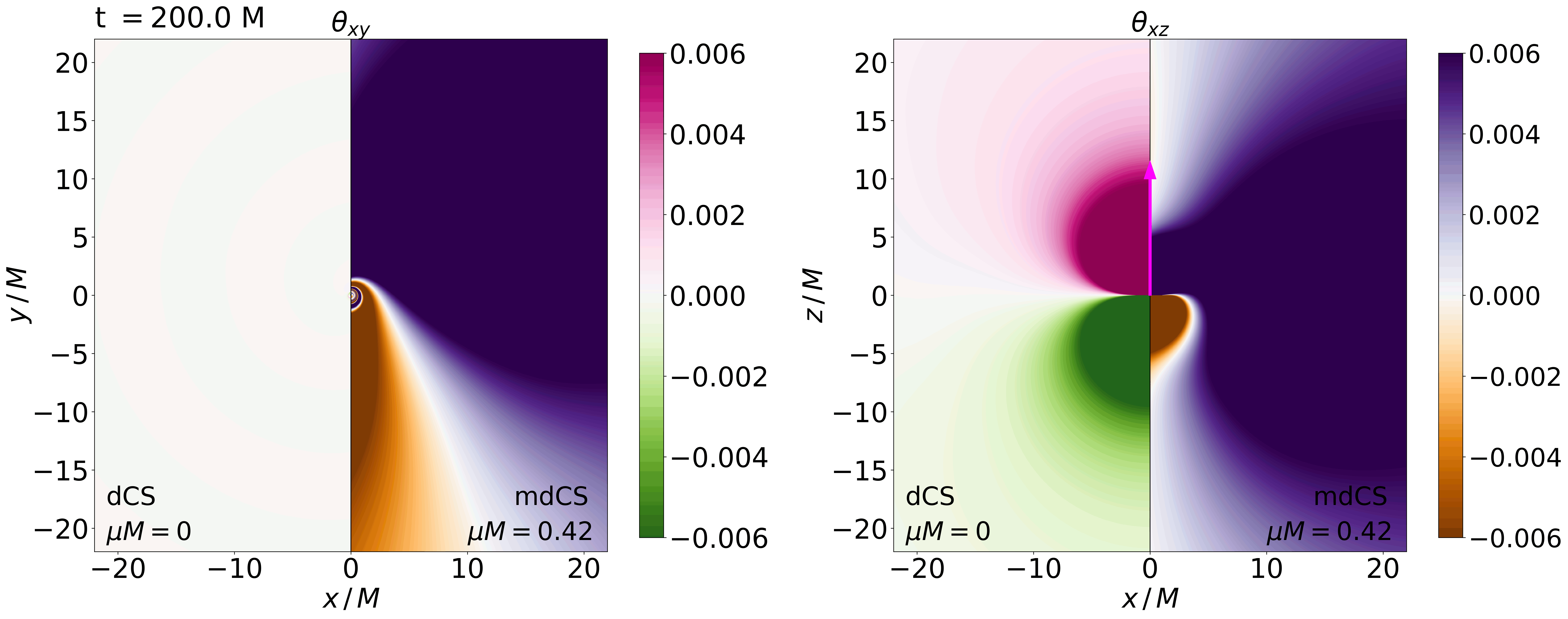}
    \caption{\textbf{\dCS vs massive \dCS \,-- closeup}. Snapshots of massive scalar field $\Theta$ with $\mu M = 0.42$ coupled to a BH with $a/M = 0.99$ in the $xy$ plane (left) and $xz$ plane (right) where z is the axis of rotation.
    }\label{fig:snaps_dCSvsmdCS_closeup}
\end{figure*}

In this section we present 2D snapshots of our simulations to depict the growth of the scalar field around a \bh with spin $a/M=0.99$
at times $t/M=\{0,100,200\}$;
see Figs.~\ref{fig:snaps_GRvsmdCS}, ~\ref{fig:snaps_GRvsmdCS_closeup},~\ref{fig:snaps_dCSvsmdCS} and ~\ref{fig:snaps_dCSvsmdCS_closeup}.
These snapshots are complementary to those shown in Sec.~\ref{Sec:results}
(namely Figs.~\ref{fig:xyxz_t250_GRvsmdCS_out},~\ref{fig:xyxz_t250_GRvsmdCS_in},~\ref{fig:xyxz_t250_dCSvsmdCS_out},~\ref{fig:xyxz_t250_dCSvsmdCS_in})
that displayed the late-time evolution of the field.
In each figure we present a set of frames with subpanels.
The frames on the left-hand side show the amplitude of the scalar in the equatorial (i.e., x-y) plane while the frames on the right-hand side correspond to the x-z plane (along the rotation axis).

In Figs.~\ref{fig:snaps_GRvsmdCS} and~\ref{fig:snaps_GRvsmdCS_closeup},
we present the snapshots of a massive scalar field with $\mu M=0.42$ evolving around a \bh with spin $a/M=0.99$
to demonstrate how the initial configuration slowly grows into the oscillating scalar cloud presented in Fig.~\ref{fig:xyxz_t250_GRvsmdCS_out}.
The frames are in the order described above.
The left panel of each frame corresponds to a massive scalar field minimally coupled to gravity, i.e., in \GR{.}
The right panel of each frame depicts the massive \dCS scalar.
At large scales, one can hardly observe any difference between the two simulations.
On the other hand, the close-up in Fig.~\ref{fig:snaps_GRvsmdCS_closeup} shows how an axis-symmetric dipole forms early on in massive \dCS.
The growth of this dipole is absent in \GR and qualitatively resembles the dipolar scalar hair that \bh{s} grow in massless \dCS.
Thus, we conclude that the dipole configuration is due to the nonminimal coupling between the massive scalar and gravity via the Pontryagin density.
The latter
sources the axis-symmetric scalar modes close to the \bh, where the curvature is strongest.

In Figs.~\ref{fig:snaps_dCSvsmdCS} and ~\ref{fig:snaps_dCSvsmdCS_closeup}, we compare snapshots of a massless (left panels in each frame) and a massive (right panels in each frame) \dCS field.
In the massless case, we observe how the initial Gaussian perturbation with an $l=m=1$ profile
propagates outwards while a static, axi-symmetric dipole hair is formed.
In the massive case, instead, the mass term traps scalar multipoles with
$\bar{\omega} \lesssim \mu$
in the vicinity of the \bh{,}
and thus induces the formation of an oscillating scalar cloud that connects smoothly to the inner dipole.

The full animations made with these snapshots can be found at~\cite{YoutubeLinkmdCS}.

\clearpage
\bibliographystyle{apsrev4-2}
\bibliography{Manuscript_MassiveDCSGravity.bib}

\end{document}